\documentclass{ws-rv9x6}  
\usepackage{amssymb}
\usepackage{graphicx}
\newcommand{\beq}{\begin{eqnarray}}
\newcommand{\eeq}{\end{eqnarray}}
\renewcommand{\theequation}{\thesection.\arabic{equation}}
\long\def\comment#1{ }    
\newcommand{\be}{\begin{eqnarray}}
\newcommand{\ee}{\end{eqnarray}}

\newcommand{\nn}{\nonumber\\ }
\renewcommand{\theequation}{\thesection.\arabic{equation}}
\newcommand{\labe}{\label}

\def\simge{\mathrel{%
   \rlap{\raise 0.511ex \hbox{$>$}}{\lower 0.511ex \hbox{$\sim$}}}}
\def\simle{\mathrel{
   \rlap{\raise 0.511ex \hbox{$<$}}{\lower 0.511ex \hbox{$\sim$}}}}
\def\bigs{\mathrel{
   \rlap{\raise 0.531ex \hbox{$>$}}{\lower 0.531ex \hbox{$<$}}}}

\def\grad{\nabla}                               
\def\del{\partial}                              

\newcommand{\kk}{k_\perp}
\newcommand{\xx}{{\rm x}}
\newcommand{\y}{{\rm y}}
\newcommand{\munu}{\mu\nu}
\newcommand{\bb}{b_\perp}
\newcommand{\rr}{r_\perp}

\newcommand{\Tr}{\rm Tr}
\newcommand{\xp}{x_{\cal P}}

\begin{document}


\chapter*{
The Color Glass Condensate and High Energy Scattering in QCD}

\author{Edmond Iancu}
\address{Service de Physique Th\'eorique, CEA Saclay\,,\\
        91191 Gif-sur-Yvette cedex, France}


\author{Raju Venugopalan}
\address{Physics Department, Brookhaven National Laboratory\\
         and RIKEN-BNL Research Center, Upton, NY 11973, USA}




\comment{

\title{The Color Glass Condensate and High Energy Scattering in QCD}

\author{Edmond Iancu
 \\
 {\small\it Service de Physique Th\'eorique, CEA Saclay\,,}\\
 {\small\it 91191 Gif-sur-Yvette cedex, France}\\
 \\
 Raju Venugopalan\\
 {\small\it Physics Department and RIKEN-BNL Research Center,}\\
 {\small\it Brookhaven National Laboratory, Upton, NY 11973, USA}}

\maketitle
}


\abstract{At very high energies or small values of Bjorken x, the density of partons, 
per unit transverse area, in hadronic 
wavefunctions becomes very large leading to a saturation of partonic distributions. When the 
scale corresponding to the density per unit transverse area, the saturation scale $Q_s$, becomes 
large ($Q_s\gg \Lambda_{QCD}$), the coupling constant becomes weak ($\alpha_S(Q_s)\ll 1$) 
which suggests that the high energy limit of QCD may be studied using weak coupling techniques.
This simple idea can be formalized in an effective theory, the Color Glass Condensate (CGC), 
which describes the behavior of the small x components of the hadronic wavefunction in QCD. 
The Green functions of the theory satisfy Wilsonian renormalization group equations which reduce to 
the standard linear QCD evolution equations in the limit of low parton densities. The effective theory 
has a rich structure that has been explored using analytical and numerical techniques. The CGC can be  
applied to study a wide range of high energy scattering experiments from Deep Inelastic Scattering at 
HERA and the proposed Electron Ion Collider (EIC) to proton/deuterium-nucleus and nucleus-nucleus 
experiments at the RHIC and LHC colliders. 
}
\newpage
\tableofcontents

\section{Outstanding Phenomenological Questions in High Energy QCD}
\setcounter{equation}{0}
 \subsection{Introduction}

The advent of a new generation of high energy collider experiments, beginning with HERA and the Tevatron in the early 
1990's to RHIC and LHC in the new millenium (and others in the planning stages) have launched a new era in the 
study of the strong interactions. Questions which have been around since the early days of the strong interactions, such 
as the behavior of cross-sections at high energies, the universality of hadronic interactions at high energies, 
the nature of multi-particle production and the possibility of creating thermalized states of strongly interacting 
matter have acquired 
fresh vigor. For instance, it is often believed that little could be learned about the high energy limit of 
QCD since the physics is assumed to be entirely non-perturbative. On the other hand, we have learned from HERA that 
parton densities are large at high energies or equivalently, at small values of Bjorken x. 
The large densities of small x, or ``wee'' partons suggest, as we will discuss at length later, 
that semi-hard scales may be present which allow one to describe the physics of this regime using weak coupling 
techniques. QCD at high energies can thus be described as a many-body theory of partons which are weakly coupled albeit 
non-perturbative due to the large number of partons. We will call this system a Color Glass Condensate (CGC), for the following reasons:

\noindent
$\bullet$ ``Color'', since the gluons are colored.

\noindent
$\bullet$ ``Glass'' because of the strong analogy of the system to actual glasses. A glass is a disordered system
which evolves very slowly relative to natural time scales: it is like a solid on short time scales and like a liquid 
on much longer time scales. Similarly, the partons of interest are disordered and evolve in longitudinal momentum 
in a manner analogous to a glass. 

\noindent 
$\bullet$ ``Condensate'' because it contains a very high 
density of massless gluons whose momenta are 
peaked about some characteristic momentum. Increasing the energy forces 
the gluons to occupy higher momentum states (due to repulsive interactions) causing the coupling to become weaker. 
The gluon density saturates at a value of order $1/\alpha_s \gg 1$, corresponding to a multiparticle Bose 
condensate state.

We will argue in the following that the Color Glass Condensate is the effective theory describing high energy 
scattering in QCD. We will outline the rich structure of the theory and discuss how it provides insight into 
outstanding conceptual issues in QCD at asymptotically high energies. The theory can be applied to study 
phenomena at a number of existing and upcoming high energy collider facilities. 
We will discuss applications of the CGC to study the initial conditions and equilibration in heavy ion 
collisions and to describe heavy ion phenomenology. 
We will discuss applications to Deep Inelastic Scattering, to the current experiments at HERA but also 
for future experiments being discussed at HERA and at Brookhaven. In addition, we will discuss applications of 
the CGC to proton/deuteron-nucleus scattering experiments planned at RHIC and at the LHC and to 
peripheral nucleus-nucleus scattering experiments at RHIC.

  \subsection{Light cone kinematics and dynamics}

The appropriate kinematics to discuss high energy scattering are light-cone (LC)
coordinates. Let $z$ be the longitudinal 
axis of the collision. For an arbitrary 4-vector
$v^\mu=(v^0,v^1,v^2,v^3)$ ($v^3=v_z$, etc.), 
we define its LC coordinates as
\be
v^+\equiv {1 \over \sqrt{2}} (v^0+v^3),\qquad
v^-\equiv {1 \over \sqrt{2}} (v^0-v^3),\qquad v_\perp\equiv (v^1,v^2).
\ee
In particular, we shall refer to $x^+=(t+z)/\sqrt{2}$ 
as the LC ``time'', and to $x^-=(t-z)/\sqrt{2}$ 
as the LC ``longitudinal coordinate''. The
invariant scalar product of two four-vectors reads:
\be
        p \cdot x \,= \,p^- x^++ p^+x^-  - p_\perp \cdot x_\perp\,,
\ee
which suggests that $p^-$ should be
interpreted as the LC energy,
and $p^+$ as the (LC) longitudinal momentum. In particular,
since $p^\pm  = (1/ \sqrt{2})(E \pm p_z)$
with $E=(m^2+{\bf p}^2)^{1/2}$, the light cone dispersion relation takes the form
\be
p^- \,= {1 \over 2} \frac{ p_\perp^2 +m^2}{p^+} \,= \,
 {1 \over 2} \, \frac{m_\perp^2}{p^+} \, ,
\label{dispersion}
\ee 
where the transverse mass $m_\perp$ is defined as $m_\perp^2=p_\perp^2+m^2$. 
The momentum space rapidity is further simply as: 
\be\label{y-DEF}
        y \,\equiv\, {1 \over 2} 
\ln\frac{p^+}{p^-} = {1 \over 2} \ln{2p^{+2}\over m_\perp^2}.
\ee
These definitions are useful, among other reasons, because of their simple
properties under longitudinal Lorentz boosts:  $p^+ \rightarrow
\kappa p^+$, $p^- \rightarrow (1/\kappa)p^-$,
where $\kappa$ is a constant.  Under boosts,
the rapidity is just shifted by a constant: $y\rightarrow
y+ \kappa$.  

The utility of light cone kinematics is not merely that of a convenient coordinate transformation. 
The Hamiltonian dynamics of 
quantum field theories quantized on the light cone have several remarkable features as was first 
pointed out by Dirac~\cite{Dirac}. Firstly, the LC Hamiltonian $p^-$ (which is the generator of 
translations in the light cone time $x^+$) can be written in the form $p^-=p^{-}_0 + V$, 
where $p^{-}_0$ is the free
Hamiltonian, corresponding to a complete set of non-interacting Fock eigenstates, and 
$V$ is the light cone potential. Second, the LC vacuum is trivial, namely, the vacuum state is an 
eigenstate of both the free and the full Hamiltonian. As a consequence of these properties, multi-parton 
Fock states can be constructed as eigenstates of the QCD Hamiltonian. 
Thus, in LC quantization (and in the light-cone gauge $A^+=0$)
the quark-parton picture of QCD becomes manifest. Finally, we note that 
the apparent non-relativistic structure of the light cone 
Hamiltonian suggested by the dispersion relation in Eq.~(\ref{dispersion}) is not accidental 
but is a consequence of an exact isomorphism between the Galilean subgroup 
of the Poincar{\'{e}} group and the symmetry group of two dimensional quantum mechanics~\cite{Susskind}. Thus in LC quantization, the Rayleigh-Schr{\"{o}}dinger 
perturbation theory with off-shell energy denominators 
can be used instead of the more usual Feynman rules. 
For a more detailed discussion of the light cone 
formalism and its application to high energy scattering, 
we direct the reader to Ref.~\cite{lightcone}.

 \subsection{High energy behavior of total cross-sections}

We now return to the outstanding phenomenological questions we mentioned in the introduction.
Clearly, computing total cross-sections as $E \rightarrow \infty$ is one of the
great unsolved problems of QCD.  
Unlike processes which are computed in perturbation theory,
it is not required that any energy transfer become large as the total 
collision energy $E \rightarrow \infty$.  Computing a total cross-section for 
hadronic scattering therefore appears to be an intrinsically non-perturbative procedure.
In the 60's and early 70's, Regge
theory was extensively developed in an attempt to understand the total
cross-section.  The results of these analyses were, to our mind,
inconclusive, and at any rate, certainly cannot be claimed to be understood from 
first principles in QCD.

On the basis of very general arguments invoking unitarity, analyticity and crossing, 
Froissart has shown that the total cross-section 
for the strong interactions grows at most as fast as $\ln^2
E$ as $E \rightarrow \infty$~\cite{Froissart,Martin}. 
Several questions arise in this regard. Is the coefficient of $\ln^2 E$ universal
for all hadronic processes?  Can this coefficient be computed from first principles in 
QCD? How do we understand the saturation of the unitarity limit dynamically in QCD?
Or is the Froissart bound an intrinsically non-perturbative phenomenon?

  \subsection{Multi-particle production in QCD}

\begin{figure} 
\begin{center} 
\includegraphics[width=0.5\textwidth]
{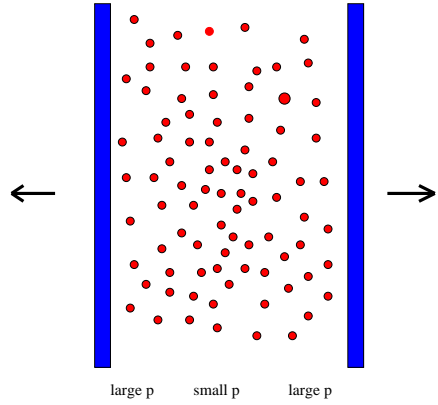} 
\caption{A hadron-hadron collision.  The produced particles are shown
as circles. } 
\label{collision} 
\end{center}
\end{figure}

Can we compute $N(E)$, the total multiplicity of
produced particles as a function of energy in QCD ? 
By this we mean not only the multiplicity of particles in jets (which is 
fairly well understood in perturbative QCD) but also the total number of
particles, at least, for semi-hard momenta.
Consider the collision of two identical hadrons
in the center of mass frame, as shown in Fig. 
\ref{collision}.  The colliding hadrons are ultrarelativistic and 
therefore Lorentz contracted in the direction of their motion.
Furthermore, we assume that the typical transverse momenta of the produced particles is
large compared to $\Lambda_{QCD}$. We know from experiments that the leading 
particles (valence partons) typically lose only some finite fraction of their momenta in the collision.
The produced particles, which are mostly mesons, are produced in the ``wake'' 
of the nuclei as they pass through each other.  
In light cone coordinates, the right moving particle (``the projectile'') has 
a 4-momentum $p_1^\mu=(p_1^+,p_1^-,0_\perp)$
with $p^+_1\simeq \sqrt{2} p_z$  and
$p_1^- \simeq M^2/2\sqrt{2}p_z$ 
(since $p_z\gg M$ and $m_\perp=M$, with $M=$ the projectile mass). 
Similarly, for the left moving hadron (``the target''),
we have $p^+_2 = p^-_1$ and $p^-_2 = p^+_1$.
The invariant energy squared is $s=(p_1+p_2)^2=2p_1\cdot p_2
\simeq 2 p^+_1 p^-_2 \simeq 4p_z^2$, and coincides, at it should, with
the total energy squared $(E_1+E_2)^2$ in the center of mass frame.

\begin{figure} 
\begin{center} 
\includegraphics[width=0.5\textwidth]
{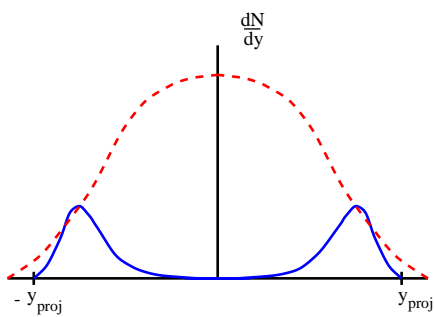} 
\caption{The rapidity distribution of particles produced in a hadronic 
collision. } 
\label{dndy} 
\end{center}
\end{figure}

Consider a pion produced in this collision and which is moving
in the positive $z$ direction. For such a pion,
we define the longitudinal momentum fraction,
or Feynman $\xx$, as :
\be
       \xx \,\equiv\,\frac{p^+_\pi}{p_1^+}\, \qquad {\rm (right\,\,mover)},
\ee
which implies ${m_\perp}/{\sqrt{2}p_1^+} \le \xx \le 1$.
The {\it rapidity} of the pion is then
\be
        \y \,=\, {1 \over 2} 
\ln{p^+_\pi \over p^-_\pi} = {1 \over 2} \ln
{2p^{+2}_\pi\over m_\perp^2}\,=\,
\y_{proj} - \ln{1\over \xx} + \ln{M\over m_\perp},
\ee
($\y_{proj} = \ln(\sqrt{2}{p^+_1/M})\simeq \ln(\sqrt{s}/M)$),
and lies in the range $0 \le \y \le \y_{proj} + \ln(M/m_\perp)$.
For a left moving pion ($p^z_\pi <0$), we use similar definitions 
where $p^+$ and $p^-$ are exchanged. This gives a symmetric range
for y, as  in Fig. \ref{dndy}.
All the pions are produced in a
distribution of rapidities within this range. 

In Fig. \ref{dndy}, $dN/d{\rm y}$ is the number of produced particles (say,
pions) per unit rapidity.
The leading particles are shown in the solid line and are clustered
around the projectile and target rapidities.  For example, in a heavy ion
collision, this is where the nucleons would be.  In the dashed
line, the distribution
of produced mesons is shown.

Several theoretical issues arise in multiparticle production.  Can we
compute $dN/d{\rm y}$?  Or even $dN/d{\rm y}$ at $\y = 0$ (``central
rapidity'') ? How does the average transverse
momentum of produced particles $\langle p_\perp\rangle$ 
behave with energy?  What is the
ratio of produced strange/nonstrange mesons, and corresponding ratios of
charm, top, bottom etc at $\y = 0$ as the center of mass energy
approaches infinity?
Does multiparticle production as $s \rightarrow \infty$ at
$\y = 0$ become simple, understandable and computable?

Note that $\y = 0$ corresponds to particles with $p_z=0$ or
$p^+=m_\perp/\sqrt{2}$, for which 
$\xx=m_\perp/(\sqrt{2}p^+_1) = m_\perp/\sqrt{s}$ 
is small, $\xx\ll 1$, in the high-energy limit of interest.
Thus the multiparticle production at central
rapidity reflects properties of the  small-x degrees of freedom
in the colliding hadron wavefunctions.

\begin{figure} 
\begin{center} 
\includegraphics[width=0.5\textwidth]
{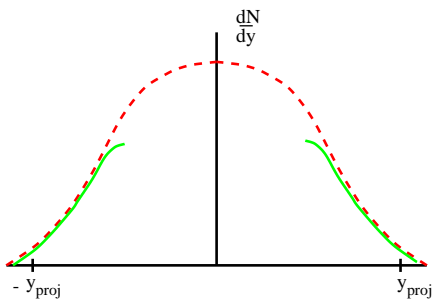} 
\caption{Feynman scaling of rapidity distributions. The two different lines
correspond to rapidity distributions at different energies.} 
\label{feynman} 
\end{center}
\end{figure}

There is a remarkable feature of rapidity distributions of produced 
hadrons, generally referred to as either Feynman scaling or Limiting Fragmentation. 
If we plot rapidity distributions of produced hadrons at different energies, then as function of 
$\y-\y_{proj}$,
the rapidity distributions are to a good
approximation independent of energy.  
This is illustrated in Fig. \ref{feynman}, where the rapidity distribution
measured at one energy is shown with a solid line and the rapidity
distribution at a different, higher, energy is shown with a dotted line.  In this plot,
the rapidity distribution at the lower energy has been shifted by an amount
so that particles of positive rapidity begin their distribution at 
the same $\y_{proj}$ as the high energy particles, and correspondingly
for the negative rapidity particles.  This of course leads to a gap
in the center for the low energy particles due to this mapping.

This means that as we go to higher and higher energies, the new physics is 
associated with the additional degrees of freedom at small rapidities
in the center of mass frame (small-x degrees of freedom).  
The large x degrees of freedom do not change much.
This suggests that there may be some sort of renormalization group description
in rapidity where the degrees of freedom at larger x are held fixed as we go to
smaller values of x.  We shall see that in fact these large x degrees of
freedom act as sources for the small x degrees of freedom, and the
renormalization group is generated by integrating out degrees of 
freedom at relatively large x to generate these sources.
With this understanding, one may be able to compute the number and distribution of particles produced 
at central rapidities. 

\subsection{Deep Inelastic Scattering}
\label{SEC-DIS}

\begin{figure}
\begin{center}
\includegraphics[width=0.5\textwidth] {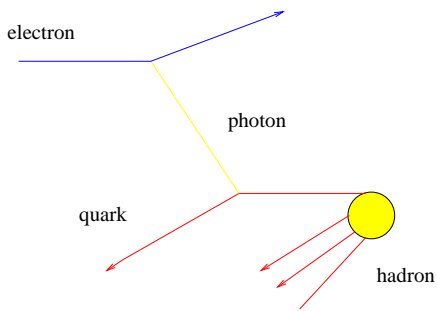}   
\caption{Deep inelastic scattering of an electron on a hadron.}
\label{electron}
\end{center}
\end{figure}

In the previous section, we discussed hadron-hadron collisions in which large numbers of particles are produced. 
Here we will discuss Deep Inelastic Scattering (DIS) of a lepton scattering off a hadronic 
target~\cite{TB-DIS}. 
Fewer particles are produced in DIS,
so this provides a relatively clean environment to study QCD at high energies. 
In Fig. \ref{electron} is shown the cartoon of a DIS experiment. 

To describe quark distributions, it is convenient to work in a
reference frame where the hadron has a large light-cone
longitudinal momentum
$P^+\gg M$ (``infinite momentum frame'').  In this frame, one can
describe the hadron as a collection of constituents (``partons''),
which are nearly on-shell excitations carrying some fraction
x of the total longitudinal momentum $P^+$. (The correct mathematical formulation 
of this picture  involves the light cone quantization mentioned previously: 
the hadron can then be expressed in  a ``quark-parton''
basis.) Thus, the longitudinal momentum of a parton is $p^+=\xx P^+$, 
with $\,0\le \xx  <1$.

For the struck quark in Fig. \ref{electron}, this x variable
(``Feynman x'') is equal (modulo target mass corrections) to the empirical Bjorken variable $\xx_{Bj}$, 
which is defined in a frame independent way as
$\xx_{Bj} = Q^2/2P\cdot q$. 
In this definition, $Q^2=-q^\mu q_\mu$, with $q^\mu$ the 
(space-like) 4-momentum of the exchanged photon.
The Bjorken variable scales like $\xx_{Bj}\sim Q^2/s$, with
$s=$ the invariant energy squared. Thus,  in deep inelastic
scattering at high energy (large $s$ at fixed $Q^2$)
we measure the quark distributions $dN_{quark}/d\xx$ at small x
($\xx\ll 1$).

It is again useful to think about these distributions as a function of rapidity.
We define this for deep inelastic scattering as $\y = \y_{hadron} - \ln(1/\xx)$,
and the invariant rapidity distribution as
${dN/ d{\rm y}} = \xx {dN/ d{\rm x}}$.
At high $Q^2$, the measured quark and gluon distribution functions are simply 
related (at least, in a  leading order approximation)
to the number of partons per unit rapidity in the hadronic wavefunction. 

The typical $dN/d{\rm y}$ distribution for  constituent gluons
of a hadron as measured in DIS is similar to the rapidity distribution of produced
particles in hadron-hadron collisions (see Fig.~\ref{dndy}). 
This suggests  a relationship between the structure functions in DIS and the
rapidity distributions for particles produced in hadronic collisions.  We expect 
the gluon distribution function to be proportional
to the pion rapidity distribution. This relation is plausible
(since the degrees of freedom of the gluons should not be lost) and is 
incorporated in many models of multi-particle production.

\begin{figure} 
\begin{center} 
\includegraphics[width=0.5\textwidth]
{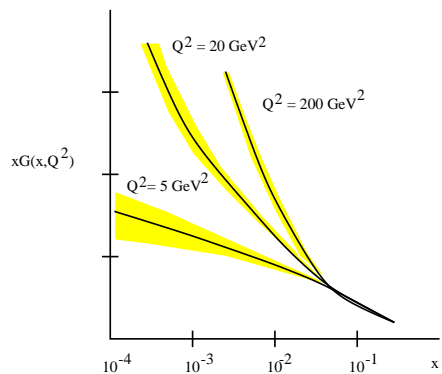} 
\caption{The Zeus data for the gluon structure functions. } 
\label{gluon} 
\end{center}
\end{figure}

The small x problem is that in experiments at HERA, the rapidity
distribution functions for quarks and gluons 
grow rapidly as the rapidity difference
\be\label{tau-Intro}
\tau\,\equiv\,\ln(1/\xx)\,=\,\y_{hadron} - \y\ee
between the quark and the hadron increases \cite{z}.  
In Fig. \ref{gluon}, the ZEUS data for the gluon distribution are
plotted for $Q^2 = 5~ {\rm GeV}^2$, 
$20~ {\rm GeV}^2$ and $200~{\rm GeV}^2$ \cite{z}.
The gluon distribution is the number of gluons per unit rapidity
in the hadron wavefunction, $\xx G(\xx,Q^2)=dN_{gluons}/d\y\;$.
Experimentally, it is extracted from the data for the quark structure
functions, by analyzing the dependence of the latter upon the resolution $Q^2$ of
the probe.

The growth seen in Fig.~\ref{gluon} appears to be more rapid
than $\tau$ or $\tau^2$. Perturbative considerations of the high energy limit
in QCD by Lipatov and colleagues lead to an evolution equation 
commonly called the BFKL equation~\cite{BFKL}
which suggests that distributions may grow as an exponential in $\tau$ \cite{BFKL,TB-BFKL}. 
Alternatively, the double logarithmic DGLAP evolution equation \cite{DGLAP}
predicts a less rapid growth, like an exponential in $\sqrt{\tau}$. Both of these evolution 
equations would predict asymptotically a growth of the distributions which would exceed the 
Froissart unitarity bound discussed previously. 

\begin{figure} 
\begin{center} 
\includegraphics[width=0.5\textwidth]
{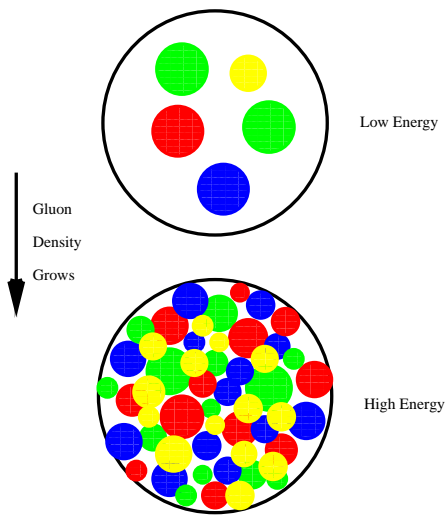} 
\caption{Saturation of gluons in a hadron.  A view of a hadron head on as
x decreases.} 
\label{saturation} 
\end{center}
\end{figure}

How do we understand in QCD the problem of the rapid rise of gluon distributions at small x?
Consider Fig.~\ref{saturation}, where we view the hadron head on. The constituents are the
valence quarks, gluons and sea quarks, all shown as colored circles.  As we add more and more constituents,
the hadron becomes more and more densely populated.  If one attempts to resolve
these constituents with an elementary probe, as in DIS, then, at sufficiently
small x (for a given transverse resolution), the density of the constituents becomes
so large that one cannot neglect their mutual interactions any longer. One expects such
interactions to give ``shadowing'' by which we imply a decrease of the scattering
cross-section relative to what is expected from incoherent independent scattering.

More precisely, we shall see later that, as a effect of these interactions,
the parton distribution functions at fixed $Q^2$ {\it saturate}, in the sense of
showing only a slow, logarithmic, increase with $1/\xx$\cite{GLR,MQ,BM87,MV94,AM2,SAT}. 
(See also Refs.~\cite{Larry01,Levin,AMCARGESE,CARGESE} for recent reviews
and more references.) For a given $Q^2$, this
saturation occurs if $\xx$ is low enough, lower than some critical value $\xx_s(Q^2)$.
Converserly, for given $\xx$, saturation occurs for transverse momenta below some
critical value $Q^2_s(\xx)$, defined as
\be\label{Qs-INTRO}
   Q_s^2(\xx)\, =\, \alpha_sN_c\,{1 \over {\pi R^2}} {{dN} \over {d{\rm y}}}\,,
\ee 
where $dN/d\y$ is the gluon distribution at $\y = \y_{hadron} - \ln(1/\xx)$. Only
gluons are included, since, at high energy, the gluon density grows much faster than the quark density,
and is the driving force towards saturation. This explains why in the following we shall
 focus primarily on the gluons. In Eq.~(\ref{Qs-INTRO}),
$\pi R^2$ is the hadron area in the impact parameter space (or transverse plane).
This is well defined provided the wavelength of the probe is small
compared to $R$, which we assume throughout.
Finally, $\alpha_s N_c$ is the color charge squared of a single
gluon. Thus, the ``saturation scale'' (\ref{Qs-INTRO}) has the meaning
of the average color charge squared of the gluons per unit transverse area per unit rapidity.

Since the gluon distribution increases rapidly with the energy,
as the HERA data suggests, so does the saturation scale. 
For high enough energy, or small enough x, 
 \be
        Q_s^2(\xx)  \gg \Lambda_{QCD}^2\,,
\ee
and $\alpha_s( Q_s^2) \ll 1$. This suggests that weak coupling techniques can be used 
to study the high energy regime in QCD.

However, weak coupling does not necessarily mean that the
physics is perturbative.  There are many examples of nonperturbative
phenomena at weak coupling.  An example is instantons in electroweak
theory, which lead to the violation of baryon number. Another example
is the atomic physics of highly charged nuclei. Although the electromagnetic 
coupling constant is very weak, $\alpha_{\rm em}\ll 1$, the large charge $Z$ of the atoms 
introduces a scale $Z\alpha_{\rm em}\sim 1$ producing a strong nuclear Coulomb field in 
which the electron propagates. Nevertheless, there is a systematic  
technique which enables one to compute the non-perturbative properties of high $Z$ atoms.
Yet another example is QCD at high temperature $T\gg \Lambda_{QCD}$; this is
a weakly coupled quark-gluon plasma, but exhibits nonperturbative
phenomena on large distances $r\gg 1/T$
due to the collective behaviour of many quanta \cite{BI01}.

Similarly, the small-x gluons with
transverse momenta $Q^2\le Q_s^2(\xx)$ make a high
density system, in which the interaction probability $\sigma\times n$ (where $\sigma\sim \alpha_s/{Q^2}$ is 
the typical parton cross-section and $n$ is the gluon density, $n={dN\over dy}/\pi R^2$) 
is of order one \cite{GLR,MQ,BM87} (cf. Eq.~(\ref{Qs-INTRO})).
That is, although the coupling is
small, $\alpha_s( Q^2_s) \ll 1$, the effects of the interactions
are amplified by the large gluon density and ordinary perturbation theory breaks down.
A resummation of the high density effects is therefore necessary.
Our strategy, to be developed in the following sections, will be to construct 
an {\it effective theory} --- the Color Glass Condensate \cite{MV94,Kovchegov96,JKMW97,PI,CARGESE} --- in which
the small-x gluons are described as the classical color
fields radiated by ``color sources'' at higher rapidity. Physically,
these sources are the ``fast'' partons, i.e., the hadron constituents 
with larger longitudinal momenta $p^+\gg \xx P^+$. The properties
of the color sources will be obtained via a 
renormalization group analysis, in which the ``fast'' partons 
are integrated out in steps of rapidity and in the background
of the classical field generated at the previous step \cite{JKLW97,PI}.

  \subsection{Nucleus-Nucleus and Proton-Nucleus Collisions}

\begin{figure} 
\begin{center} 
\includegraphics[width=0.5\textwidth]
{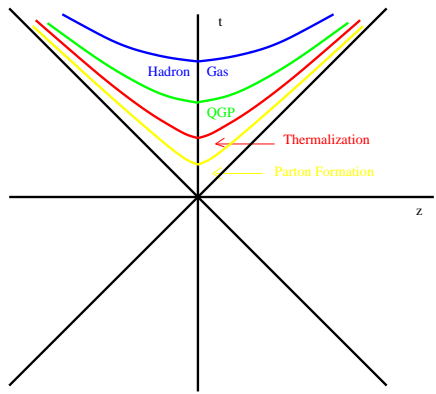} 
\caption{A space-time figure for ultrarelativistic heavy ion collisions. } 
\label{spacetime} 
\end{center}
\end{figure}

In Fig. \ref{spacetime}, we plot a cartoon of the space-time evolution of 
a heavy ion collision~\cite{bj}.
Imagine we have two Lorentz contracted
nuclei approaching one another at the speed of light.  
We choose coordinates such as the collision takes place at
$z=t=0$, or $x^+=x^- = 0$. Since the two nuclei are well
localized in the longitudinal direction, 
they can be thought of as sitting at $z\simeq t$ ($x^- = 0$)
for the right mover, respectively at $z\simeq - t$ (or $x^+ = 0$)
for the left mover. To analyze this problem for $t \ge 0$, namely, after the
collision takes place, it is convenient to
introduce a time variable which is Lorentz invariant under longitudinal
boosts\footnote{This should not be confused with the rapidity variable
introduced previously, in eq.~(\ref{tau-Intro}), and which will not
appear in this subsection.} $\tau = \sqrt{t^2 - z^2}$
(the ``proper time'') and a space-time rapidity variable
\be
        \eta\, =\, {1 \over 2}\, \ln\left( {{t+z} \over {t-z}} \right)
\, =\, {1 \over 2}\, \ln\frac{x^+}{x^-}\,.\ee
For free streaming particles with velocity $v_z$ ($z = v_zt = {p_z \over E} t$),
the space-time rapidity equals the momentum space rapidity (\ref{y-DEF}): $\eta = \y$.
We shall see later that this identification remains approximately true also for the
off-shell quantum fluctuations (the partons) with relatively large longitudinal momenta.

At high energies, in the central rapidity region, particle distributions vary slowly and 
it should be a good approximation to take them to
be rapidity invariant and therefore also independent of $\eta$. 
Therefore distributions are the same on the lines of constant proper time $\tau$, which are shown in 
Fig. \ref{spacetime}.

An outstanding problem is to formulate the initial conditions for 
a heavy ion collision and to study the subsequent evolution of the produced partons. 
Can one argue from first principles that the partonic matter will thermalize into a quark-gluon 
plasma ?

There are two separate classes of problems one has to understand for the
initial conditions.  Firstly, the two nuclei which are colliding are coherent quantum mechanical wavepackets.  
Therefore, for some early time, the degrees of freedom must be quantum mechanical.  This means,
in particular,
\be
        \Delta z \Delta p_z \ge 1\,,
\ee
which particularly constrains the small-x gluons, which are
delocalized over large longitudinal distances $\Delta z\sim 1/p_z$,
and thus overlap with each other. Such degrees
of freedom cannot be described by semi--classical transport theory for {\it particles}.
(Classical particles
are characterized by a distribution function $f(\vec{p}, \vec{x}, t)$, which is a 
simultaneous function of momenta and coordinates.)
However, fortuitously, quantum coherent states can be described as classical {\it fields} because 
they have large occupation numbers $\sim 1/\alpha_s \gg 1$. 
Heisenberg commutators between particle creation and annihilation
operators become negligible in this limit:
\be
[a_k,\,a^\dagger_k]\,=\,1\,\ll\,a^\dagger_ka_k=N_k\,.
\ee
Classical field theory is also the appropriate language to describe
another important feature of the initial conditions, namely the
classical charge coherence. At very early times, we have a tremendously large number of particles
packed into a longitudinal size scale of less than a fermi due
to the Lorentz contraction of the nuclei.  We know that such particles
cannot interact incoherently.  For example, if we measure the field due
to two opposite charges (a dipole) at a distance scale $r$ large compared to their
separation, the field  falls off as $1/r^2$, not $1/r$.  On the other
hand, in parton cascade models, interactions are taken into account by cross-sections 
which involve matrix elements squared leaving no room for
classical charge coherence. These models should therefore not be applied at very early times. 

As an effective theory at small x, the Color Glass Condensate can be
applied to study the initial stages of a heavy ion
collision~\cite{KMW,AR99,AYR01}. The only scales in the problem are
the saturation scale $Q_s$ and the transverse size of the system
$R$. On a time scale $\tau\sim 1/Q_s$, the initial energy and number
distributions of gluons can be computed by solving classical field
equations. At later times $\tau \gg 1/Q_s$, the system becomes dilute
and the classical field approximation breaks down, but one in terms of
transport equations becomes valid. There is presumably an overlap
region where the two descriptions can be both
correct~\cite{MuellerSon}. At early times, the distribution of gluons
in momentum space is primarily transverse. As the system becomes
dilute, the gluons begin to scatter ``off the transverse
plane''. Baier, Mueller, Schiff and Son~\cite{BMSS} have argued that
2$\rightarrow$ 3 processes are those which lead to the most efficient
thermalization.  Whether on not the initial non-equilibrium gluon
distributions thermalize to form a quark gluon plasma at the energies
of interest is of outstanding phenomenological interest at RHIC and
LHC energies. It is also of interest to understand how much of the
RHIC data can be understood purely from initial state
effects~\cite{KN,KLN,JDLR,DLL} as opposed to the final state effects
which are important for thermalization.  Proton (or
deuterium)--nucleus collisions should be very helpful in helping to
isolate initial state effects from final state rescattering effects.
In a proton-nucleus collision, one does not expect final state
rescattering to dominate the measured particle spectrum.  Initial
state effects, on the other hand, should be especially important in
the proton fragmentation region and have been computed recently within
the CGC framework~\cite{KM98,DumitruMcLerran,DumitruJJM,GelisJJM}. The
upcoming experiments at RHIC on deuterium-gold collisions will
hopefully clarify the role of initial/final state effects in high
energy scattering.

  \subsection{Universality of High Energy Scattering}

In pion production, it is observed that, with the exception of globally conserved quantities like the
energy and the total charge, the rapidity correlations are of
short range. If the theory is local in rapidity, the only dimensionful parameter which can
determine the physics at a given rapidity is $Q_s^2(\xx)$. 
For an approximately scale invariant theory such as QCD, a typical
transverse momentum of a constituent will therefore be of order $Q_s^2$.
If $Q_s^2 \gg 1/R^2$, where $R$ is the radius of the hadron, the
finite size of the hadron becomes irrelevant.  Thus at small
enough x, all hadrons become the same---specific properties of the hadrons
(like their size or atomic number $A$) enter only via the 
saturation scale $Q_s^2(\xx, A)$.

Hence there should be some equivalence between nuclei and 
protons: when their $Q_s^2$ values are the same, their physics
should be the same.  Eq.~(\ref{Qs-INTRO}) suggests the following
empirical parametrization of the saturation momentum:
\be
Q_s^2(\xx,A)\,\sim\, {{A^{1/3}} \over \xx^\lambda}
\label{universal}
\ee
where the value $\lambda \sim 0.2 - 0.3$ seems to be preferred by both the data 
\cite{z,GBW99} and the most recent theoretical calculations~\cite{DT02}.
There should thus be the following correspondence:

\noindent
$\bullet$ RHIC with nuclei $\sim$ HERA with protons

\noindent
$\bullet$ LHC with nuclei $\sim$ HERA with nuclei

Estimates of the saturation scale for nuclei at RHIC energies give
$Q_s\sim 1-2 ~{\rm GeV}$, and at LHC $2-3 ~{\rm GeV}$.

This further suggests that for relatively simple processes like deep inelastic
scattering (at least), the observables should be universal functions of the ratio 
between the transferred momentum $Q^2$ and the saturation scale $Q_s^2(\xx, A)$.
This feature, called ``geometric scaling'', is rather well satisfied by the proton 
structure functions measured at HERA,  for all values of Bjorken x 
smaller than $0.01$ and in a broad region of $Q^2$ 
(between $0.045$ and $450\,{\rm GeV}^2$) \cite{geometric}. Interestingly, the observed
scaling extends up to relatively large values of $Q^2$, well above the saturation scale
$Q_s^2(\xx)$, which suggests that the phenomenon of gluon saturation at $Q^2 \le 
Q_s^2(\xx)$ has a rather strong influence also on the physics at much larger $Q^2$~\cite{KIIM}. 

The previous considerations suggest how to reconcile unitarity with the growth of 
the gluon distribution function at small x.  The point is that the smaller x is the larger 
$Q_s(\xx)$ is and thus and the typical partons are smaller. Therefore, when decreasing
x, although we are increasing the number of gluons, we do it by adding in more gluons of
smaller and smaller size.  A probe of transverse
size resolution $\Delta x_\perp \sim 1/Q$  
will not see partons smaller than this resolution size.  
Thus when $Q<Q_s$, newly created partons will not contribute to the cross-section.

\section{The Effective Theory for the Color Glass Condensate}
\setcounter{equation}{0}

\subsection{The hadron wavefunction at small x}

As discussed in Sect. 1.2, a relativistic quantum field theory quantized on the lightcone 
has a simple structure in terms of the bare quanta or ``partons'' of the theory. For instance, in 
QCD, the eigenstates of the light-cone Hamiltonian $P_{QCD}^-$ can be expressed as a linear superposition of the 
eigenstates of the non-interacting part of the Hamiltonian 
$P_{QCD,0}^- =P_{QCD}^- - V_{QCD}$, where
$V_{QCD}$ is the LC potential. The proton wavefunction for instance can be written in this 
parton basis as 
\be
|\psi> = c_1 |qqq> + c_2|qqqg> + \cdots + c_n|qqqgggg...q{\bar q}gg> + \cdots \, .
\ee
It is {\it a priori} not obvious what the advantage of such a decomposition is. The magic of the parton model is 
that, at high energies, it is apparent from the structure of $V_{QCD}$ that 
the interactions of the partons with each other are time dilated. The very complicated picture 
of the scattering of the proton off an external potential can be replaced with the simple picture of individual 
partons scattering off the external potential. (In the eikonal approximation, each of these partons acquires 
a simple phase in the scattering~\cite{BjKogutSoper}.) At lower energies, the Fock states involving large numbers of 
partons are not very important but at higher energies they are increasingly important,
and involve predominantly gluons. Since the total LC
momentum of the proton $P^+$ is divided among a large number of partons, 
a typical parton will carry a momentum $k^+<<P^+$, i.e., only a small fraction 
$\xx=k^+/P^+$ of the proton total momentum. Understanding the physics of QCD at high 
energies or small x 
thus requires that we understand the properties of the 
n-gluon components of the hadron's LC wavefunction. 

A very interesting approach to computing the
 properties of the n-gluon component of the hadronic wavefunction is 
Mueller's color dipole approach for heavy quarkonia~\cite{Mueller94}. 
This is valid in the limit where the number of colors $N_c$ is large, so 
the gluons can be effectively replaced by $q\bar q$ pairs (``color dipoles'').
It will be shown later that this approach and the CGC formalism 
give identical results for the evolution of distributions in $\xx$, 
in this large-$N_c$ limit.

The CGC approach that we shall follow here 
is to construct a coarse grained effective theory for the small-x 
component of the  hadron LC wavefunction.
We shall first consider a large nucleus, for which this construction is most intuitive.
Then, we shall argue that, at sufficiently large energy, or small enough x, a similar 
theory can be constructed for any hadron, via weak coupling calculations in QCD.

\subsection{The McLerran-Venugopalan model for a large nucleus}
\label{EFT-MV}

Consider a nucleus in the infinite momentum frame (IMF)
with momentum $P^+\rightarrow \infty$. 
We will assume that the nucleus is of nearly infinite transverse extent 
with a uniform nuclear matter distribution. As we will discuss later, 
the model can be extended to include 
realistic nuclear density profiles. In the IMF, partons which carry
large fractions of the nuclear momentum (``valence'' partons), 
are Lorentz contracted to a distance $\sim 2R/\gamma$, with $\gamma=P^+/m_p$ and $m_p$ the 
mass of the proton. The ``wee'' partons with momentum fractions $x\ll 1$ are delocalized 
in the $x^-$ direction 
over much larger distances: a simple kinematic estimate suggests that partons with momentum fraction $x\ll A^{-1/3}$ 
are delocalized over distances larger than the nuclear diameter. These partons then ``see'' the partons at large 
x as infinitely thin sources of color charge. The model assumes a simple kinematic distinction between 
wee and valence partons. The reality is more complex as we will discuss later but certain key features of this 
simple model will survive.

Another salient feature of wee partons is their very short lifetimes.
 Their lifetime on the lightcone, $\Delta x^+$, 
is, from the uncertainity principle, conjugate to their light cone energy: $\Delta x^+ \sim 1/k^-$. From the light 
cone dispersion relation, $k^- = m_\perp^2/2k^+ \equiv m_\perp^2/2\xx P^+$. 
Thus $\Delta x^+ \sim \xx P^+/m_\perp^2$ is proportional to x.
These estimates imply that wee parton lifetime is much shorter 
than that of the valence partons: on the scale of 
the lifetime of the former, the latter appear to live forever. 
The valence parton sources are thus {\it static sources of color charge}. 
Since their momenta are large, they are unaffected by absorbing or emitting soft quanta: they are 
recoilless sources of color charge. In this ``eikonal'' approximation, 
the wee parton cloud couples only to the ``plus'' component of the LC current, which, 
from the discussion here, can be written as (see also Sect. 2.3 below) :
\be\label{eikonalJ}
J^{\mu,a} = \delta^{\mu +}\delta(x^-)\rho^a(x_\perp) \, ,
\ee
where $\rho^a(x_\perp)$ is the valence quark color charge density in the transverse plane. 
The $\delta$-funtion in $x^-$ assumes an infinitely 
thin sheet of color charge. The assumption can and must be relaxed; namely, $\rho^a(x_\perp)\rightarrow \rho^a(x_\perp,x^-)$ 
as we will discuss later. Note that $\rho^a$ is {\it static}, i.e.,
 independent of the LC time $x^+$, for the reasons explained previously. 

%

We now turn to the color charge density $\rho^a (x_\perp)$ and how it is
generated. We assume that the nucleus is interacting with an external
probe which can resolve distances of size $\Delta x_\perp$ in the transverse plane
that are much smaller than the nucleon size $\sim \Lambda_{QCD}$. 
Now, in the longitudinal
direction, the small probe which has $x\ll A^{-1/3}$ simultaneously
couples to partons from nucleons all along the nuclear diameter. Since
its transverse size is much smaller than the nucleon size, it sees
them as sources of color charge.  If the density $n\equiv N_c A/\pi R^2_A
\simeq \Lambda_{QCD}^2 A^{1/3}$ ($R_A=R_0A^{1/3}$ is the radius of the nucleus) of the valence
quarks in the transverse plane is large, $n \gg \Lambda_{QCD}^2$, and if
$n^{-1}\ll \Delta S_\perp \ll 1/\Lambda_{QCD}$, with $\Delta S_\perp \sim (\Delta x_\perp)^2$,
then the number $\Delta N = n \Delta S_\perp$
of color charges within the tube of transverse area $\Delta S_\perp$
crossing the nucleus is large: $\Delta N \gg 1$. These sources belong typically to
different nucleons, so they are uncorrelated with each other, because of confinement. 
They are thus {\it random} sources of color charge and the total
color charge ${\cal Q}^a$ in the tube is the incoherent sum of the
color charges of the individual partons. Thus, 
\be\label{MV-Q}
\langle {\cal Q}^a\rangle\,=\,0, \qquad \langle {\cal Q}^a {\cal Q}^a
\rangle\,=\,g^2C_f\Delta N\,=\,\Delta S_\perp\, \frac{g^2C_f N_c
A}{\pi R^2_A},\ee where we have used the fact that the color charge
squared of a single quark is $g^2t^at^a=g^2C_f$. One can treat this
charge as classical since, when $\Delta N$ is large enough, we can
ignore commutators of charges: \\
$ \mid [{\cal Q}^a, {\cal Q}^b] \mid \,=\, \mid i f^{abc} {\cal Q}^c \mid \,\ll\, {\cal Q}^2\,.$

Let us introduce the color charge density $\rho^a(x^-,x_\perp)$ in such a way that:
$\rho^a(x_\perp)\equiv \int dx^- \rho^a(x^-,x_\perp)$. Then,
\be\label{Qtot}
{\cal Q}^a\,=\,\int_{\Delta S_\perp} d^2x_\perp\,\rho^a(x_\perp)\,=\,
\int_{\Delta S_\perp} d^2x_\perp\int dx^-\,\rho^a(x^-,x_\perp),\ee
and eqs.~(\ref{MV-Q}) imply (recall that $C_f=(N_c^2-1)/2N_c$) :
\be\label{MV-corr}
\langle \rho_a(x_\perp)\rho_b(y_\perp)\rangle_A&=&
\delta_{ab}\delta^{(2)}(x_\perp-y_\perp)\,\mu_A^2,\qquad
\mu_A^2\equiv \frac{g^2A}{2\pi R^2_A}\,,\nn
\langle \rho_a(x^-,x_\perp)\rho_b(y^-,y_\perp)\rangle_A&=&
\delta_{ab}\delta^{(2)}(x_\perp-y_\perp)\delta(x^- -y^-)\,
\lambda_A(x^-),\nn  \int dx^-\,\lambda_A(x^-)&=&\mu_A^2\,.
\ee
Here, $\mu_A^2\sim A^{1/3}$ is the average color charge squared
of the valence quarks per unit transverse area and per
color, and $\lambda_A(x^-)$ is the corresponding
density per unit volume. The latter has some dependence upon
$x^-$, whose precise form is, however, not important since
the final formulae will involve only the integrated density $\mu_A^2$.
There is no explicit dependence upon $x_\perp$ in $\mu_A^2$ or
$\lambda_A(x^-)$ since we assume transverse homogeneity within
the nuclear disk of radius $R_A$. Finally, the correlations are local
in $x^-$ since, as argued before, color sources at different values
of  $x^-$ belong to different nucleons, so they are uncorrelated.
All the higher-point, connected, correlation functions of
$\rho_a(\vec x)$ are assumed to vanish. 
The non-trivial correlators (\ref{MV-corr}) are generated
by the following weight function \cite{MV94,Kovchegov96} (with $\vec x=(x^-,x_\perp)$) :
\be\label{MV-W}
W_A[\rho]\,=\,{\cal N}{\rm exp}\left\{-\,{1\over 2}\, \int d^3x
\frac{\rho_a(\vec x)\rho_a(\vec x)}{\lambda_A(x^-)}\right\},\ee
which is a Gaussian in $\rho_a$, with a local kernel. 
This is gauge-invariant (since local), so the variable $\rho_a$ in this expression
can be the color source in any gauge. The choice of a gauge will however soon become
an issue when we shall study the dynamics of the gluons radiated by this
random distribution of color charges.

The local Gaussian form of the weight function in Eq.~(\ref{MV-W}) is valid, by construction,
for a large nucleus, and within some restricted kinematical range that we spell
here again, for more clarity. As already discussed, this is correct for a transverse resolution
$Q^2 \equiv 1/\Delta S_\perp$ within the range $\Lambda_{QCD}^2 \ll Q^2 \ll \Lambda_{QCD}^2 A^{1/3}$.
But, clearly, the assumption that the valence quarks are uncorrelated must fail for 
transverse separations of order $R_0 \sim 1/\Lambda_{QCD}$ or larger, since the 
$N_c=3$ valence quarks within the same nucleon are confined in a color singlet state.
Thus, the total color charge, together with its higher multipolar moments, must vanish
when measured over distances of the order of the  nucleon size $R_0$, or larger.
As emphasized by Lam and Mahlon~\cite{LM00} (see also Ref.~\cite{RajGavai} for an earlier discussion), 
the requirement of color neutrality can be included in the Gaussian weight function by replacing
the $\delta$-function in eq.~(\ref{MV-corr}) with
$\langle \rho^a (x^-,x_\perp)\rho^b (0)\rangle = \lambda(x^-,x_\perp)\delta^{ab}$, 
where $\lambda(x^-,x_\perp)$ is such that its Fourier transform $\lambda(x^-,\kk)$ 
vanishes rapidly at momenta $\kk \simle \Lambda_{QCD}$.

Consider also the validity range of Eq.~(\ref{MV-W}) in longitudinal momenta. As 
explicit in the previous analysis, the color fields that we are computing have small values of
$x\ll A^{-1/3}$, that is, they carry longitudinal momenta $k^+$  much lower
than those of their sources, the valence quarks. Thus, the weight function (\ref{MV-W})
can be seen as part of an {\it effective theory} for gluon correlations at momenta
$k^+$ smaller than some upper cutoff $\Lambda^+$, of the order of the typical
longitudinal momentum of the valence quarks. (This theory will be
completely specified in Sects. 2.3--2.4.) Note however that, for the classical approximations
underlying Eq.~(\ref{MV-W}) to be valid, the value $k^+$ of interest should be not {\it much}
smaller than  $\Lambda^+$. Indeed, as we shall see, new color sources with momenta $ p^+ < \Lambda^+$
are produced by radiation from the original sources at $p^+ \ge \Lambda^+$. If the gap between $k^+$ and 
$\Lambda^+$ is relatively large --- the precise condition is
that $\alpha_s N_c \ln (\Lambda^+/k^+) \ge 1$ --- these new sources, which are mostly gluons,
will completely dominate the physics at the scale $k^+$ of interest.
As we shall in Sect. 3, these new sources can be explicitly constructed by integrating
out layers of quantum fluctuations in a renormalization group analysis, but the ensuing
weight function is generally {\it not} a Gaussian. Still, the Gaussian (\ref{MV-W})
may be a good initial condition for this {\it quantum evolution}. Moreover,
rather remarkably, it turns out that this is also 
a good approximation in the quantum theory, but with a non-trivial transverse
momentum dependence for
the 2-point function $\lambda$ that will be specified in Sect. 3.

\subsection{The Color Glass}

Once the weight function for the classical color charge 
configurations associated with the large-x partons is known, it is possible
to write down an {\it effective theory for the small-x gluons}. The 
generating functional for the correlation functions of the small-x gluons reads \cite{PI}:
\be\labe{ZCGC} 
Z[j]\,=\,\int {\cal D}\rho\,W_{\Lambda^+}[\rho]\left\{
\frac{\int^{\Lambda^+} {\cal D}A\,\delta(A^+)\,{\rm e}^{\,iS[A,\,\rho] -\int j\cdot A}}
{\int^{\Lambda^+} {\cal D}A\,\delta(A^+)\,\,{\rm e}^{\,iS[A,\,\rho]}}
\right\},\ee 
where the external current $j^\mu_a$ is a formal device to generate
Green's functions via differentiations, and $S[A,\,\rho]$ is the action that describes
the dynamics of the wee gluons in the presence of the classical color charge $\rho$
(see below). The path integral over the gluon fields is written
in the light cone (LC) gauge $A^+_a=0$, since this is the gauge which allows for the most
direct partonic interpretation. Correspondingly, $\rho_a$ is the color charge density
in the LC gauge.

Note the dependence upon the intermediate longitudinal momentum cutoff $\Lambda^+$
in the integrals in Eq.~(\ref{ZCGC}). As explained at the end of the previous subsection,
 $\Lambda^+$ is the scale which separates the `fast' partons ($p^+ > \Lambda^+$), 
which have been `integrated out' and replaced by the classical color charge $\rho_a$,
from the `wee' gluons ($k^+ < \Lambda^+$), for which the effective theory is meant, and
which are still explicit in the path integral, as the gauge fields $A^\mu_a$. Since obtained
after integrating out the modes with $p^+ > \Lambda^+$ (see Sect. \ref{SRGE}),
the weight function $W_{\Lambda^+}[\rho]$ depends upon the separation scale  $\Lambda^+$. 
Clearly, $\Lambda^+$ must be chosen such that $\Lambda^+/P^+\ge \xx$, with x the longitudinal
fraction of interest.

For instance, for a large nucleus at not so high energies (e.g., a gold nucleus at RHIC)
we have $\xx\simeq 10^{-1} - 10^{-2}$, so we can neglect the quantum evolution in a first
approximation, and identify $\rho$ with the color charge of the valence quarks. Then, we can
use the MV model described previously, in which case $W_{\Lambda^+}[\rho]\equiv W_A[\rho]$
is the Gaussian weight function given by Eqs.~(\ref{MV-W}) and (\ref{MV-corr}).

But for smaller values of x (say $\xx < 10^{-3}$), as relevant for DIS at HERA, and also
for nucleus-nucleus scattering at LHC, quantum effects are essential, so $\rho$ must
include the color sources generated by quantum evolution down to  $\Lambda^+$.
As we shall see in Sect. \ref{SRGE}, these sources are predominantly gluons.
Thus, in what follows, we shall restrict the quantum evolution to the gluonic sector; 
that is, the only fermions to be
included among the color sources are the valence quarks in the initial condition.

If the scale $k^+=\xx P^+$ of interest is of order $\Lambda^+$, or slightly below it,
the correlation functions at that scale can be computed in the {\it classical approximation}, 
i.e., by evaluating the path integral over $A^\mu$ in Eq.~(\ref{ZCGC}) in the saddle
point approximation. There are two reasons for that: a) The quantum corrections due to gluons
in the intermediate range $k^+ < p^+ < \Lambda^+$ are, at most, of order $\alpha_s\ln(\Lambda^+/k^+)$,
and thus are truly higher order effects as long as $\ln(\Lambda^+/k^+) \simle 1$.
b) The small-x gluons have large occupation numbers $\gg 1$ (as we shall see, at saturation
the occupation numbers are parametrically of order $1/\alpha_s$), so their mutual interactions
can be treated in the classical approximation, i.e., by solving the classical field equations
$\delta S/\delta A^\mu=0$, or, more explicitly:
\be
(D_{\nu} F^{\nu \mu})_a(x)\, =\, \delta^{\mu +} \rho_a(\vec x)\,,
\label{cleq0}
\ee
where $D_{\nu} = \partial_\nu - ig A_\nu^a T^a$ with $(T^a)_{bc}= -if_{abc}$.

Note however that, precisely because the occupation numbers are so large, the corresponding
classical fields are strong ($A^i\sim 1/g$ at saturation), and thus the classical non-linearities
must be treated {\it exactly}. In particular, we need the
exact solution to the classical equations of motion (EOM) (\ref{cleq0}), that we shall
construct in the next subsection. Once this solution is known as an explicit
functional of $\rho$, the correlation functions of interest are obtained
by averaging over $\rho$, with weight
function $W_{\Lambda^+}[\rho]$. For instance, the 2-point function is computed as:
\be\label{clascorr}
\langle A^\mu_a(x^+,\vec x)A^\nu_b(x^+,\vec y)\rangle_{\Lambda^+}\,=\,
\int {\cal D}\rho\,\,W_{\Lambda^+}[\rho]\,{\cal A}_a^\mu({\vec x})
{\cal A}_b^\nu({\vec y})\,,\ee
where ${\cal A}_a^\mu\equiv {\cal A}_a^\mu[\rho]$ is the solution to Eq.~(\ref{cleq0}),
and is independent of the LC time $x^+$ (because so is the source $\rho_a(\vec x)$).
This means that only {\it equal-time} correlators can be computed in this
way; but these are precisely the correlators of interest for small-x scattering.

The remaining question is: What is the  weight
function $W_{\Lambda^+}[\rho]$ for $ \Lambda^+\ll P^+$ ? As we shall see
in Sect. \ref{SRGE}, this can be answered via a quantum calculation, the result
of which will allow us to compute 
$W_{\Lambda^+}$ in terms of the initial condition $W_{\Lambda^+_0}$, by integrating out
the gluons with momenta $p^+$ in the range $\Lambda^+_0 < p^+ <\Lambda^+$.
For the purpose of this calculation, we need to also 
specify the action $S[A,\,\rho]$ in Eq.~(\ref{cleq0}). The simple guess
$S[A,\,\rho]=S_{\rm YM}[A] + \int_x \rho_a A^-_a$
(which would generate the classical EOM (\ref{cleq0})) cannot
be right since the second term, involving $\rho$, is not gauge-invariant.
This reflects the fact that, written as in Eq.~(\ref{cleq0}), the classical EOM 
is not manifestly gauge-covariant either. In fact,  Eq.~(\ref{cleq0}) is correct as
written only for field configurations having $A^-_a=0$. This is not a limitation
for {\it classical} calculations, since it is always possible to construct a classical
solution having this property. But Eq.~(\ref{cleq0}) is not sufficient to determine
$S[A,\,\rho]$, which is explicitly needed for the {\it quantum} calculation.

To find the general equation which replaces Eq.~(\ref{cleq0}) in some arbitrary gauge
(where $A^-_a\ne 0$), notice that, in general, the current $J^\mu_a$ in the r.h.s. of 
the Yang-Mills equations must satisfy the covariant conservation law $D_\mu J^\mu=0$
(since we also have $D_\mu D_{\nu} F^{\nu \mu}=0$). For the eikonal current (\ref{eikonalJ}),
this implies $D^- J^+\equiv ({\del\over \del x^+} - igA^-)J^+ =0$, which reduces indeed
to $\del^-\rho =0$ when $A^-=0$. But in general, this is satisfied by
$J^+(x^+, \vec x)\,=\,W(x^+,\vec x)\,\rho(\vec x)
\,W^\dagger(x^+,\vec x)$.
We have introduced here the temporal Wilson line:
\be\labe{WLINE1}
     W[A^-](x^+,\vec x)\,\equiv\,{\rm T}\, \exp\left\{\,
ig\int_{-\infty}^{x^+} dz^+ A^-(z^+, \vec x) \right\},
\ee
with T denoting the ordering of the color matrices in the exponential w.r.t. their
$x^+$ arguments.
The action generating the EOM with current (\ref{WLINE1}) reads \cite{JKLW97,PI}
\be\label{ACTION}
S[A,\rho]=- \int d^4x \,{1 \over 4} \,F_{\mu\nu}^a F^{\mu\nu}_a
+{i \over {gN_c}} \int d^3 \vec x\, {\rm Tr}\,\bigl\{ \rho(\vec x)
\,W[A^-](\vec x)\bigr\},\,\,\,
\ee
where $W[A^-](\vec x) $ is given by eq.~(\ref{WLINE1}) with $x^+\to \infty$.
This action is gauge-invariant indeed \cite{PI}.
(Another gauge invariant generalization for $\rho A^-$, namely
 ${\rm Tr}\,\bigl\{ \rho(\vec x)\,\ln(W[A^-])(\vec x)\bigr\}$, 
has been checked in Ref. \cite{JSR} to give equivalent results.)

The mathematical structure of the average over $\rho$ in Eqs.~(\ref{ZCGC})
and (\ref{clascorr}) is that of a {\it Color Glass} \cite{PI,RajGavai}.
Note indeed the special structure of the 2-point function that follows from  Eq.~(\ref{ZCGC}):
\be\labe{2point} 
\langle {\rm T}A^\mu(x)A^\nu(y)\rangle_{\Lambda^+}
=\!\int\! {\cal D}\rho\,W_{\Lambda^+}[\rho]\left\{
\frac{\int^{\Lambda^+} {\cal D}A\,
\,A^\mu(x)A^\nu(y)\,{\rm e}^{\,iS[A,\,\rho]}}
{\int^{\Lambda^+} {\cal D}A\,\,\,{\rm e}^{\,iS[A,\,\rho]}}
\right\}.\ee
This is {\em not} the same as :
\be\label{AV-ORD}
\frac{\int {\cal D}\rho\,\,W_\Lambda[\rho]\,
\int^\Lambda {\cal D}A
\,A^\mu(x)A^\nu(y)\,\,{\rm e}^{\,iS[A,\,\rho]}}
{\int {\cal D}\rho\,\,W_\Lambda[\rho]\,
\int^\Lambda {\cal D}A\,\,{\rm e}^{\,iS[A,\,\rho]}}\,.\ee
The physical reason for this is 
the fundamental separation in time between the rapidly varying wee
gluons and the comparatively `frozen' large-x partons. One 
thus solves the dynamics of the wee gluons at a {\it fixed} distribution of color charges,
and only then averages over the latter. There is no feedback from the {\it evolution}
of the sources on the wee gluon fields. And there is no interference between successive
configurations of the color sources.
These features, together with the large fluctuations in the color charge density, 
are the ultimate reasons for treating the large-x partons
as forming a {\it classical random distribution}. 

The prototype of a glass is the ``spin glass''--a collection of magnetic impurities 
randomly distributed on a non-magnetic lattice.
The dynamical degrees of freedom, which are rapidly varying, are the
magnetic moments of the impurities (the ``spins''), while the slowly varying disorder
refers to the positions of these spins in the host lattice. To study the
thermodynamics of such a system, one first computes the free-energy (= the logarithm
of the partition function) of the spin system for a {\it fixed} disorder
(namely, for a given spatial configuration of the impurities), and subsequently makes an 
average over all such configurations, with some weight function. 
The final average over the configurations is not a
thermal one: what is averaged is the free-energy computed separately for each configuration.

Similarly, the {\em connected} correlation functions of the small-x
gluons in the present effective theory are obtained from 
the following generating functional:
\be\labe{PART}
{F}[j^\mu_a]\,=\,\int {\cal D}\rho\,\,W_\Lambda[\rho]
\,\,\ln\left(\int^\Lambda {\cal D}A\,
\delta(A^+)\,\,{\rm e}^{\,iS[A,\,\rho]-i\int j\cdot A}\right),\ee
where the logarithm is taken {\it inside} the integral over $\rho$.
That is, the free-energy reaches its extremum as a function of the external
source $j$ for a fixed distribution of the color sources. The {\it measured}
free-energy (or correlation function) is finally obtained by also averaging
over $\rho$.

Note that the presence of a non-trivial color charge background 
breaks gauge symmetry explicitly. But this symmetry is restored in the
process of averaging over $\rho$ provided the weight function 
is {\it gauge invariant}, which we shall assume in what follows.

\subsection{The classical color field}
\label{CLASSMV}

In this subsection, we shall construct the solution to the classical EOM 
(\ref{cleq0}). We note first that, for a large class of gauges,
one can always find a solution with the following properties \cite{CARGESE}:
\be\label{YMprop}
F^{ij}_a=0,\qquad A^-_a=0,\qquad A^+_a,\,A^i_a\,:\,\,{\rm static}\,,
\ee
where ``static'' means independent of $x^+$.
This follows from
the specific structure of the color source which has just a ``$+$''
component, and is static. Since $F^{ij}=0$, the transverse fields $A^i$
form a two-dimensional pure gauge; that is, there exists a gauge rotation
$U(x^-,x_\perp)\in {\rm SU}(N)$ such that:
\be
{A}^i(x^-,x_{\perp})\,=\,{i \over g}\,
U(x^-,x_{\perp})\,\partial^i U^{\dagger}(x^-,x_{\perp})\,.
\labe{tpg} 
\ee
(in matrix notations appropriate for the adjoint representation:
${A}^i={ A}^i_a T^a$, etc).
Thus, the requirements (\ref{YMprop}) leave just two independent
field degrees of freedom, $A^+(\vec x)$ and $U(\vec x)$, which are
further reduced to one (either $A^+$ or $U$)
by imposing a gauge-fixing condition.

We consider first the covariant gauge 
$\partial_\mu {A}^\mu =0$. By eqs.~(\ref{YMprop}) and (\ref{tpg}),
this implies $\partial_i A^i=0$, or $U=0$. Thus, in this gauge, 
$\tilde {\cal A}^\mu_a(x)\,=\,\delta^{\mu +}\alpha_a(x^-,x_{\perp})$, 
with $\alpha_a(\vec x)$ linearly related to the color source
$\tilde\rho_a$ in the COV-gauge :
\be\labe{EQTA}
- \nabla^2_\perp \alpha_a({\vec x})\,=\,{\tilde \rho}_a(\vec x)\,.
\ee
The only non-trivial field strength is the electric field
$\tilde {\cal F}^{+i}_a\,=\,-\partial^i\alpha_a$. Eq.~(\ref{EQTA})
has the solution :
\be\labe{alpha}
\alpha_a (x^-,{ x}_\perp)&=&\int d^2y_\perp\,
\langle x_\perp|\,\frac{1}{-\grad^2_\perp}\,|y_\perp\rangle\,
\tilde\rho_a  (x^-,{ y}_\perp)\nn&=&\int \frac{d^2y_\perp}{4\pi}\,
\ln\frac{1}{({x}_\perp - {y}_\perp)^2\mu^2}\,
\tilde\rho_a  (x^-,{ y}_\perp),\ee
where the infrared cutoff $\mu$ is necessary to invert the
Laplacian operator in two dimensions, but it will eventually
disappear from (or get replaced by the confinement scale
$\Lambda_{QCD}$ in) our subsequent formulae.

We shall need later
the classical solution in the LC-gauge $A^+=0$. This is of
the form ${\cal A}^\mu_a=\delta^{\mu i}{\cal A}^i_a$ with
${\cal A}^i_a(x^-,{ x}_\perp)$ a ``pure gauge'', cf. eq.~(\ref{tpg}).
The gauge rotation $U(\vec x)$ can be most simply obtained by a gauge rotation
of the solution in the COV-gauge:
\be\label{COVLC}
{\cal A}^\mu\,=\,U\bigl(\tilde {\cal A}^\mu+\frac{i}{g}\partial^\mu
\bigr)U^\dagger,\ee
where the gauge rotation $U(\vec x)$ is chosen such that ${\cal A}^+=0$,
i.e.,
\be
U^{\dagger}(x^-,x_{\perp})=
 {\rm P} \exp
 \left \{ig \int_{-\infty}^{x^-} dz^-\,\alpha_a (z^-,x_{\perp}) T^a
 \right \}.\labe{UTA}
\ee
The lower limit $x^-_0\to -\infty$ in the integral over $x^-$ in 
eq.~(\ref{UTA}) has been chosen such as to impose
the ``retarded'' boundary condition:
\be\label{retAi}
A^i_a(x)\,\to\,0 \quad {\rm as} \quad x^-\,\to\,-\infty,\ee
which will be useful in what follows.
(Note that the ``retardation'' 
property refers here to $x^-$, and not to time.)

Together, eqs.~(\ref{tpg}), (\ref{alpha}) and (\ref{UTA}) provide 
an explicit expression for the  LC-gauge solution ${\cal A}^i$ in 
terms of the color source ${\tilde \rho}\,$ in the { COV}-gauge.
This is sufficient for the purpose of computing observables since 
the average in Eq.~(\ref{clascorr}) can be re-expressed as a functional 
integral over the covariant gauge color source $\tilde \rho$ by 
a change of variables:
\be\labe{COVclascorr}
\langle A^i(x^+,\vec x)A^j(x^+,\vec y)
\cdots\rangle_{\Lambda^+}\,=\,
\int {\cal D} \tilde\rho\,\,W_{\Lambda^+}[\tilde\rho]\,{\cal A}^i_{x}
[\tilde \rho]\,{\cal A}^j_{y}[\tilde \rho]\cdots\,.
\ee
\begin{figure} 
\begin{center} 
\includegraphics[width=0.6\textwidth]
{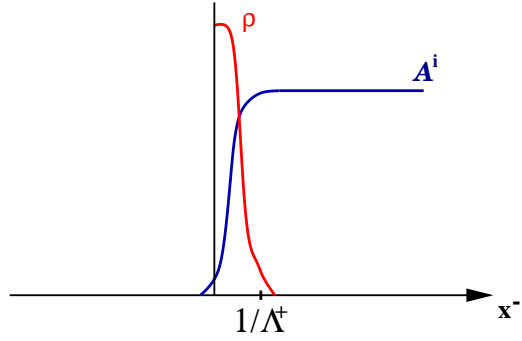} 
\caption{The longitudinal structure of the color source $\rho$
and of the classical field solution ${\cal A}^i$ for the effective
theory at the scale $k^+\,$. As functions of $x^-$,
$\alpha$ and ${\cal F}^{+i}$ are
as localized as $\rho$.}
\label{YM-PLOT} 
\end{center}
\end{figure}

Up to now, the longitudinal structure of the source 
has been arbitrary: the solutions written
above hold for any function $\rho^a(x^-)$. 
For what follows, however, it is useful to recall,
from Sect. \ref{EFT-MV}, that $\rho$ has is localized near $x^-=0$.
More precisely, the quantum analysis in Sect. \ref{SRGE}
will demonstrate that the classical source at the longitudinal
scale $\Lambda^+$ has support at
$x^-$ within the range $0\le x^-\le 1/\Lambda^+$. From 
Eq.~(\ref{alpha}), it is clear that
this is also the longitudinal support of the ``Coulomb''-field $\alpha$.
Thus, integrals over $x^-$ as that in eq.~(\ref{UTA}) receive
their whole contribution from $x^-$ in this limited range.
Any probe with momenta $q^+\ll \Lambda^+$, and therefore a
 much lower longitudinal resolution, will not be able to 
discriminate the internal structure of the source.
Rather, it will see a source/field structure 
which is singular at $x^-=0$ : 
$\rho_a(x^-,x_{\perp})\approx \delta(x^-)\rho_a(x_{\perp})$
(see Fig.~\ref{YM-PLOT}). In particular:
\be\labe{APM}
{\cal A}^i(x^-,x_\perp) \approx \theta(x^-)\,
\frac{i}{g}\,V(\del^i V^\dagger)(x_{\perp}),\ee
where $V$ and $V^\dagger$ are obtained by letting
$x^-\to\infty$ in  Eq.~(\ref{UTA}) :
\be\label{v}
V^\dagger(x_{\perp})\,\equiv\,{\rm P} \exp
 \left \{
ig \int_{-\infty}^{\infty} dz^-\,\alpha (z^-,x_{\perp})
 \right \}.\ee

   \subsection{The gluon distribution} 

We denote by $G(\xx,Q^2) d\xx$ the number of gluons in the
hadron wavefunction having longitudinal momenta between 
$\xx P^+$ and $(\xx+d\xx)P^+$, and a transverse size $\Delta x_\perp \sim 1/Q\,$.
In other terms, the {\em gluon distribution}
$\xx G(\xx,Q^2)$ is the number of gluons with transverse momenta  
$k_\perp \simle Q$ per unit rapidity (see Refs. \cite{AMCARGESE,CARGESE}
for more details) :
\be\labe{GDFdef}
\xx G(\xx,Q^2)&=&\int^{Q^2} {d^2k_\perp}\,
k^+\frac{dN}{dk^+d^2k_\perp}\bigg|_{k^+=xP^+}\nn&=&
\int {d^3k}\,\Theta(Q^2-k_\perp^2)\,
\xx\delta(\xx-k^+/P^+)\,\frac{dN}{d^3 k}\,,\ee
where $\Theta(x)$ is the step function, $\vec k \equiv (k^+,{\bf k}_\perp)$ and 
\be
\label{TPS}
\frac{dN}{d^3 k}\,=\langle   
a^{i\dagger}_{c}(x^+, \vec k)\,a_{ c}^i(x^+, \vec k)  \rangle
=  \frac{2k^+}{(2 \pi)^3}\, \langle    A^i_c(x^+,\vec k)
A^i_c(x^+,-\vec k)\rangle\,\,,\ee
is the Fock space gluon density, namely, the
number of gluons per unit of volume in momentum space. 

The difficulty is, however, that this number depends upon the
gauge, so in general it is not a physical observable. Still, as we will 
shortly argue, this quantity can be given a gauge-invariant
meaning when computed in the light-cone gauge  $A^+_a=0$.
Using the fact that, in this gauge, $F_a^{i+}(k)=ik^+ A_a^i(k)$, one obtains
(with $k^+=\xx P^+$ from now on):
\be\labe{GDF}
\xx G(\xx,Q^2)=\frac{1}{\pi}\int {d^2k_\perp \over (2 \pi)^2}\,\Theta(Q^2-
k_\perp^2)\bigl\langle F^{i+}_a(\vec k)
F^{i+}_a(-\vec k)\bigr\rangle,\ee
which so far does not look gauge invariant.  A manifestly gauge invariant 
operator can be constructed by appropriately inserting Wilson lines \cite{PI,CARGESE}. 
In LC gauge, this gauge 
invariant expression reduces to Eq.~(\ref{GDF}) once the residual gauge freedom of the 
transverse components of the gauge field is fixed by imposing the 
``retarded'' boundary condition (\ref{retAi}) \cite{CARGESE}. This particular gauge fixing
in the classical field problem has important consequences for the quantum calculation
in Sect. \ref{SRGE}, in that it fixes the $i\epsilon$ prescription to be used
for the `axial pole' in the LC-gauge gluon propagator \cite{PI}.

We shall need later also the gluon density in the transverse phase-space 
(also referred to as the ``unintegrated gluon distribution'', or the ``gluon
occupation number''). This is defined as:
\be\label{phidef}
\varphi_\tau(\kk)\,\equiv\,\frac{4\pi^3}{N_c^2-1}\,
\frac{1}{\pi R^2}\,\frac{d^3 N}{d\tau d^2k_\perp}\,=\,\frac{1}{\pi R^2}\,
\frac{
\langle F^{i+}_a(\vec k)F^{i+}_a(-\vec k) \rangle}{N_c^2-1}\,,\ee
where $\tau=\ln(1/\xx)=\ln(P^+/k^+)$. Up to the factor $4\pi^3$, this
is the number of gluons of each color per unit rapidity per unit of transverse phase--space.
(As before, we assume a homogeneous distribution in the transverse plane, for simplicity.)

For illustration, let us compute the gluon distribution of a nucleus in the MV model.
We start with the low density regime, valid when the atomic number $A$ is not too high, 
so the corresponding classical field is weak and
can be computed in the linear approximation.
By expanding the general solution (\ref{tpg}) to linear order in 
$\rho$, or, equivalently, by directly solving the linearized version
of Eq.~(\ref{cleq0}), one easily obtains:
\be
{\cal A}^i_a(k)\,\simeq\, -{k^i 
\over k^+ +i \varepsilon} {\rho_a(k^+,k_{\perp}) 
\over k_{\perp}^2}\,,\qquad {\cal F}^{+i}_a(k)
\,\simeq\,i{k^i \over k_{\perp}^2}\,
{\rho_a(\vec k)}\,,
\labe{aaimom}
\ee
which together with  Eq.~(\ref{MV-corr}) implies:
\be\label{linphi}
\langle{\cal F}^{i+}_a(\vec k){\cal F}^{i+}_a(-\vec k)\rangle_A
\simeq\frac{1}{ k_{\perp}^2}\,\langle\rho_a(\vec k)\rho_a(-\vec k)\rangle_A
\,=\,\pi R^2_A(N_c^2-1)\frac{\mu_A^2}{ k_{\perp}^2}\,.\,\,\,\,\ee
By inserting this approximation in Eqs~(\ref{phidef}) and (\ref{GDF}),
one obtains the following estimates for the gluon density and
distribution function:
\be\label{MV-LOW}
\varphi_A(k_\perp)&\simeq&\frac{\mu_A^2}{ k_{\perp}^2}\,,
\\
\xx G_A(\xx,Q^2)&\simeq&\frac{(N_c^2-1)R_A^2}{4\pi}\,\mu_A^2\int^{Q^2}_{\Lambda_{QCD}^2}
\frac{dk_\perp^2}{ k_{\perp}^2}\,=\,\frac{\alpha_s AN_cC_f}{\pi}\,
\ln\frac{Q^2}{\Lambda_{QCD}^2}\,,\nonumber\ee
(with $\alpha_s=g^2/4\pi$). The integral over $ k_{\perp}$
in the second line has a logarithmic infrared divergence which
has been cut off at the scale $\Lambda_{QCD}$ since we know that,
because of confinement, there is color neutrality on the nucleon size
$R_0\sim 1/\Lambda_{QCD}$ \cite{LM00}. We will argue later that, after
taking into account quantum evolution, the actual scale for the 
screening of the infrared physics is not $\Lambda_{QCD}$ but the saturation 
scale $Q_s$.

Eqs.~(\ref{MV-LOW}) are in fact the expected results, which
could have been obtained also by a direct analysis of the gluon 
radiation by a single quark, together
with the assumption that gluons radiated by different quarks do
not interact with each other, so that the total gluon distribution
is simply the sum of independent contributions from 
the $A\times N_c$ valence quarks. This is the Weizs\"acker--Williams approximation 
for radiation off independent quarks.

   \subsection{Gluon saturation in a large nucleus}

According to eq.~(\ref{MV-LOW}), the gluon density in the
transverse phase-space is proportional to $A^{1/3}$, and becomes
arbitrarily large when $A$ increases. This is however an artifact
of our previous approximations which have neglected the interactions
among the radiated gluons, i.e., the non-linear effects in the
classical field equations. To see this, one needs to recompute
the gluon distribution by using the exact, non-linear solution
for the classical field, as obtained in Sect. \ref{CLASSMV}.
By using  ${\cal F}^{+i}_a (\vec x)= U_{ab}^\dagger(-\partial^i\alpha^b)$,
one can express the relevant LC-gauge field-field correlator in terms
of the color field in the COV-gauge:
\be\label{FF0}
\langle {\cal F}^{+i}_a(\vec x) {\cal F}^{+i}_a
(\vec y) \rangle_A &=&
\bigl\langle\bigl (U_{ab}^\dagger\partial^i\alpha^b\bigr)_{\vec x}\,
\bigl(U_{ac}^\dagger\partial^i\alpha^c\bigr)_{\vec y}\bigr\rangle_A. \ee
One can show that the RHS of this expression can be written as 
\be\label{CONTRACTION}
\bigl\langle\bigl (U_{ab}^\dagger\partial^i\alpha^b\bigr)_{\vec x}\,
\bigl(U_{ac}^\dagger\partial^i\alpha^c\bigr)_{\vec y}\bigr\rangle
=\left\langle \partial^i\alpha^b(\vec x)
\partial^i\alpha^c(\vec y)\right\rangle
\left\langle U_{ab}^\dagger 
(\vec x) U_{ca}(\vec y)\right\rangle\nn
=\delta(x^- -y^-)\langle{\rm Tr}\,U^\dagger(\vec x) U(\vec y)
\rangle\big(-\grad^2_\perp\gamma_A(x^-,x_\perp-y_\perp)\big),\quad\ee
where we have used $U_{ac}^\dagger=U_{ca}$ in the adjoint
representation. Here we have made use of the following correlation function 
\be\label{MV-aa}
\langle \alpha_a(\vec x)\alpha_b(\vec y)
\rangle_A&=&
\delta_{ab}\delta(x^- -y^-)\,\gamma_A(x^-,x_\perp-y_\perp),\nn
\gamma_A(x^-,k_\perp)&\equiv&\frac{1}{ k_{\perp}^4}\,\lambda_A(x^-)\, ,\ee
which follows from
 $\tilde\rho^a(x^-, k_\perp)=k_{\perp}^2\alpha^a(x^-, k_\perp)$ together with Eq.~(\ref{MV-corr}).

\begin{figure} 
\begin{center} 
\includegraphics[width=0.7\textwidth]
{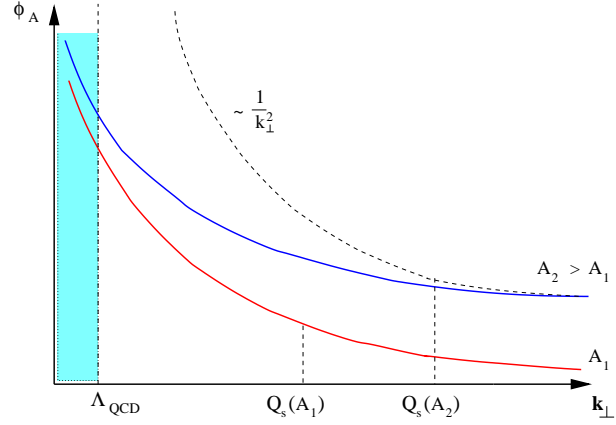} 
\caption{The gluon phase-space density ${\varphi}_A(k_\perp)$
of a large nucleus
(as described by the MV model) plotted as a function of $k_\perp$
for two values of $A$. Notice the change from 
a $1/k_\perp^2$ behaviour at large
momenta $k_\perp > Q_s$ to a logarithmic behaviour at small 
momenta $k_\perp < Q_s$.
}
\label{SATURATION-MV} 
\end{center}
\end{figure}

Eq.~(\ref{CONTRACTION}) can be proven \cite{CARGESE} using rotational
symmetry, the path ordering of the Wilson lines in $x^-$, and
 the fact that the 2-point function of the color fields,
Eq.~(\ref{MV-aa}), is local in $x^-$. The trace in Eq.~(\ref{CONTRACTION}),  
\be\label{SAdef}
S_A(x^-, x_\perp-y_\perp)\,\equiv\,\frac{1}{N_c^2-1}\,
\langle{\rm Tr}\,U^\dagger(x^-,x_\perp) U(x^-,y_\perp)\rangle_A\,\,,\ee
can be explicitely computed as \cite{JKMW97,CARGESE} 
\be\label{Stau}
S_A(x^-, r_\perp)={\rm exp}\big\{\!-g^2N_c\!\int_{-\infty}^{x^-}\! dz^-
[\gamma_A(z^-, 0_\perp)-\gamma_A(z^-, r_\perp)]\big\},\,\ee
where (cf. Eq.~(\ref{MV-aa}))
\be\label{mux}
\gamma_A(x^-,0_\perp)-\gamma_A(x^-,r_\perp)\,=\,\lambda_A(x^-)\int 
{d^2k_\perp\over (2\pi)^2}\,
\frac{1}{k_\perp^4}\,\bigl[1-
{\rm e}^{ik_\perp\cdot r_\perp}\bigr]\,.
\label{xiox}\ee
The above integral over $k_\perp$ is dominated by soft momenta,
and has a logarithmic infrared divergence which, in this classical context,
can be screened only by confinement at the scale $\Lambda_{QCD}$.
To leading-log accuracy, i.e., by keeping only terms enhanced by the large logarithm
$\ln(1/r_\perp^2\Lambda^2_{QCD})$, the precise value of the infrared cutoff is not
important, and we can also expand the integrand as:
\be\label{xi1}
\int {d^2k_\perp\over (2\pi)^2}
\frac{1-
{\rm e}^{ik_\perp\cdot r_\perp}}{k_\perp^4}\simeq
\int\limits^{1/r_\perp^2} {d^2k_\perp\over (2\pi)^2}
\frac{1}{k_\perp^4}{(k_\perp\cdot r_\perp)^2\over 2}
\simeq{r_\perp^2\over 16\pi}
\ln{1\over r_\perp^2\Lambda^2_{QCD}}\,.\nonumber\ee
This gives, with $\mu_A^2(x^-) \equiv \int_{-\infty}^{x^-}dz^-\lambda_A(z^-)$, 
\be\label{SMV}
S_A(x^-, r_\perp)\,\simeq\,
{\rm exp}\left\{-\,\frac{\alpha_s N_c}{4}\, r_\perp^2 \,\mu_A^2(x^-)\,
\ln{1\over r_\perp^2\Lambda^2_{QCD}}\right\},\ee
which together with Eq.~(\ref{CONTRACTION}) can be used to finally
evaluate the gluon density in Eq.~(\ref{phidef}). After simple manipulations, one
obtains \cite{JKMW97,KM98} 
\be\label{NMV}
\varphi_A(k_\perp)=
\int d^2r_\perp {\rm e}^{-ik_\perp\cdot r_\perp}\,
\frac{1-{\rm exp}\big\{-\frac{1}{4}\, r_\perp^2 Q_A^2
\ln{1\over r_\perp^2\Lambda^2_{QCD}}\big\}}{\pi \alpha_s N_cr_\perp^2 }\,,\ee
where 
\be\label{QAMV}
Q_A^2\,\equiv\,\alpha_s N_c \mu_A^2\,=\,\alpha_s N_c 
\int dx^-\lambda_A(x^-)\,\sim\,A^{1/3}.\ee
To study the $k_\perp$--dependence of Eq.~(\ref{NMV}),
one must still perform the Fourier transform,
but the result can be easily anticipated: 

Let us first introduce the {\it saturation momentum} $Q_s(A)$ which, as we shall see,
is the scale separating between linear and non-linear behaviours. This is defined
by the condition  that, for $r_\perp = 2/Q_s(A)$, the exponent
in Eq.~(\ref{NMV}) becomes of order one, which gives:
\be\label{QSMV}
Q_s^2(A)\,\simeq\,\alpha_s N_c \mu_A^2\,\ln{Q_s^2(A)\over \Lambda^2_{QCD}}\,\sim\,
A^{1/3}\ln A\,.\ee
Note that this is larger than $Q_A$, eq.~(\ref{QAMV}), since we work in the hypothesis
that $Q_A\gg \Lambda_{QCD}$. Then we distinguish between two regimes:

i) At high momenta
$k_\perp\!\gg\! Q_s(A)$, the integral is dominated by small
$r_\perp\!\ll \!1/ Q_s(A)$, and can be evaluated by
expanding out the exponential. To lowest non-trivial order (which
corresponds to the linear approximation), one
recovers the bremsstrahlung spectrum of eq.~(\ref{MV-LOW}):
\be\label{High-N}
\varphi_A(k_\perp)\,\approx\,
\frac{ 1}{\alpha_s N_c }\,\frac{Q_A^2}{k_\perp^2}
\,=\,\frac{\mu_A^2}{k_\perp^2}\qquad {\rm for}\quad k_\perp\gg Q_A.\ee
ii) At small momenta, $k_\perp\!\ll \! Q_s(A)$, the dominant contribution
comes from large distances $r_\perp\!\gg\! 1/ Q_s(A)$, where one
can simply neglect the exponential in the numerator and recognize
$1/r_\perp^2$ as the Fourier transform\footnote{The saturation scale
provides the ultraviolet cutoff for the logarithm in eq.~(\ref{Low-N})
since the short distances $r_\perp\ll 1/ Q_s(A)$ are cut off
by the exponential in eq.~(\ref{NMV}).} of $\ln k_\perp^2\,$:
\be\label{Low-N}
\varphi_A(k_\perp)\,\approx\,
\frac{ 1}{\alpha_s N_c }\ln\frac{Q_s^2(A)}{k_\perp^2}
\qquad {\rm for}\quad k_\perp\ll Q_A.\ee
Unlike the linear distribution (\ref{High-N}), which
grows like $A^{1/3}$, and is strongly infrared dominated (as it goes like 
$1/k_\perp^2$), the distribution in Eq.~(\ref{Low-N}),
which takes into account the non-linear effects in the classical Yang-Mills
equations,  rises only logarithmically as a function of both $A$ and
$1/k_\perp^2$. This is {\it saturation}. At saturation, the gluon occupation factor
is parametrically of order 
$1/\alpha_s$, which corresponds to a {\it Bose condensate}, and
is the maximum density allowed by the repulsive interactions
between the strong color fields $\bar A^i=\sqrt{\langle A^iA^i\rangle}\sim
1/g$. When increasing the atomic number $A$, the new gluons are
produced preponderently at large transverse momenta $\simge Q_s({A})$.
where this repulsion is less important. This is illustrated in Fig. 
\ref{SATURATION-MV}.

To clarify the physical interpretation of the saturation scale,
note that, at short-distances $r_\perp\!\ll \!1/ Q_s({A})$,
\be\label{QSMV1}
\mu_A^2\,
\ln{1\over r_\perp^2\Lambda^2_{QCD}}\,\propto\,
\frac{\xx G(\xx, 1/r_\perp^2)}{(N^2_c-1)\pi R_A^2}\,\ee
is the number of gluons (of each color) having tranverse
size $r_\perp$ per unit of transverse area (cf. eq.~(\ref{MV-LOW})).
Since each such a gluon carries a color charge squared $(gT^a)(gT^a)
=g^2N_c$, we deduce that
$\alpha_s N_c\,\mu_A^2\, \ln{1\over r_\perp^2\Lambda^2_{QCD}}$ 
is the average color charge squared of the
gluons having tranverse size $r_\perp$ per unit area and per color. 
Then, eq.~(\ref{QSMV}) is the condition that the total color
charge squared within the area occupied by each gluon is of
order one. This is the original criterion of saturation by
Gribov, Levin and Ryskin \cite{GLR}, 
for which the MV model offers an explicit realization.

Let us finally compute the (integrated) gluon distribution in the saturation
regime, i.e., for $Q^2 \ll Q_s^2({A})$, and compare with the corresponding result
in the linear regime at high $Q^2$, Eq.~(\ref{MV-LOW}). By using (\ref{Low-N}) in
Eqs.~(\ref{GDF})--(\ref{phidef}), one immediately finds:
\be\label{xGxsat}
xG_A(x,Q^2)\,\simeq\,\frac{N^2_c-1}{4\pi N_c}\,\frac{1}{\alpha_s}\,R^2_AQ^2\,\ln{Q_s^2({A})\over
Q^2}\,.\ee
This shows strong {\it nuclear shadowing} : it scales like $A^{2/3}\ln A$, unlike the
linear result (\ref{MV-LOW}), which scales like $A$. 

\subsection{Dipole-hadron scattering at high energy}
\label{DHS}

Although the previous definition of the gluon distribution in terms of the Fock space
gluon density is useful for a conceptual
discussion of saturation, it is on the other hand less clear whether it
corresponds to something that could be directly measured in experiments.
Recall that, in the lowest-order analysis of deep inelastic scattering where one neglects
non-linear effects in the hadron wavefunction, the gluon distribution is related
to the scaling violation in the  hadron structure function $F_2$ :
$\del F_2/\del\ln Q^2 \propto \alpha_s \xx G(\xx,Q^2)$.
It is therefore interesting to compute this quantity also in the presence of
non-linear effects, and identify some measurable consequences of saturation.
This is what we shall do starting with this section, first in the framework of the MV model,
then by including the effects of the quantum evolution in going towards smaller values
of $\xx$ (in Sects. 3 and 4). As a general conclusion, we shall find that saturation effects
in the hadron wavefunction correspond to {\it unitarity} effects in the high-energy
virtual-photon--hadron collision, whose total cross-section is related to $F_2$ via:
\be \label{F2gamma}
\sigma_{\gamma^* H}(\xx,Q^2)\,= \,{4 \pi^2 \alpha_{em}\over Q^2}\,F_2(\xx,Q^2)\,.
\ee

As we have seen in Sect. \ref{SEC-DIS}, when viewed in the infinite momentum frame (IMF)
of the hadron, DIS appears as the scattering of the virtual photon off a quark (with longitudinal
fraction equal to the Bjorken x of the collision) in the hadron wavefunction. At very small x,
this quark is typically not a valence quark, but rather a see quark which is emitted, most probably,
off the small-x gluons. It is then convenient to disentangle this final quark emission from
the quantum evolution which involves mostly gluons. This can be done by performing a Lorentz boost 
in such a way to pull the $\gamma^* q \bar q$ vertex  out of the hadron. That is, instead of the
hadron IMF, it is preferable to use the so-called {\it dipole frame} \cite{AMCARGESE} 
(and references therein) in which most of the energy
is still carried by the hadron (so that the high density effects are again associated with the
hadron wavefunction), 
but the virtual photon moves in the negative $z$ (or positive $x^-$) direction with
enough energy to dissociate before scattering into a quark--antiquark pair in a color singlet
state (a {\it color dipole}), which then scatters off the hadron.
This sequential picture of DIS is appropriate at high energy, since the lifetime
of the $q\bar q$ pair is much larger than the interaction time between
this pair and the hadron.

More precisely, if $\tau\equiv \ln(1/x) = \y_{H} - \y_{\gamma^*}$ is the (boost-invariant)
rapidity gap, with $\tau\gg 1$ at small x, then the dipole frame corresponds to choosing
$\y_{\gamma^*}\ll \y_{H}$ and such that $\alpha_s \y_{\gamma^*}\ll 1$, which ensures that
one can neglect additional gluon radiation from the quark and the antiquark in the dipole.

This physical picture translates into the following factorization formula 
for $\sigma_{\gamma^* H}$ \cite{AM0,NZ91,MV99}, that we shall derive in Sect. 4.1:
\be\label{sigmagamma}
\sigma_{\gamma^* H}(\tau,Q^2)\,=\,
\int_0^1 dz \int d^2r_\perp\,|\Psi(z,r_\perp;Q^2)|^2\,
\sigma_{\rm dipole}(\tau,r_\perp).\ee
Here, $\Psi(z,r_\perp;Q^2)$ is the light-cone wavefunction for the photon
splitting into a $q\bar q$ pair with transverse size $r_\perp$ and
a fraction $z$ of the photon's longitudinal momentum carried
by the quark\cite{AM0,NZ91}. Furthermore,
$\sigma_{\rm dipole}(\tau,r_\perp)$ is the dipole--hadron total cross section
for a dipole of transverse size $r_\perp=x_{\perp}-y_{\perp}$ (with the quark
located at $x_{\perp}$ and the antiquark at $y_{\perp}$), 
and is obtained by integrating the (imaginary part) of the scattering amplitude
${\cal N}_\tau(x_{\perp}, y_{\perp}) = {\cal N}_\tau(r_{\perp},b_{\perp})$
over all the impact parameters $b_\perp=(x_{\perp}+y_{\perp})/2\,$ :
\be\label{sigmadipole}
\sigma_{\rm dipole}(\tau,r_\perp)\,=\,2\int d^2b_\perp\,
{\cal N}_\tau(r_{\perp},b_{\perp})\,.\ee
At high energy, the dipole-hadron scattering can be treated
in the eikonal approximation\cite{AM0,BH,Balitsky1996}.
This amounts to neglecting the recoil of the quark (or the antiquark) during its
scattering off the color field in the target: the whole effect of the scattering 
consists in a color precession. Then, the scattering amplitude reads
${\cal N}_\tau(x_{\perp}, y_{\perp})= 1- S_\tau(x_{\perp},y_{\perp})$, with
the following $S$-matrix element:
\be\label{Ntau}
S_\tau(x_{\perp},y_{\perp})\,\equiv\,\frac{1}{N_c}\,
\big\langle {\rm tr}\big(V^\dagger(x_{\perp}) V(y_{\perp})\big)
\big\rangle_\tau,\ee
where $V^\dagger$ and $V$ are the Wilson lines describing the eikonal interaction between
the quark (or the antiquark) and the color field at rapidity $\tau$
due to color sources within the hadron: 
\be\label{Wilson0}
V^\dagger(x_{\perp})\,=\,{\rm P}\,{\rm exp}\left(ig\int_{-\infty}
^\infty dx^- A^+_a(x^-,x_{\perp}) t^a\right).\ee
That is, this is the same as Eq.~(\ref{v}), but rewritten
in the fundamental representation. The average in Eq.~(\ref{Ntau}) is over all the
configurations of the color fields in the hadron. In the CGC formalism, where
$A^+_a=\alpha_a$, cf. Eq.~(\ref{EQTA}), this average
is computed as in Eq.~(\ref{COVclascorr}).

In what follows we shall focus on the computation of the $S$-matrix element 
(\ref{Ntau}), which encodes all the information about the hadronic scattering, and
thus about the non-linear and quantum effects in the hadron wavefunction.
[Once this is known, $F_2$ can be immediately obtained by using 
Eqs.~(\ref{sigmagamma}) and (\ref{F2gamma}).]

In the MV model, to which we shall restrict in the remaining part of this subsection,
this  $S$-matrix element is already known, as obvious when comparing 
Eqs.~(\ref{Ntau}) and (\ref{SAdef}). By translating Eq.~(\ref{SMV}) to the 
fundamental representation ($N_c \to C_F = (N_c^2-1)/2N_c$) and letting $x^-\to\infty$
there, one obtains:
\be\label{SMV2}
S_A(r_{\perp})\,\simeq\,{\rm exp}\bigg\{-\frac{r_\perp^2 \bar Q_A^2}{4}
\ln{1\over r_\perp^2\Lambda^2_{QCD}}\bigg\},\ee
where $\bar Q_A^2=\alpha_s C_F\mu_A^2$ differs only via a color factor from Eq.~(\ref{QAMV}).
As in the previous discussion of the gluon distribution, we distinguish between
a small--$r_{\perp}$ and a large--$r_{\perp}$ regime, with the separation between the two
regimes given by the saturation scale $\bar Q_s^2(A)$, defined by analogy with Eq.~(\ref{QSMV}).

i) A small dipole, with $r_\perp\!\ll \!1/\bar Q_s(A)$, is only
weakly interacting with the hadron:
\be\label{transp}
{\cal N}_A(r_{\perp})\equiv 1-S_A(r_\perp)\,\approx\,\frac{1}{4}\, r_\perp^2 \bar Q_A^2
\ln{1\over r_\perp^2\Lambda^2_{QCD}}\,\,\ll\,\,1,\ee
a phenomenon usually referred to as ``color transparency'' \cite{CTRAN}.

ii) A relatively large dipole, with  $r_\perp\! \gg \!1/ \bar Q_s$ (but 
$r_\perp\ll 1/\Lambda_{QCD}$ though, for the present perturbative treatment to apply)
is strongly absorbed : $S_A\ll 1$ or
${\cal N}_A(r_\perp)\,\approx\,1$
which corresponds to the unitarity (or ``black disk'') limit.
This is in sharp contrast with the single--scattering approximation in Eq.~(\ref{transp}),
which would predict a scattering amplitude growing indefinitely with the area $r_\perp^2$
spanned by the dipole, and also with the atomic number $A$ of the target. Eq.~(\ref{SMV2}) 
shows that, when $r_\perp$ and/or $A$ are large enough, the effects of multiple
scattering become non-negligible, and eventually ensure the unitarization of the 
scattering amplitude at fixed impact parameter\footnote{The impact parameter dependence
was omitted in writing Eq.~(\ref{SMV2}) since trivial for the case of
a homogeneous target. This will be reintroduced later, when needed.}. Note that,
according to Eq.~(\ref{SMV2}), the multiple scatterings are {\it higher-twist} effects,
i.e., their contributions are of higher order in $r_\perp^2$.

We see that, remarkably, it is the same scale --- the saturation momentum
--- which sets the critical transverse size for both gluon saturation and 
the unitarization of the dipole-hadron scattering. 
This conclusion, that we have found here in the framework of the MV model,
will be seen in Sect. 3 to remain valid after including the quantum evolution.
Physically, this can be understood as follows: Both saturation and unitarization 
(when the scattering is seen in the dipole frame) 
require strong color fields in the hadron wavefunction,
such that $g\int dx^- A^+ \sim 1$. $Q_s$ is the critical transverse scale 
at which this strong field condition begins to be satisfied.

What is specific to the present MV model (and, more generally, to any approximation in 
which the color sources are only weakly correlated with each other,
like the gluonic sources in the DGLAP approximation \cite{DGLAP,MQ}) 
is that the dipole scatters {\it independently}
off the color sources in the hadron (here, the valence quarks).  This is best seen
by noticing that Eq.~(\ref{SMV2}) can be rewritten as a Glauber formula :
\be\label{SMV3}
S_A(r_{\perp})\,\simeq\,{\rm exp}\bigg\{- \alpha_s r_\perp^2\,
\frac{\pi^2 C_F}{N_c^2-1}\,\frac{ A x G_N(x,1/r_\perp^2)}{\pi R^2_A}\bigg\},\ee
where $x G_N(x,Q^2)$ is the gluon distribution of a {\it nucleon},
and is given in the present approximation
by the second line in Eq.~(\ref{MV-LOW}) with $A \to 1$.

As we shall see in the next section, the previous picture changes quite substantially
after including quantum corrections, due to the fact that the evolution towards small x 
induces correlations among the color sources. As a result, not only the general Glauber
formula (\ref{SMV3}) becomes inapplicable (the successive scatterings 
are not independent any longer), but even its linearized `leading-twist' approximation,
corresponding to a single scattering, fails to apply when $1/r_\perp$ is close enough to 
$Q_s$, while still above it. This is the BFKL regime where `higher-twists' effects appear
already in the linear evolution. 

Most interestingly, we shall see that gluon saturation at small x holds independently of
the non-linear effects in the classical Yang-Mills equations. Rather, this is
the consequence of the correlations among the color sources induced by {\it non-linear
effects in the quantum evolution}. These same correlations will be shown to ensure color
screening already over the perturbative scale $1/Q_s \ll 1/\Lambda_{QCD}$, which thus
eliminates the infrared sensitivity of the classical MV model to the non-perturbative
physics of confinement (see, e.g., Eqs.~(\ref{NMV}) or (\ref{SMV2})).

\section {The Quantum Evolution of the Color Glass Condensate}
\label{SRGE}

In this section, we shall explain how to construct the CGC effective theory
at small x by integrating out the gluons with $\xx' > \xx$ in perturbation theory, 
in the presence of high density effects. The central result of this analysis
will be a renormalization group equation for the weight function
$W_{\Lambda^+}[\rho]$ in Eq.~(\ref{ZCGC}), which generalizes the BFKL equation
by including non-linear effects, and has important physical consequences
among which gluon saturation.

\setcounter{equation}{0}
\subsection{The BFKL evolution and its small-x problem}
\label{SEC-BFKL}

\begin{figure}[htb]
\centering
\resizebox{.80\textwidth}{!}{%
\includegraphics*{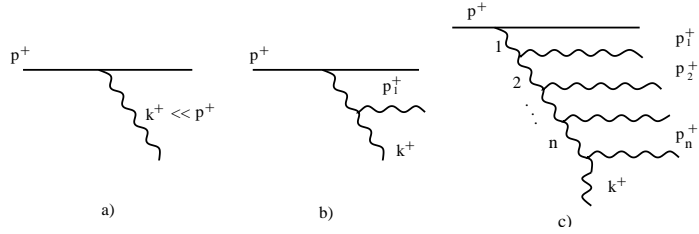}}
   \caption{a) Small--x gluon emission by a fast parton;
b) the lowest-order radiative correction; c) a gluon cascade.}
\label{gluoncascade}
\end{figure}

Within perturbative QCD, the enhancement of the gluon distribution at small x 
proceeds via the gluon cascades depicted in Fig.~\ref{gluoncascade}.
Fig.~\ref{gluoncascade}.a shows
the direct emission of a soft gluon with longitudinal momentum  $k^+ = \xx P^+
\ll P^+$ by a fast moving parton (say, a valence quark) with $p^+ = \xx_0 P^+$
and $1> \xx_0\gg \xx$. Fig.~\ref{gluoncascade}.b displays the lowest-order
radiative correction\footnote{At the same level of accuracy, a complete
calculation must include also the appropriate virtual corrections (self-energy
and vertex renormalization); but for the present, qualitative purposes,
it is sufficient to consider the real gluon emission.}
which is of the order (with $\bar\alpha_s\equiv\alpha_s N_c/\pi$) 
\beq\label{one-gluon}
\frac{\alpha_s N_c}{\pi}
\int_{k^+}^{p^+}\,\frac{d p_1^+}{p_1^+}
\,=\,\frac{\alpha_s N_c}{\pi}
\,\ln\,\frac{p^+}{k^+}\,= \,\bar\alpha_s\ln\frac{\xx_0}{\xx}\,
\eeq
relative to the tree-level process in Fig.~\ref{gluoncascade}.a.
This correction is enhanced by the large { rapidity} interval
$\Delta \tau= \ln(\xx_0/\xx)$ available for the emission of the additional
gluon. 

A similar enhancement holds for the gluon cascade  
in Fig.~\ref{gluoncascade}.c, in which the succesive gluons are strongly ordered
in longitudinal momenta: $p^+\gg p_1^+\gg p_2^+ \gg\,\cdots\,\gg 
p_n^+\gg k^+$. This gives a contribution of relative order
\beq\label{more-gluon}
\frac{1}{n !}
\left(\bar\alpha_s\ln\frac{\xx_0}{\xx}\right)^n\,,\eeq
where the factorial comes from the  ordering in $p^+$.
Clearly, when $\xx$ is so small that $\ln(\xx_0/\xx)\sim 1/\bar\alpha_s$,
all such quantum ``corrections'' become of order one, and must be
resummed for consistency.  A calculation which
includes effects of order $(\alpha_s\ln(1/{\rm x}))^n$ to all orders
in $n$ is said to be valid to ``leading logarithmic accuracy'' (LLA).

The gluon cascades in Fig.~\ref{gluoncascade} contribute all
to the production of (virtual) gluons with longitudinal  
fraction $\xx$. Thus, by resumming these cascades, one can compute the 
number of such gluons per unit rapidity, i.e., the gluon distribution
(\ref{GDFdef}). One can recognize in 
Eqs.~(\ref{one-gluon})--(\ref{more-gluon}) the expansion of an exponential.
Therefore:
\beq\label{expgr}
{dN\over d\tau}\,\equiv \,x G(x,Q^2)\,\sim\,
{\rm e}^{\omega \bar\alpha_s\tau}\,=\,
{x^{-\omega \bar\alpha_s}}\,,\eeq
with $\omega$ a pure number. We have tacitly assumed that all the gluons
in the cascade have transverse momenta of the same order, namely of
order $Q$. A more refined treatment, based on the BFKL equation \cite{BFKL}, 
allows one to compute $\omega$ and specifies the $Q^2$--dependence, and also
the subleading $\tau$--dependence (beyond the exponential behaviour shown
in Eq.~(\ref{expgr})) of the gluon distribution. 

To describe the effects of the BFKL evolution in more detail,
it is instructive to consider the dipole-hadron scattering 
introduced in Sect. \ref{DHS}.
With increasing energy, the gluon fields change in the hadron wavefunction, 
and therefore so does also the cross-section for
the dipole which couples to these fields (cf. Eq.~(\ref{Ntau})).
Specifically, to LLA, and in the linear regime where one can neglect multiple 
scattering,  the amplitude 
${\cal N}_\tau(x_{\perp}, y_{\perp}) = 1- S_\tau(x_{\perp}, y_{\perp}) \equiv {\cal N}_{xy}$ obeys to:
\beq\label{linearBK}
{\del \over {\del \tau}} {\cal N}_{xy}
=\bar\alpha_s\int \!{d^2z_\perp \over 2\pi}\,
\frac{(x_\perp-y_\perp)^2}{(x_\perp-z_\perp)^2(y_\perp-z_\perp)^2 } 
\big\{ {\cal N}_{xz} + {\cal N}_{zy} - {\cal N}_{xy}\big\},\,\,
\eeq
which is the coordinate form of the BFKL equation \cite{BFKL,Mueller94}.

\begin{figure}[htb]
\centering
\resizebox{.8\textwidth}{!}{%
\includegraphics*{{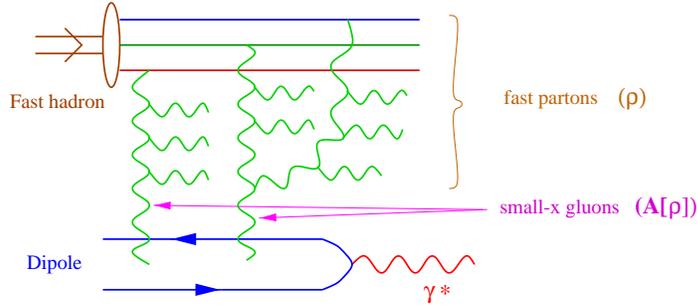}}}
\caption{Deep inelastic scattering in the dipole frame}
\label{DIS2}
\end{figure}

The physical interpretation of this  equation depends upon the Lorentz frame that we
choose to visualize the process. When using the dipole frame of Sect. \ref{DHS}, the quantum 
evolution is put entirely in the wavefunction of the hadron (which is boosted
to higher and higher energies with increasing $\tau$), while the dipole remains
a simple $q\bar q$ pair. In this frame, Eq.~(\ref{linearBK}) describes the
dipole-hadron scattering as the exchange of a BFKL ladder; 
the dipole couples to the last gluon (with the smallest value
of x) in a gluon cascade which develops fully inside the hadron. (See also Fig. \ref{DIS2}.)

Alternatively, by a change of frame, one can
use the increase in the total energy to accelerate the {\it dipole},
and study the evolution of its wavefunction with $\tau$. 
Under an increment $d\tau$ such that $\bar\alpha_s d\tau \sim 1$, the dipole
evolves by emitting one gluon (from either the quark or the antiquark), and the
ensuing $q\bar q g$ state scatters off the hadronic target.
It is convenient (although not necessary) to view this evolved state in the 
large--$N_c$ limit, in which the radiated gluon is effectively replaced by
a $q\bar q$ pair in a color octet state. Then, the evolution looks like the
splitting of the original dipole into two new dipoles, each of them made of
a quark (or antiquark) from the initial dipole and an antiquark (or a quark) 
from the emitted gluon. From this perspective, the various terms in Eq.~(\ref{linearBK})
have a simple interpretation: the quantity 
\be\label{probdip}
\frac{\bar\alpha_s}{2\pi}\,\frac{(x_\perp-y_\perp)^2}{(x_\perp-z_\perp)^2(y_\perp-z_\perp)^2 }\ee
is the differential probability for the initial dipole $(x_\perp,y_\perp)$ to decay
into a pair of dipoles $(x_\perp, z_\perp)$ and $(x_\perp, z_\perp)$,
while ${\cal N}_{xz}$ and ${\cal N}_{zy}$ are
the amplitudes for the scattering between any one of these final dipoles and the target.
Finally, the negative contribution proportional to $ - {\cal N}_{xy}$ represents the decrease
in the scattering amplitude of the original dipole due to its dissociation
(this term is necessary for the conservation of the probability).

One should mention here that this different perspective, in which the quantum evolution is put
in the dipole and studied in the large-$N_c$ limit,
lies at the basis of an approach originally developed by Mueller
\cite{Mueller94} --- the `Color Dipole approach' ---, in which the
wave function of a very energetic color dipole (an ``onium'') is constructed in the
BFKL approximation. In this approach, the scattering between two ``onia'' (physically,
this corresponds, e.g., to the $\gamma^*\gamma^*$ scattering) can be treated as the product of 
the number of dipoles in each onium times the dipole-dipole cross-section. The ensuing
scattering amplitude has been shown to satisfy equation (\ref{linearBK}).
 
The BFKL equation (\ref{linearBK}) can be solved by standard techniques \cite{TB-BFKL}.
At high energy, $\bar\alpha_s \tau \gg \ln (1/r_{\perp}^2\Lambda_{QCD}^2)$, and for a homogeneous target
(e.g., for a large nucleus, and impact parameters near the center of the target), the solution
reads (with $r_{\perp} = x_\perp-y_\perp$)
\be\label{BFKLsol1}
{\cal N}_\tau(\rr)\simeq \sqrt{\rr^2 Q_0^2}\,
{{\rm e}^{\omega\bar\alpha_s\tau}\over\sqrt{2\pi \beta \bar\alpha_s \tau} } 
\exp \left\{ \frac{- \ln^2 \left(\rr^2 Q_0^2\right) }{2\beta \bar\alpha_s\tau} 
\right\}.
\ee
where the reference scale $Q_0$ is introduced by the initial conditions at low energy, 
and thus is of order $\Lambda_{QCD}$ (for a nucleus, this carries the dependence on $A$).
Furthermore, $\omega\!=\!4\ln 2\approx 2.77$ and $\beta\!=\!28\zeta(3)\approx 33.67$.
In writing  Eq.~(\ref{BFKLsol1}) 
we have assumed a fixed coupling $\alpha_s$, as appropriate at leading-order BFKL accuracy.
The modifications due to the running of the coupling  will be discussed in Sect. ??.

Eq.~(\ref{BFKLsol1}) exhibits two essential features of the BFKL approximation,
which eventually provoke its failure in the high energy limit :

{ ({\it a}) {\it Violation of the unitarity bound}} : The solution 
(\ref{BFKLsol1}) increases exponentially with $\tau$, that is, as a power of the energy. At high
energy, such a behaviour violates both the unitarity bound ${\cal N}_\tau(\rr,b_\perp)\le 1$
on the scattering amplitude at fixed impact parameter, 
and the Froissart bound $\sigma_{\rm dipole}(s) \le \sigma_0\ln^2 s$ \cite{Froissart}
on the total dipole cross-section (\ref{sigmadipole}).

{ ({\it b}) {\it Infrared diffusion}} : When seen as a function of $\ln\left(\rr^2 Q_0^2\right)$, 
the solution (\ref{BFKLsol1}) shows a diffusive behaviour, with diffusion `time' proportional
to  $\tau$. With increasing energy, the typical
transverse momenta carried by the gluons within the BFKL ladder
diffuse away from the hard external scale $1/r_{\perp}^2 \gg \Lambda_{QCD}^2$,
and eventually enters the non-perturbative region at $\kk^2 \simle \Lambda^2_{QCD}$, which
contradicts the use of perturbation theory.

The first difficulty is similar to that with the linear approximation
to the scattering amplitude in the MV model, Eq.~(\ref{transp}), which
violates unitarity\footnote{We mean here the unitarity of the
scattering amplitude at fixed impact parameter. The discussion of the
total cross-section is more involved, and deferred to Sect. 4.4.} for
very large $A$. 
As in that case, we expect unitarity to be restored by multiple
scatterings, i.e., by keeping terms of all orders in $gA^+$ in the
Wilson lines in Eqs.~(\ref{Ntau})--(\ref{Wilson0}). 
Both Eqs.~(\ref{transp}) and (\ref{BFKLsol1}) correspond to a single
scattering approximation--they are obtained by retaining
terms which are, at most, quadratic in $gA^+$ in the expansion of the
Wilson lines in Eq.~(\ref{Ntau}).)

However, unlike in the MV model, at small x we expect an {\it additional} source of 
non-linearities which arise from the interactions among the gluonic sources. Such 
interactions lead to the fusion of gluons from different parton cascades (``gluon recombination''), 
a phenomenon which should tame 
the rapid growth of the number of partons. Thus saturation arises from the competition of 
two effects: the growth of the gluonic density due to radiation and its depletion due to 
recombination effects~\cite{GLR,MQ}. Since now the dominant color sources are themselves gluons, 
the saturation we speak of here refers simultaneously to the 
source and the fields radiated by them. This is because what we call ``sources'' and ``radiated
fields'' is only relative, as it depends upon the scale $\Lambda^+ = \xx P^+$ at which we consider
the effective theory. 

Both the multiple scattering and gluon recombination mechanisms are
illustrated in Fig. \ref{DIS2}. We expect both mechanisms to become
important at the same scale, $Q^2_s(\tau,A)$, which is the critical
gluon density at which the non-linear effects become of order one.
This {\it saturation scale} is also the typical transverse momentum of
the gluons in the hadron wavefunction at small x. As anticipated in
the Introduction, and will be verified explicitly in what follows,
this scale increases rapidly with $\tau$ and $A$. The emergence of
such a {\it hard} intrinsic momentum scale can also solve the
`infrared diffusion' problem of the BFKL approximation and therefore
restore the applicability of perturbation theory to high energy
processes in QCD.

In the dipole frame, gluon recombination is seen as the merging of two gluon cascades,
as illustrated in the r.h.s. of Fig. \ref{DIS2}. It is interesting to see this process
also from the boosted frame in which the quantum evolution 
proceeds via the dissociation of the incoming dipole into two dipoles.
After the boost, the final gluon in the cascade
in Fig. \ref{DIS2}---the one which couples to the dipole---gets incorporated
within the dipole wavefunction, so the merging of two cascades now happens
{\it inside the dipole}. Thus, from the boosted frame, the non-linear process is
seen as the {\it simultaneous} scattering of the two final dipoles off
the hadronic target.

These considerations suggest the following simple equation which generalizes
Eq.~(\ref{linearBK}) by taking non-linear effects into account:
\be\labe{BK}
{\del \over {\del \tau}} {\cal N}_{xy}
&=&{\bar \alpha_s}\int {d^2z_\perp\over 2\pi}
\frac{(x_\perp-y_\perp)^2}{(x_\perp-z_\perp)^2(y_\perp-z_\perp)^2 } \nn
&{}&\qquad\times
\left\{ {\cal N}_{xz} + {\cal N}_{zy} - {\cal N}_{xy}
- {\cal N}_{xz}{\cal N}_{zy}\right\}.\ee
This is the equation derived by Kovchegov \cite{K} within
Mueller's `Color Dipole approach' \cite{Mueller94}. 
The main assumption used in its derivation
was the fact the two final dipoles scatter {\it independently} off the target; it is
correct only in the large--$N_c$ limit. Deriving this equation, (together with
its generalizations to finite $N_c$ originally obtained by Balitsky \cite{Balitsky1996}
within a different formalism) within the framework of the CGC effective theory,
will be a main objective of the forthcoming developments in this section.

\subsection{Non-linear evolution for the CGC}

The CGC provides a natural framework for the description of the
non-linear effects in the quantum evolution towards small x, and of the phenomenon of
saturation. The main observation is that, to LLA, all the quantum corrections described
previously---both the exponentially developing BFKL cascade, and gluon recombination
which tames this rapid growth---can be incorporated into a change of the classical color charge and its correlations,
namely, into a renormalization of the weight function $W_{\Lambda^+}[\rho]$ in Eq.~(\ref{ZCGC}).

\begin{figure}[htb]
\centering
\resizebox{.75\textwidth}{!}{%
\includegraphics*{{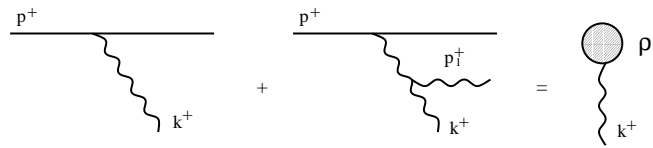}}}
   \caption{Effective color source after including
the lowest-order radiative correction.}
\label{SOURCE}
\end{figure}

To see this at an intuitive level, let us reconsider the first radiative correction, 
the one--gluon emission in  Fig.~\ref{gluoncascade}.b, and note that, to LLA, the
typical contributions to the integral in Eq.~(\ref{one-gluon}) come from momenta $p_1^+$
such that $p^+\gg p_1^+\gg k^+$. That is, the condition of separation of scales is indeed
satisfied for the intermediate gluon with momentum $p_1^+$ to be treated as a `frozen'
color source for the final gluon  with momentum $k^+$. The effect of this quantum 
correction is therefore simply to renormalize the {\it effective} color source at scale $k^+$, as
pictorially illustrated in  Fig. \ref{SOURCE}.

\begin{figure}[htb]
\centering
\resizebox{.6\textwidth}{!}{%
\includegraphics*{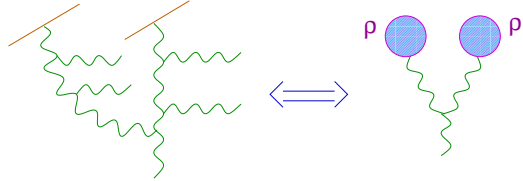}}
\caption{The fusion of two gluon cascades and its interpretation in the CGC theory.}
\label{GLRfig}
\end{figure}

By iterating this argument, it is quite clear that a whole BFKL cascade 
(see Fig.~\ref{gluoncascade}.c) can be included in the definition of the classical
color source at the scale $\Lambda^+ = \xx P^+$ of interest. It is furthermore clear
that the fusion between two gluon cascades, as illustrated in the l.h.s. of Fig. \ref{GLRfig},
can be represented in the CGC theory as a non-linear effect in the {\it classical} dynamics
of the color fields generated by this effective source (see the r.h.s. of Fig. \ref{GLRfig}).

\begin{figure}[htb]
\centering
\resizebox{.9\textwidth}{!}{%
\includegraphics*{{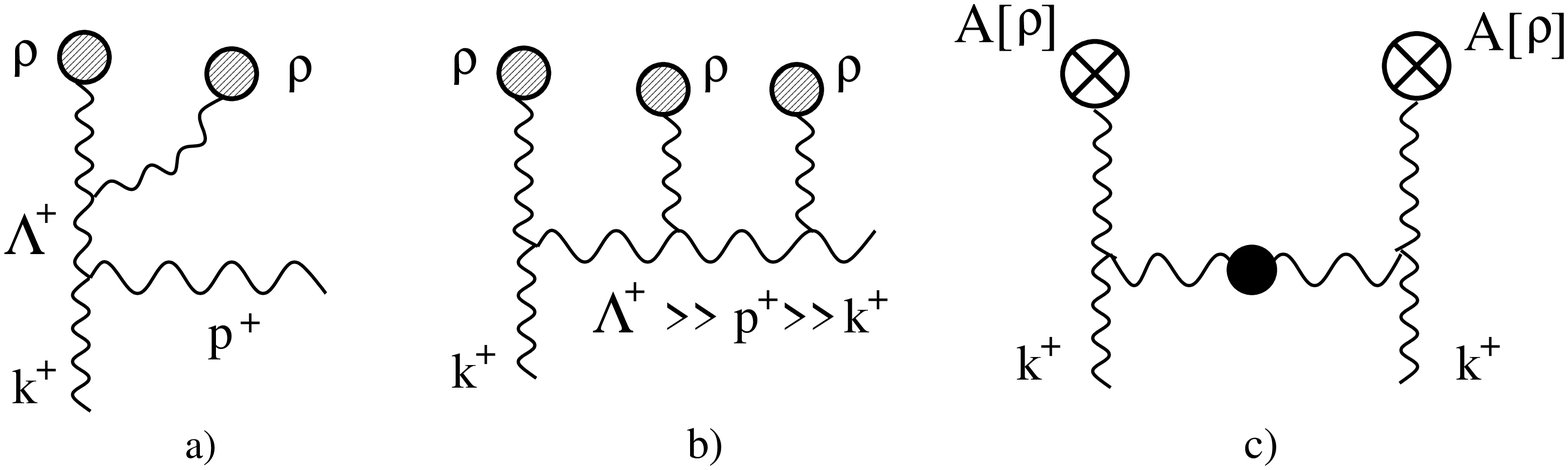}}}
   \caption{Some typical non-linear effects in the quantum evolution}
\label{CHIfig}
\end{figure}

But non-linear effects are important also in the {\it quantum} evolution, and
actually interfere with it, as illustrated in  Fig. \ref{CHIfig}. Fig. \ref{CHIfig}.a
is an immediate generalization of the one--gluon emission in  Fig.~\ref{gluoncascade}.b.
It is clear that what is renormalized by the scattering off the ``semi-fast''
($\Lambda^+ \gg p^+ \gg k^+$) quantum fluctuation is the classical field
${\cal A}^i[\rho]$ at scale $\Lambda^+$, which in turn is non-linear in $\rho$.
(The Feynman rules for evaluating diagrams like those in Fig. \ref{CHIfig}, and
also the present discussion, are adapted to the LC gauge $A^+=0$, in which the
quantum effective theory is written, cf. Eq.~(\ref{ZCGC}).) Fig. \ref{CHIfig}.b
shows an additional source of non-linearity, arising from the propagation 
of the radiated gluon in the classical `background' field 
${\cal A}^i[\rho]$.
If $\Lambda^+ = \xx P^+$ 
is small enough ($\xx\ll 1$), the classical field is very strong, ${\cal A}^i\sim 1/g$,
and gluon rescatterings must be included to all orders in ${\cal A}^i$. The diagrams
in both Figs. \ref{CHIfig}.a and b can be taken into account as the cut of the diagram
in Fig. \ref{CHIfig}.c. The classical field that enters the vertices is the
fully non-linear solution ${\cal A}^i[\rho]$ constructed in Sect. \ref{CLASSMV}, and
the propagator of the quantum gluon is computed to {\it all} orders in this background 
field, the resummation indicated here by a blob.

The diagram in Fig. \ref{CHIfig}.c is manifestly a quantum correction to the 2-point
function of the gauge fields at scale $k^+$, and is of order $\alpha_s \ln(\Lambda^+/k^+)$.
Thus, for this to be computable in perturbation theory,
the separation of scales between  $\Lambda^+$ and $k^+$ must not be too large:
$\Lambda^+ \gg k^+$, but $\alpha_s \ln(\Lambda^+/k^+) \ll 1$. The quantum
modes must be integrated out in layers of $p^+$, within a renormalization group procedure
\cite{JKLW97,PI}. At each step in this procedure, one has to perform a one-loop quantum
calculation, but with the exact background field propagator for the ``semi-fast'' gluons
(the quantum gluons that are integrated out in that particular step).
Such an all-order inclusion of the classical field effects
permits one to resum not only the {\it large energy logarithms}, namely the terms
$\sim (\alpha_s\ln 1/x)^n$, but also the dominant {\it high density effects}--
the non-linear effects (like gluon recombination) which become of order one at saturation.

The corresponding analysis is technically quite involved and has been described in detail in previous publications
\cite{PI,CARGESE}. Here we shall present only the final results and their consequences.
As shown in Ref. \cite{PI}, the condition that the new
correlations induced by integrating out quantum fluctuations be
reproduced by the CGC effective theory 
leads to a functional {\it renormalization group equation} (RGE) for the 
weight function $W_{\Lambda^+}[\rho]\equiv W_\tau[\rho]$, which is most succinctly
written as \cite{PI}
\be\label{RGE}
{\del W_\tau[\rho] \over {\del \tau}}\,=\,
 {1\over 2} \int_{x_\perp,y_\perp}\,{\delta \over {\delta
\rho_\tau^a(x_\perp)}}\,\chi_{ab}(x_\perp, y_\perp)[\rho]\, 
{\delta \over \delta \rho_\tau^b(y_\perp)}\,W_\tau[\rho]\,,\ee
in notations that we shall shortly explain.
Early versions of this equation can be found in the pioneering works of Refs. \cite{JKLW97,JKW99}.
A formally similar, and physically equivalent~\cite{BIW}, functional evolution equation has 
been obtained  by Weigert \cite{W}, within a different formalism \cite{Balitsky1996}.
We shall discuss this latter approach in the next subsection. 

Let us now discuss the meaning and structure of the terms in Eq.~(\ref{RGE}).  The rapidity variable 
$\tau = \ln(1/x) = \ln(P^+/\Lambda^+)$ indicates the dependence
of the effective theory upon the separation scale. This is convenient since (as 
illustrated by the BFKL evolution discussed in Sect. \ref{SEC-BFKL}), $\tau$ is the
natural ``evolution time''. The contribution of the quantum modes within a 
small layer in $p^+$ (say $\Lambda^+ > p^+ > k^+$) is proportional to the rapidity extent 
$\Delta\tau = \ln(\Lambda^+/k^+) $ of that layer.

The kernel $\chi[\rho]$ is a positive definite non--linear functional of $\rho$ 
(Eq.~(\ref{RGE}) is a {\it diffusion equation}) 
and is highly non--local in both longitudinal and transverse coordinates. 
The non--linearity in  $\rho$ and the non--locality in $x^-$
are strongly correlated, since they have a common origin: 
$\chi[\rho]$ depends upon $\rho$ via the Wilson lines (\ref{v}).
Physically, $\chi[\rho]d\tau$ is the charge-charge correlator
induced when integrating out quantum gluons within
a rapidity interval $d\tau$ and in the presence of a classical color charge
distribution with density $\rho_a(\vec x)$ \cite{PI,CARGESE}.
Technically, this is computed by evaluating the diagram
in Fig. \ref{CHIfig}.c with the Feynman rules in Eq.~(\ref{ZCGC}). Note that,
in addition to this real-gluon emission diagram, there are also virtual 
(self-energy and vertex) corrections which must be similarly computed~\cite{JKLW97,PI}.
Such corrections are already included in the RGE (\ref{RGE}), where they correspond
to the functional derivative of the kernel $\chi[\rho]$ \cite{PI}.

The argument $\rho_\tau^a(x_\perp)$ of the functional derivatives in
Eq.~(\ref{RGE}) denotes the color charge density
$\rho^a(x^-,x_\perp)$ at $x^-=x^-_\tau$.
The color source generated by the quantum evolution up to rapidity
$\tau$ has support within a limited interval in $x^-$, namely at $0 <
x^- < x^-_\tau$, with $x^-_\tau\propto {\rm e}^{\tau}$.  The reason
this is so follows from the uncertainty principle: since the classical
source at rapidity $\tau$ is obtained by integrating out quantum modes
with large longitudinal momenta $p^+ \gg \Lambda^+\equiv {\rm
e}^{-\tau}P^+$, it must be localized near $x^-=0$, within a distance
$\Delta x^-\sim {\rm e}^{\tau}x^-_0$ (with $x^-_0\equiv 1/P^+$).

However, eq.~(\ref{RGE}) shows that the correlation between the quantum evolution in $\tau$ 
and the longitudinal distribution of the resulting color source is even stronger.
When the rapidity is further increased, say from $\tau$ 
to $\tau+{\rm d}\tau$, the additional contribution to the color source
which is generated in this way has {\it no overlap} in $x^-$ with
the original source at rapidity $\tau$. Instead, this new contribution
makes a new layer in $x^-$, which is located
between $ x^-_\tau$ and  $ x^-_{\tau+d\tau}$.
This is why the functional derivatives in  eq.~(\ref{RGE}) 
involve just the color source $\rho_\tau\equiv 
\rho(x^-_\tau)$ in this outermost layer.

This correlation is most simply formulated if one uses
 the {\it space--time rapidity} y,
\be {\rm y}\,\equiv\, \ln(x^-/ x^-_0),\quad 
x^-_0\equiv 1/P^+\,,\quad -\infty < {\rm y} < \infty\,,\ee
to indicate the longitudinal
coordinate of a field. For example, 
\be\label{rhotau}
\rho_{{\rm y}}^a(x_\perp)&\equiv&
 x^-\rho^a(x^-,x_\perp)\qquad {\rm for}\quad
x^-=x^-_{\rm y}\equiv x^-_0{\rm e}^{{\rm y}},\nn
\int d{\rm y}\,\rho_{{\rm y}}^a(x_\perp)&=&\int dx^-\,\rho^a(x^-,x_\perp),
\ee
and similarly for $\alpha$, Eq.~(\ref{alpha}), or any other field. 
Eq.~(\ref{v}) can be rewritten as :
\be\label{vy}
V^\dagger(x_{\perp})\,=\,{\rm P} \exp
 \left \{ig \!\int\! d{\rm y}\,\alpha^a_{\rm y} (x_{\perp})t^a
 \right \}.\ee
The previous discussion shows that the space--time rapidity y
of a given layer in $\rho$ is identical to the usual (momentum) rapidity
of the fast gluons that have produced that layer. In particular, the
color source created by the quantum evolution up to $\tau$ has support
at space-time rapidities ${\rm y}\le \tau$ (in agreement with the simple
argument based on the uncertainty principle). Formally:
\be\label{deltatau} W_\tau[\rho] \,\propto\, \delta_\tau[\rho]\,,\ee
where $\delta_\tau[\rho]$ is a $\delta$--functional enforcing that
$\rho_{\rm y}\equiv 0$ for any ${\rm y}> \tau$. As we shall see, this constraint is important 
because the  $\tau$--dependence of the observables in the effective theory comes precisely from the upper limit on the
longitudinal support of $\rho$. The Color Glass evolves by expanding
in y.

Since the Wilson lines (\ref{vy}) and many interesting quantities (like the 
$S$-matrix element (\ref{Ntau}), or the gluon distribution (\ref{FF0})) are more
directly expressed in terms of the COV-gauge field $\alpha_{{\rm y}}^a(x_\perp)$, 
rather than the color charge $\rho$, it is often preferable to use the
`$\alpha$--representation', whose weight function
$W_\tau[\alpha]\equiv W_\tau[\tilde\rho= -\grad_\perp^2\alpha]$ satisfies the following RGE,
obtained after a change of variables in Eq.~(\ref{RGE}) :
\be\labe{RGEA}
{\del W_\tau[\alpha] \over {\del \tau}}\,=\,{1 \over 2}
{\delta \over {\delta \alpha_\tau^a(x_\perp)}}\,
\eta^{ab}(x_\perp,y_\perp)[\alpha] \,{\delta W_\tau\over 
{\delta \alpha_\tau^b(y_\perp)}}\,.\ee
We use compact notations
in which repeated color indices (and coordinates) are understood to be summed (integrated) over.  
The relation of the kernel here to that in Eq.~(\ref{RGE}) is 
\be
\eta^{ab}(x_\perp,y_\perp)\equiv \int_{z_\perp,u_\perp}
\langle x_\perp|\frac{1}{-\grad^2_\perp}|z_\perp\rangle\,
\chi^{ab}(z_\perp,u_\perp)\,
\langle u_\perp|\frac{1}{-\grad^2_\perp}|y_\perp\rangle.\nonumber 
\ee

The analysis in Ref. \cite{PI} yields (see also \cite{W}) :
\be\label{eta}
\eta^{ab}(x_\perp,y_\perp)
&=&{1\over \pi}\!\int 
{d^2z_\perp\over (2\pi)^2}\,
\frac{(x^i-z^i)(y^i-z^i)}{(x_\perp-z_\perp)^2(y_\perp-z_\perp)^2 }\nn
&{}&\qquad\quad\times\,\,
\bigl\{1
+ V^\dagger_x V_y- V^\dagger_x V_z -  V^\dagger_z V_y\bigr\}^{ab},
\ee
with $V^\dagger_x= V^\dagger(x_{\perp})$, etc.
This is real and symmetric ($\eta^{ab}(x_\perp,y_\perp)=\eta^{ba}(y_\perp,x_\perp)$),
and also positive definite, as anticipated, since:
\be
\bigl\{1
+ V^\dagger_x V_y- V^\dagger_x V_z -  V^\dagger_z V_y\bigr\}^{ab} =
(1-V^\dagger_z V_x)_{ca}(1-V^\dagger_z V_y)_{cb},\ee
and the color matrix $1-V^\dagger_z V_x$ is hermitian.
The transverse kernel in Eq.~(\ref{eta}) is similar to the `dipole kernel' in the BFKL
equation (\ref{linearBK}). Their relation will be discussed in the next subsection.

The RGE in eq. (\ref{RGEA}) [or eq. (\ref{RGE})] has the structure of a Fokker--Planck equation. In the CGC formalism, 
the quantum evolution towards small x is 
a {\it random walk} in the space of Wilson lines \cite{BIW}. The random process is one 
by which the Wilson lines are built. The physical random variable in the evolution is the
elementary contribution $\alpha_\tau^a(x_\perp)$ to the classical field in the hadron arising
from integrating out quantum fluctuations in the rapidity strip $[\tau, \tau+d\tau]$. Such
a contribution changes the Wilson lines according to:
\be\label{Uy}
U^\dagger_{\tau+d\tau}(x_{\perp})\,=\,{\rm e}^{ig d\tau\alpha_\tau^a(x_\perp) T^a}\,
U^\dagger_{\tau}(x_{\perp})\,,\ee
whose iteration defines a path in the space of the $U$ fields. (This path is unambiguously defined only
after discretizing the rapidity variable; see Ref. \cite{BIW} for details.)
By exploiting this representation, an exact but formal solution to Eq.~(\ref{RGEA}) 
has been constructed in the form of a path integral \cite{BIW}. This random walk 
can be equivalently reformulated as a Langevin equation \cite{BIW}, a formulation which is
better suited for numerical simulations on a two-dimensional lattice. 

 \subsection{The Balitsky-Kovchegov equation}  

In addition to the numerical simulations on a lattice, Eq.~(\ref{RGEA}) can be made tractable
via two strategies. Both involve some approximations. The first strategy consists in using
this functional equation to deduce ordinary differential equations for quantities of interest.
Because of the non-linearity of Eq.~(\ref{RGEA}), the ensuing equations will generally not be closed,
but rather form an infinite hierarchy of coupled equations. Nevertheless some progress can be made 
in various approximations, particularly in the large $N_c$ limit, where we shall see that 
a closed equation emerges. The other strategy, to be developed in the next subsection, is to 
search directly for approximate solutions to the functional equation (\ref{RGEA}) for the weight function.

If  $\langle O[\alpha] \,\rangle_\tau$ is any observable which
can be computed as an average over $\alpha\,$ (cf. eq.~(\ref{COVclascorr})) :
\be\label{OBSERV}
\langle O[\alpha] \,\rangle_\tau&=&
\int\,{\cal D}\alpha\,O[\alpha] \,W_\tau[\alpha],\ee
then its evolution with $\tau$ is governed by the following equation:
\be\labe{evolO}
{\del \over {\del \tau}}\langle O[\alpha] \,\rangle_\tau&=&\int
{\cal D}\alpha\,O[\alpha] \,{\del W_\tau[\alpha] \over {\del \tau}}\nn
&=&\left\langle {1 \over 2}\,{\delta \over {\delta
\alpha_\tau^a(x_{\perp})} }\,\eta_{xy}^{ab}\,
{\delta \over {\delta \alpha_\tau^b(y_{\perp})}}\,O[\alpha]\right
\rangle_\tau\,,\ee
where, in writing the second line,
we have used eq.~(\ref{RGEA})  and 
then integrated twice by parts within the
functional integral over $\alpha$. The last expression involves the functional
derivative of $\eta[\alpha]$, which is easily computed by using
[with $\delta_{xy}\equiv \delta^{(2)}(x_{\perp}-y_\perp)$] :
\be\label{DIFFU}
{\delta V^\dagger (x_{\perp})\over \delta \alpha^a_\tau(y_\perp)}=
ig\delta_{xy}T^a V^\dagger (x_{\perp}),\quad
{\delta V(x_{\perp})\over \delta \alpha^a_\tau(y_\perp)}=
-ig\delta_{xy}V(x_{\perp})T^a\,.\ee
Here $\alpha_{{\rm y}}=0$ for $ {\rm y} > \tau$, cf. Eq.~(\ref{deltatau}).

The 2-point function $S_\tau(x_{\perp},y_{\perp})$
of the Wilson lines, Eq.~(\ref{Ntau}) (which physically represents the $S$-matrix element 
for dipole-hadron scattering) can straightforwardly be computed using repeatedly Eq.~(\ref{DIFFU}) 
(see \cite{PI} for details):
\be\labe{evolV}
{\del \over {\del \tau}}\langle {\rm tr}(V^\dagger_x V_y)
\rangle_\tau&\!\!=\!\!&-{\alpha_s\over 2 \pi^2}\!\int\! d^2z_\perp
\frac{(x_\perp-y_\perp)^2}{(x_\perp-z_\perp)^2(y_\perp-z_\perp)^2 }\nn
&{}&\,\,\times \left\langle N_c {\rm tr}(V^\dagger_x V_y)
- {\rm tr}(V^\dagger_x V_z){\rm tr}(V^\dagger_z V_y)\right\rangle_\tau.\,\,\ee
This equation was originally derived by Balitsky \cite{Balitsky1996},
within a formalism based on the evolution of observables
(in high-energy dipole-hadron scattering) which are built from Wilson lines. This
is similar in spirit to the `Color Dipole approach' by Mueller \cite{Mueller94},
but it is not restricted to the large--$N_c$ limit. It is better suited for an asymmetric
collision, like that between an ``onium'' (= a high-energy dipole) and a dense hadronic target,
like a nucleus. In this respect, Balitsky's formalism is closer to the CGC formalism, where the focus is
fully on the target wavefunction.

As anticipated, the above equation is not closed: It relates
the 2-point function to the 4-point function $\langle
{\rm tr}(V^\dagger_x V_z){\rm tr}(V^\dagger_z V_y)\rangle$. 
Physically, it is so since, except at large $N_c$, the $q\bar q g$ system formed after radiating 
one gluon from the original dipole is not exactly the same as a system of two dipoles
(recall the discussion around Eq.~(\ref{probdip})).
One can similarly derive an evolution equation for the 4-point function
\cite{Balitsky1996}, but this will in turn couple the 4-point function  to a 6-point function, and so on. 
Eq.~(\ref{evolV}) is merely the first in an 
infinite hierarchy of coupled equations \cite{Balitsky1996}. 
In Ref. \cite{W}, Weigert managed to reformulate Balitsky's hierarchy 
as a single {\it functional} evolution equation for the generating functional
of the $n$-point functions of the Wilson lines. As shown in Ref.
\cite{BIW}, Weigert's equation is equivalent to the RGE (\ref{RGEA}) as 
far as the correlations of the Wilson lines are concerned.
More recently, Mueller used a similar approach to give a simple derivation~\cite{AM01} for
Eq.~(\ref{RGEA}).

A closed equation can still be obtained in
the large $N_c$ limit in which the 4-point function in eq.~(\ref{evolV}) factorizes:
\be
\left\langle {\rm tr}(V^\dagger_x V_z)\,
{\rm tr}(V^\dagger_z V_y)\right\rangle_\tau
\longrightarrow 
\left\langle {\rm tr}(V^\dagger_x V_z)\right\rangle_\tau\,
\left\langle{\rm tr}
(V^\dagger_z V_y)\right\rangle_\tau\quad {\rm for}\,\,N_c\to\infty.
\nonumber\ee
Then Eq.~(\ref{evolV}) reduces to a closed equation for the 2-point function,
which, when rewritten in terms of the scattering amplitude 
${\cal N}_\tau(x_{\perp}, y_{\perp}) = 
\frac{1}{N_c}\langle {\rm tr} (1-V^\dagger_x V_y)\rangle_\tau$, 
 is recognized as the Kovchegov equation (\ref{BK}) \cite{K}.
An early version of this equation has been 
proposed by Gribov, Levin and Ryskin \cite{GLR}, and proven by Mueller and Qiu
\cite{MQ} in the `double-logarithmic approximation'. More recently,
Braun has rederived Eq.~(\ref{BK}) by directly resumming `fan' diagrams \cite{B00}.
Following the recent literature, we shall refer to Eq.~(\ref{BK}) as the
``Balitsky-Kovchegov (BK) equation''

Clearly, in the weak scattering approximation ${\cal N}_\tau(r_{\perp})\ll 1$
(which corresponds to a very small dipole, or, equivalently, to a relatively
low gluon density in the hadronic target), the non-linear term can be neglected in
the r.h.s. of Eq.~(\ref{BK}), which then reduces to the BFKL equation (\ref{linearBK}).
But in general, the feedback provided by this non-linear term ensures that
the  solution ${\cal N}_\tau(r_{\perp})$ to Eq.~(\ref{BK}) respects the unitarity
bound ${\cal N}_\tau(r_{\perp})\le 1$. Thus, Eq.~(\ref{BK}) is a simple 
QCD--based non-linear equation consistent with unitarity. This explains the large interest 
in this equation in the recent literature, with important progress towards its resolution
via both analytic \cite{K,LT99,KIIM} and
numerical methods \cite{LT99,AB01,LL,GB01}. The conclusions reached in 
this approach are equivalent to those obtained from direct investigations
of the RGE (\ref{RGEA}) \cite{SAT,GAUSS}, and will be described in the next sections.

\subsection{Saturation momentum and geometric scaling}

\begin{figure}[htb]
\centering
\resizebox{.40\textwidth}{!}{%
\includegraphics*{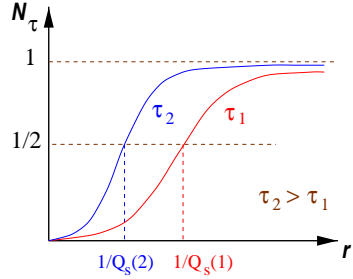} }
   \caption{The solution to the BK equation as a function of $\rr$ 
for two values of $\tau$.}
\label{NFIG}
\end{figure}

The solution to Eq.~(\ref{BK}) is shown qualitatively
in Fig.~\ref{NFIG}, which displays ${\cal N}_\tau(r_{\perp})$ as
a function of $r_\perp=x_{\perp}- y_{\perp}$ for two different rapidities. 
The scattering amplitude vanishes as $r_\perp\to 0$, as it should from its definition
${\cal N}_\tau(r_{\perp}) = \frac{1}{N_c}\langle {\rm tr} (1-V^\dagger_x V_y)\rangle_\tau$, 
and the fact that the Wilson lines are unitary matrices. For small $r_\perp$,
$ {\cal N}_\tau(r_{\perp})$ remains small (``color transparency''), and
is well approximated by the BFKL solution (\ref{BFKLsol1}). 
For large $r_\perp$, it approaches the unitarity bound ${\cal N}_\tau(r_{\perp})= 1$.
The transition between ``color transparency'' at small $r_\perp$ and ``blackness'' at
large $r_\perp$ takes place at a characteristic value of $r_\perp$ that we shall identify
with the {\it saturation length} $1/Q_s(\tau)$. More precisely, 
we shall define $Q_s(\tau)$ by the following convention:
\be\label{Nsat}
{\cal N}_\tau(r_\perp)\,=\, 1/2\qquad{\rm for}\quad r_\perp\,=\,1/Q_s(\tau)\,.\ee
As shown in Fig.~\ref{NFIG}, this saturation length decreases with $\tau$.

The result illustrated in Fig.~\ref{NFIG} clearly shows that the non-linear 
BK equation (\ref{BK}) solves the unitarity problem of the BFKL equation (at least,
at a fixed impact parameter; see also Sect. 4.4 below). Moreover, the emergence
of an intrinsic ``saturation scale'' $Q_s(\tau)$, which increases with $\tau$, 
also solves the problem of ``infrared diffusion'', as convincingly demonstrated 
by the numerical analysis in Ref. \cite{GB01}. In addition to the numerical studies
in Refs. \cite{LT99,AB01,LL,GB01}, the behaviour shown in Fig.~\ref{NFIG} is
supported also by analytic investigations focusing on qualitative features 
like the energy dependence of the saturation scale \cite{AM2,KIIM,MuellerTrianta,DT02},
the ``geometric scaling'' behaviour  \cite{KIIM,MuellerTrianta,DT02},
or the approach towards the blackness with increasing $r_\perp$ \cite{K,LT99,SAT,KIIM,GAUSS}.
We shall describe here some of these analytical studies, whose results follow from general arguments,
such as the validity of the BFKL dynamics at small $r_\perp$. and the emergence (via non-linear 
effects) of an intrinsic momentum scale, the saturation momentum $Q_s(\tau)$.

Consider first the calculation of the saturation scale. Even though the BFKL solution
(\ref{BFKLsol1}) is valid only at small $r_\perp$, well below
the saturation length, it is nevertheless possible to compute the energy dependence of 
the saturation scale by extrapolating Eq.~(\ref{BFKLsol1}) up to $r_\perp\sim 1/Q_s(\tau)$
and then imposing the {\it saturation condition} (\ref{Nsat}) \cite{AM2,KIIM}. More
precisely, we shall see shortly that $Q_s^2(\tau)$ is increasing exponentially with $\tau$.
The BFKL computation alluded to above should then correctly reproduce the value
of the exponent, but not necessarily the (slowly varying) prefactor as well (see however
\cite{MuellerTrianta}).

Towards this end, the solution of the BFKL equation in Eq.~(\ref{BFKLsol1}) is first re-written as ($\rho\equiv 
\ln(1/r_\perp^2 Q_0^2)$):
\be\label{BFKLsol2}
{\cal N}_\tau(\rr)\simeq \exp \left\{ \omega\bar\alpha_s\tau - \frac{\rho}{2}
- \frac{\rho^2}{2\beta \bar\alpha_s\tau} 
\right\},
\ee
where we have kept only the dominant terms in the regime $\rho\gg 1$ and 
$\bar\alpha_s\tau\gg 1$, with $\bar\alpha_s\tau \gg \rho$. Since the BFKL equation
is now seen as just an approximation to more general non-linear equations like
the BK equation (\ref{BK}), the solution (\ref{BFKLsol2}) is acceptable only
as long as $r_\perp\ll 1/Q_s(\tau)$, or $\rho > \rho_s(\tau)$, with
$\rho_s(\tau)\equiv \ln(Q_s^2(\tau)/Q^2_0)$.

If nevertheless extrapolated down to $\rho\sim \rho_s(\tau)$, the saturation condition
 (\ref{Nsat}) amounts to the vanishing of the exponent in  Eq.~(\ref{BFKLsol2}).
This gives a second-order algebraic equation for $\rho_s(\tau)$ with
the physical solution \cite{AM2,KIIM}:
\be\label{Qs0}
Q_s^2(\tau)\,=\,Q_0^2 \,
{\rm e}^{c\bar\alpha_s \tau}\,,\quad c\,=\,
\big[-{\beta}+\sqrt{\beta(\beta+8\omega)}\,\big]/2\,=\,4.84...\ee
This estimate is consistent with the numerical solutions to the
BK equation, which have found  $c\simge 4$ \cite{LT99,AB01,GB01}, but not 
with the phenomenology of DIS at HERA, which suggests rather a significantly lower
value for the exponent \cite{GBW99} (see Sect. 4.3 below), namely, 
$\lambda \approx 0.3$ instead of $c\bar\alpha_s \sim 1$.

The factor in front of the exponential in eq.~(\ref{Qs0})
is not under control in present approximations. In Ref. \cite{MuellerTrianta},
a more refined treatment was proposed, where the BFKL equation was solved
with an absorbtive boundary condition at $r_\perp\sim 1/Q_s(\tau)$, and a weak dependence on $\tau$ for this prefactor 
was obtained. It would be interesting to test their results against more accurate numerical solutions to Eq.~(\ref{BK}).

It is also interesting to study the behaviour of the scattering amplitude
(\ref{BFKLsol2}) for $r_\perp$ below but relatively close to $1/Q_s(\tau)$--for $\rho$ slightly above $\rho_s(\tau)$. 
Since:
\be\label{LQS}
\rho\,\equiv\, \ln \frac{1}{r _{\perp}^2Q_0^2}\, = \,\rho_s(\tau)+
\ln \frac{1} {r _{\perp}^2 Q^2_s(\tau)} \,\equiv\,\rho_s\,+\,\delta\rho\,,\ee
a simple calculation yields:
\be\label{BFKLsol3}
{\cal N}_\tau(\rr)\simeq \exp \left\{ - \gamma\,\delta\rho
- \frac{(\delta\rho)^2}{2\beta \bar\alpha_s\tau} 
\right\},
\ee
where $\gamma\equiv 1/2+c/\beta\approx 0.64$. Eq.~(\ref{BFKLsol3})
suggests a remarkable simplification: 
Assume that $r_{\perp}$  is sufficiently close to $1/Q_s(\tau)$
(although still below it) for $\delta\rho/\bar\alpha_s\tau \ll 1$.
Then,  the second term in the exponent can be neglected
compared to the first one, and we are left with
\be\label{Nscaling}
{\cal N}_\tau(r_{\perp})\,\approx\, 
\left(r^2_{\perp}Q_s^2(\tau)\right)^\gamma,\ee
which shows {\it geometric scaling} \cite{geometric,KIIM} : it
depends upon the two kinematical variables
$r_{\perp}$  and $\tau$ only via the combination $r_{\perp}^2 Q_s^2(\tau)$.
Numerically, such a scaling behaviour has been seen in the solutions to the
BK equation \cite{LT99,GB01}.

Since Eq.~(\ref{Nscaling}) is the first term in an expansion in powers of
$\delta\rho/\rho_s$, with $\rho_s=c \bar\alpha_s\tau$,
this approximation is correct for 
\be\label{swindow1}
1\,<\, \ln\frac{1}{r _{\perp}^2 Q^2_s(\tau)}\,\ll\,c \bar\alpha_s\tau\, .\ee
The condition on the left, $r_{\perp} < 1/Q_s(\tau)$, ensures 
we are still in a linear regime. 
For a dipole transverse resolution $Q^2\equiv 1/r_{\perp}^2$, this condition 
translates to the following {\it scaling window} \cite{KIIM}:
\be\label{swindow}
Q_s^2(\tau)\,\,\ll\,\, Q^2\,\,\ll\,\,\frac{Q_s^4(\tau)}{Q^2_0}\,.\ee
Since $Q_0\sim \Lambda_{\rm QCD}$ and $Q_s(\tau)\gg \Lambda_{\rm QCD}$
for sufficiently large $\tau$, the upper boundary of this scaling window
is rather large. In particular, it is much larger than the saturation
scale itself: $Q_s^2(\tau)/Q_0\gg Q_s(\tau)$.
Remarkably, as a consequence of saturation, knowledge of an intrinsic momentum scale is propagated through
the linear evolution equations up to relatively large values of $Q^2$ well
outside the saturation regime. This is especially interesting since
such values of $Q^2$ are large enough for perturbation theory to be fully
trustworthy. A property like (\ref{Nscaling}) can be and has been
tested against the experimental data \cite{geometric,WeigertFreund,DLL}. 
We shall return to phenomenological aspects of geometric scaling in Sect. 4.3.

The previous results are obtained from the leading order
BFKL equation. Recently, there has been some
progress in including next-to-leading-order $\alpha_s$ effects 
in the physics of saturation. In Refs. \cite{KIIM,MuellerTrianta},
this was done heuristically, by simply replacing the fixed coupling $\alpha_s$ in the 
BFKL equation by the one-loop running coupling of QCD with the running scale set by the saturation momentum :
$\bar\alpha_s\longrightarrow \bar\alpha_s(Q_s^2(\tau))$, with
\be \bar\alpha_s(Q^2) 
= \frac{b_0}{ \ln(Q^2/\Lambda^2_{\rm QCD})} , \qquad 
b_0= \frac{12 N_c}{11N_c-2N_f}\,.
\label{running}\ee
The only modification due to the running coupling is in the functional form of the
saturation scale, whose growth with $\tau$ becomes somewhat milder
($\tau_0$ is an arbitrary constant, and $c$ is the same number as in
in Eq.~(\ref{Qs0})):
\be\label{Qsrun}
Q_s^2(\tau)=\Lambda_{\rm QCD}^2\ {\rm e}^{\sqrt{2 b_0 c 
(\tau +\tau_0)}}\,,
\ee
where the overall scale is now set by $\Lambda_{\rm QCD}$ rather than
the initial scale $Q_0$. All the previous results on geometric scaling
(the scaling law (\ref{Nscaling}), including the value of the ``anomalous dimension''
$\gamma$, and its range of validity (\ref{swindow})) remain unchanged, except for the
expression for the saturation scale entering these results.

For nuclei, Eq.~(\ref{Qsrun}) has an intriguing consequence 
for the dependence of the saturation scale upon the atomic number $A$ \cite{AM03}.
Assume an initial condition of the MV type at $\tau=0$, $Q_s^2(\tau=0,A)=Q_s^2(A) \sim A^{1/3}\ln A$.
For fixed coupling BFKL evolution, where Eq.~(\ref{Qs0}) applies, 
this initial condition identifies the hitherto unspecified `initial' scale $Q_0$ with the MV 
saturation scale, 
\be\label{QAfixed}
Q_s^2(\tau,A)\,=\,Q_s^2(A){\rm e}^{c\bar\alpha_s \tau}\,\quad
{\rm (fixed\,\,coupling)}\,,\ee
which preserves the $A$--dependence of the initial condition
at any later `time' $\tau$: $Q_s^2(\tau,A) \sim A^{1/3}\ln A$.

For the running coupling BFKL evolution case where Eq.~(\ref{Qsrun}) applies (with $\tau_0$ fixed by the
initial condition as $2b_0c \tau_0= [\ln (Q_s^2(A)/\Lambda_{\rm QCD}^2)]^2$), 
one obtains a very different $A$--dependence at small and large
$\tau$, respectively \cite{AM03}:

i) At relatively small energies, such that $\tau\ll \tau_0\sim \ln^2 A^{1/3}$,
\be\label{QArunningI}
Q_s^2(\tau,A)\,\approx\,
 Q_s^2(A)\,{\rm e}^{c\bar\alpha_s(Q_s^2(A)) \tau}\,,\ee
which is the `fixed-coupling'--like behaviour, with the $\bar\alpha_s$ in the exponent
being the running coupling (\ref{running}) evaluated at the initial saturation scale.

ii) At higher energies,  $\tau\gg \ln^2 A^{1/3}$, one obtains
\be\label{QArunningII}
Q_s^2(\tau,A)\,\approx\,\Lambda_{\rm QCD}^2\,{\rm e}^{\sqrt{2 b_0 c \tau}}\,
\exp\left\{\frac{1}{2\sqrt{2 b_0 c \tau}}\left(\ln\frac{Q_s^2(A)}{\Lambda_{\rm QCD}^2}
\right)^2\right\},\ee
which, for very large $\tau$, is {\it nearly independent of $A$}.

In Ref. \cite{DT02}, which did not consider the $A$--dependence,
Triantafyllopoulos presented a complete computation of the NLO effects on the energy dependence of the saturation scale.
Recall that the NLO corrections to the BFKL equation~\cite{NLBFKL}, turn out to be anomalously large and 
require resummation to obtain sensible results. Ref. \cite{DT02} used the RG--improved resummation scheme 
of Ciafaloni, Colferai, and Salam \cite{Salam99} and found that, although $Q_s^2(\tau)$ is in 
general a more complicated function than the simple exponential (\ref{Qs0}), 
it can nevertheless be represented as such for a rather wide range of rapidities 
(including those of phenomenological interest). Specifically, if one defines
$\lambda(\tau) \equiv d \ln (Q^2_s/\Lambda^2)/d\tau$, then $\lambda(\tau)$ turns out to be
a very slowly decreasing function, with $\lambda(\tau = 5 - 9) \simeq 0.30 - 0.29$.
Remarkably, this value is also favoured by the current phenomenology at HERA
\cite{GBW99} (and Sect. 4.3 below). 
It would be very interesting to compute the $A$--dependence within this fully NLO formalism.

\subsection{Gluon saturation and perturbative color neutrality}

\begin{figure} 
\begin{center} 
\includegraphics[width=0.75\textwidth]
{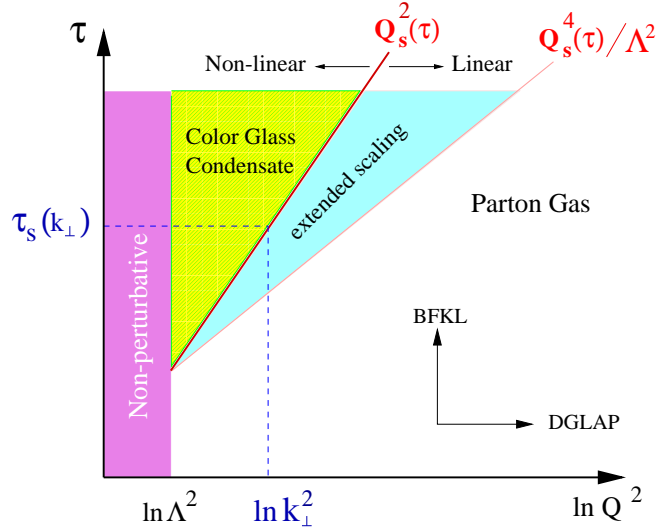}
\caption{A map of the quantum evolution
in the $\tau-k_\perp$ plane.}
\label{phase-diagram} 
\end{center}
\end{figure}

In this section, we return to the RGE (\ref{RGEA}), which describes the evolution of the
hadron wavefunction as a whole, and construct approximate solutions to it. As usual,
these approximations depend upon the transverse resolution scale $Q^2$ at which correlations are 
measured. If $Q^2$ is large enough ($Q^2 \gg Q_s^2(\tau)$),
 we probe color sources with small transverse size, which
do not overlap with each other. In this {\it dilute} regime,
a description in terms of uncorrelated sources, such as the
MV model, may be a good approximation. With increasing $\tau$ at fixed  $Q^2$ (or, equivalently,
with decreasing $Q^2$ at fixed $\tau$), spatial correlations start to develop, initially
according to the linear BFKL evolution and then, 
once the density is high enough, according to the general
non-linear RGE which predicts gluon saturation. A schematic map of the
kinematical regimes for quantum evolution is shown in Fig. ~\ref{phase-diagram}
(see the discussion below for details).

\bigskip
\noindent {\bf i) High--momentum regime ($Q^2\gg Q_s^2$): Recovering BFKL}
\smallskip

In the dilute regime at $Q^2\gg Q_s^2$,
the color charge density is low, hence the classical field is weak,
$g\alpha\ll 1$. One can therefore keep only the linear term in the expansion
of the Wilson lines in Eq.~(\ref{RGEA}) in powers of $g\alpha$ :
\be\label{vlinear}
V^\dagger(x_{\perp})&\approx & 1\,+\,
ig \!\int\! d{\rm y}\,\alpha_{\rm y} (x_{\perp})
\,\equiv\,1\,+\,ig\alpha(x_{\perp}),\nonumber\\
1-V^\dagger_z V_x&\approx & ig(x^j-x^j) \del^j \alpha(x_{\perp})
.\ee
($\alpha_a(x_{\perp})$ is the effective color field in the transverse plane.)
The kernel $\eta[\alpha]$ in Eq.~(\ref{eta}) then becomes quadratic
in $\alpha(x_{\perp})$, and the RGE  takes the generic form:
\be\label{RGE-lin}
{\del  W_\tau[\alpha] \over {\del \tau}}\,\approx\,{1 \over 2}\,
{\delta \over {\delta\alpha}}\big(\alpha\,{\cal K}\,\alpha\big)
{\delta \over {\delta\alpha}}\,W_\tau[\alpha]\,,
\ee
where the new kernel ${\cal K}$ is non-local in the transverse coordinates.
Its explicit form is easily extracted from Eq.~(\ref{eta}) using
Eq.~(\ref{vlinear}). Even 
in this {dilute} regime, the RGE is non-linear and the corresponding
weight function $W_\tau[\alpha]$ is not a Gaussian.
Nevertheless, compared to the general RGE (\ref{RGEA}), the evolution generated by
Eq.~(\ref{RGE-lin}) exhibits an important simplification:
it does not mix correlations $\langle \alpha(1)\alpha(2)\dots\alpha(n)\rangle$
with different numbers $n$ of fields \cite{PI}. Indeed, the quartic
operator acting on $W_\tau[\alpha]$ in the r.h.s. of Eq.~(\ref{RGE-lin}) is 
formally the same as a the second-quantized Hamiltonian for a non-relativistic 
many-body system, and is diagonal in the number of ``particles''.

Eq.~(\ref{RGE-lin}) provides a closed evolution equation 
for the 2--point function, which is the BFKL equation.
This equation is most commonly written for the charge-charge correlator
$\mu_\tau^2 (k_{\perp})\propto 
\langle\rho_a(k_\perp)\rho_a(-k_\perp)\rangle_\tau$. In the linear regime, it 
is the same as the unintegrated gluon distribution (\ref{phidef}). Specifically,
if one defines $\mu_\tau^2 (k_{\perp})$  as the Fourier transform of :
 \be\label{mutau}
\mu_\tau^2(x_\perp,y_\perp)\equiv 
\frac{\langle \rho^{a}(x_\perp)\,\rho^{a}(y_\perp)\rangle_\tau}{N_c^2-1}\,,\quad
\rho^a(x_{\perp})\equiv \!\int\! d{\rm y} \rho_{\rm y}^a (x_{\perp}),\ee
then Eqs.~(\ref{aaimom}) and (\ref{phidef}) imply, similar to Eq.~(\ref{linphi}),
\be\label{Fock-high}
\varphi_\tau(\kk)
\,\simeq\,\,\frac{\mu_\tau^2(k_\perp)} {\kk^2}\,
\qquad  {\rm for} \qquad k_\perp\gg Q_s(\tau),
\ee
while from the RGE (\ref{RGE-lin}) one obtains \cite{JKLW97} :
\be
 {\partial \mu_\tau^2 (k_{\perp}) \over \partial \tau}\,=\,{\bar\alpha_s}
\int \frac{d^2 p_{\perp}}{\pi}\,
 {k^2_{\perp} \over p^2_{\perp} (k_{\perp}-p_{\perp})^2}\big(
 \mu_\tau^2(p_{\perp}) -
\frac{1}{2}\, \mu_\tau^2(k_{\perp})\big)
\label{BFKL}
\end{eqnarray}
which is indeed the BFKL equation~\cite{BFKL}.

Note that $\mu_\tau^2 (k_{\perp})$ corresponds to $\mu_A^2$ of the MV model, 
but, unlike the latter, it carries a non-trivial transverse momentum dependence, and
a $\tau$ dependence, both of which are obtained by solving Eq.~(\ref{BFKL}).
In fact, given the formal similitude between Eqs.~(\ref{BFKL}) and (\ref{linearBK}),
it is clear that the corresponding solution for $\mu_\tau^2 (k_{\perp})$ can be 
obtained by replacing  $r_{\perp}^2 Q_0^2\rightarrow k_{\perp}^2/Q_0^2$ in
Eq.~(\ref{BFKLsol1}). This
shows the expected rapid exponential growth with $\tau$ and infrared diffusion as well.

\bigskip
\noindent {\bf ii) Low--momentum regime ($Q^2\ll Q_s^2$) : Saturation}
\smallskip

An external probe with low transverse resolution $Q^2\ll Q_s^2$ couples
mostly to the saturated gluons, which have momenta $\kk\simle Q_s$ and
occupation numbers $\sim 1/\alpha_s$.
The  corresponds classical fields are strong, $g\alpha(x_{\perp}) \sim 1$, 
so the Wilson lines (\ref{vy})---which are complex exponentials built with these fields---oscillate around 
zero over a characteristic distance $\sim 1/Q_s(\tau)$ in the transverse
plane. This implies that  Wilson lines which are separated by large 
distances $\gg 1/Q_s(\tau)$ are necessarily uncorrelated (since their
relative phases are random). 
Thus, when studying the dynamics over large transverse separations 
$\rr\gg 1/Q_s(\tau)$, it should be a good approximation
to neglect the correlations of the Wilson lines
(or, more generally, to treat them as small quantities).
This is the ``random phase approximation'' (RPA) introduced in
Refs. \cite{SAT,W}.

In this approximation, the RGE (\ref{RGE}) simplifies drastically
\cite{SAT}. Neglecting the Wilson lines,
the kernel $\eta$ becomes independent of $\alpha$, and the RGE reads
in momentum space,
\be\label{RGE-RPA}
{\del  W_\tau[\alpha] \over {\del \tau}}\,=\,{1 \over 2}
\int\!\frac{d^2k_\perp}{(2\pi)^2}\,\,\frac{1}{\pi\kk^2}\,
{\delta^2  W_\tau[\rho]\over {\delta\alpha_\tau^a(k_{\perp})
\delta \alpha_\tau^a(-k_{\perp})}}\,.
\ee
Being quadratic, this equation can be immediately integrated~\cite{SAT,CARGESE}
\be\label{W-low}
{\cal W}_\tau^{\,{\rm low}}[\rho]\approx {\cal N}_\tau\,
{\rm exp}\!\bigg\{-\,{\pi \over 2}
\int\limits_{-\infty}^\tau d{\rm y}
\int\limits^{Q_s({\rm y})}{d^2k_\perp\over (2\pi)^2}\,
\frac{\rho_{\rm y}^a(k_{\perp})
\rho_{\rm y}^a(-k_{\perp})}{k_\perp^2}
\bigg\},\ee
which for convenience has been written as a functional of
$\rho_{\rm y}^a(k_\perp)=k_{\perp}^2\alpha_{\rm y}^a(k_\perp)$.
Eq.~(\ref{W-low}) is a low-momentum approximation:  for a given rapidity y (with
${\rm y}\le \tau$). It is to be used only for modes $k_{\perp} < Q_s({\rm y})$.

We see an interesting {\it duality} emerging at saturation: this strong
field regime allows for a description in terms of a Gaussian weight
function, like for a free theory. But even with such a Gaussian
weight function, the CGC effective theory remains non--trivial, since the classical
solution (\ref{tpg})  and the observables for high energy scattering, like
(\ref{Ntau}), are non-linear functionals of $\rho$.

Eq.~(\ref{W-low}) 
shows that the only non-trivial correlation of the color sources with
small $k_{\perp}$ is the 2-point function, and reads:
\be\label{rhoHM}
\langle \rho_{\rm y}^{a}(x_\perp)\,
\rho_{{\rm y}'}^{b}(y_\perp)\rangle_\tau&=&\delta^{ab}\delta({\rm y}
-{\rm y}')\theta(\tau-{\rm y})\,\lambda_{\rm y}(x_\perp,y_\perp),\nonumber\\\label{lamRPA}
\lambda_{\rm y}(\kk)&\simeq&\frac{1}{\pi}\,\kk^2\,,\quad {\rm for}\quad \kk\ll Q_s({\rm y}).
\ee
This distribution is local in (space-time) rapidity y, and
homogeneous in all the (longitudinal and transverse) coordinates.
In the transverse plane, it is only a function of the
relative coordinate $x_\perp-y_\perp$, and in the longitudinal direction,
it is independent of y. For a given $\kk\ll Q_s(\tau)$, Eq.~(\ref{lamRPA}) applies
only for y in the interval $\tau_s(\kk)<{\rm y} < \tau$,
with $\tau_s(\kk)$ being the
rapidity at which the saturation scale becomes equal to the momentum
$\kk$ of interest :
\be\label{taus} 
Q_s^2(\tau)\,=\, \kk^2\qquad {\rm  for}\qquad \tau\,=\,\tau_s(\kk),
\ee
 (see Fig. ~\ref{phase-diagram}).
It follows that the integrated quantity (cf. eq.~(\ref{mutau})) :
\be\label{mu-sat}
\mu_\tau^2(k_\perp)\big |_{\rm sat}=\!
\int\limits_{\tau_s(k_\perp)}^\tau \! d{\rm y \,} \,\frac{k_\perp^2}{\pi}\,
= \big(\tau-\tau_s(k_\perp)\big)\frac{k_\perp^2}{\pi}\,,
\,\quad (\kk\ll Q_s(\tau)),\,\,\ee
which measures the density of saturated color sources (with given $k_\perp$)
in the transverse plane,
grows only {\it linearly} with $\tau$. This should be contrasted with the exponential
increase of the corresponding quantity at $\kk\gg Q_s(\tau)$, obtained  from the BFKL equation (\ref{BFKL}).
(In Ref. \cite{AM02}, the result in Eq.~(\ref{mu-sat}) was obtained from a study of the BK equation.)

We conclude that, at low momenta 
$k_\perp\ll Q_s(\tau)$, the color charge density {\it saturates}, because
of the non--linear effects in the quantum evolution \cite{SAT}. 
In contras, in the MV model, the corresponding quantity
$\mu_A^2$ increases indefinitely as a power of the atomic number $A$
(the analog of the energy in the MV model), since the respective sources
are uncorrelated. Due to the quantum evolution, correlations
arise already in the linear regime via the BFKL equation. However it is only
after including the non-linear effects associated with gluon recombination
that these correlations are such as to limit the growth of the color charge density
with $\tau$. 

Since the color sources saturate with $\tau$, so does also the gluon distribution
radiated by these sources. This is obvious if the {\it linear} 
approximation\footnote{Recall that Eq.~(\ref{Fock-high}) was obtained by using
the linearized solution to the classical EOM, i.e., Eq.~(\ref{aaimom}).},
Eq.~(\ref{Fock-high}), is used to compute this distribution. Namely, Eqs.~(\ref{Fock-high})
and (\ref{mu-sat}) imply:
\be\label{Fock-low}
\varphi_\tau(\kk)\,\simeq\,\frac{\tau-\tau_s(k_\perp)}{\pi}\,=\,
\frac{1}{\pi c\bar\alpha_s}\,\ln\frac{Q_s^2(\tau)}{\kk^2}\,,
\ee
where in writing the second equality we have used Eq.~(\ref{Qs0}) for the
saturation scale\footnote{Recall that Eq.~(\ref{Qs0}) corresponds to a fixed coupling
$\alpha_s$. In the case of a running coupling, where Eq.~(\ref{Qsrun}) applies, 
Eq.~(\ref{Fock-low}) remains formally the same, 
but $\bar\alpha_s$ in the denominator must be understood as the running coupling
(\ref{running}) evaluated at $Q^2=\kk Q_s$.},
together with the definition (\ref{taus}) of $\tau_s(k_\perp)$.
A more careful calculation, based on the non-linear solution (\ref{tpg}),
shows that the correct answer for $\varphi_\tau(\kk)$ differs from 
Eq.~(\ref{Fock-low}) only by a numerical factor $\delta \simle 1$ \cite{SAT,GAUSS}. 
This factor cannot be computed in the present approximations, as it
is sensitive to the transition regime at $k_\perp\sim Q_s(\tau)$. 
This overall factor will not affect the salient
features of Eq.~(\ref{Fock-low}), namely its dependence upon the energy and the
transverse momentum.

\begin{figure} 
\begin{center} 
\includegraphics[width=0.7\textwidth]
{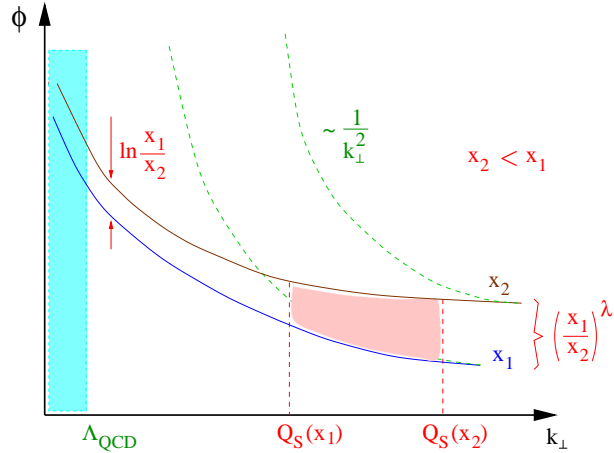} 
\caption{The gluon phase-density ${\varphi}_\tau(k_\perp)$
in the effective theory plotted as a function of $k_\perp$ for
two values of $\tau=\ln(1/\xx)$.}
\label{SAT-RGE} 
\end{center}
\end{figure}
In addition to {\it $\tau$--saturation} (the linear increase with $\tau$), 
Eq.~(\ref{Fock-low}) shows also {\it $k_\perp$--saturation}--the $k_\perp$ spectrum 
is only logarithmic in $1/k_\perp$ at low momenta. Recall that, in the 
classical MV model, a similar spectrum emerged (Eq.~(\ref{Low-N}))
only after fully taking into account the non-linear effects in the classical EOM
(\ref{cleq0}). By contrast, in the quantum case, the non-linear effects responsible
for saturation have already been encoded in the distribution of the color sources.
This is natural since, as noticed at the end of Sect. 2,  the difference
between ``sources'' and ``wee gluons'' is a matter of convention and depends
upon the resolution scale in $k^+$.
As anticipated, the gluon occupation factor at saturation, Eq.~(\ref{Fock-low}),
is parametrically of order $1/\alpha_s$, as in the MV model.

To conclude, the unintegrated gluon distribution predicted by the CGC effective theory
is illustrated in Fig.~\ref{SAT-RGE}, and looks qualitatively similar to that
in the MV model, cf. Fig.~\ref{SATURATION-MV}. 
At large $k_\perp\gg Q_s(\tau)$, the distribution in Fig.~\ref{SAT-RGE} is given by
the solution to the BFKL equation, cf. Eqs.~(\ref{Fock-high})--(\ref{BFKL}), while at
low  $k_\perp\ll Q_s(\tau)$, it is given by Eq.~(\ref{Fock-low}). The saturation
condition (\ref{Nsat}) can be also formulated in terms of the 
unintegrated gluon distribution as 
\be\label{satmu}
\varphi_\tau(\kk)
\,\sim\,\frac{1}{\bar\alpha_s}
\qquad{\rm for}\qquad \kk\sim Q_s(\tau)\,.\ee
This condition, together with BFKL evolution at higher momenta,
implies that, for $k_\perp$ within the scaling window (\ref{swindow}), 
the unintegrated gluon distribution has a scaling form similar to 
Eq.~(\ref{Nscaling}) \cite{GAUSS}:
\be\label{muhigh}
\varphi_\tau(k_\perp)\,\simeq\,\frac{\kappa}{\bar\alpha_s}
\left(\frac{Q_s^2(\tau)}{\kk^2}\right)^\gamma\,.
\ee
where $\kappa$ is a yet undetermined numerical prefactor.
We see that the scaling property characteristic of saturation (cf. Eq.~(\ref{Fock-low}))
is preserved by the linear BFKL evolution up to a relatively large moemtum $\kk \sim 
Q_s^2(\tau)/\Lambda_{\rm QCD}$ (cf. Eq.~(\ref{swindow})), which is well above the 
saturation scale. This ``extended scaling'' region \cite{KIIM} where the gluon density
is relatively low, but takes the scaling form (\ref{muhigh}), is represented on
the diagram in Fig. ~\ref{phase-diagram}.

\bigskip
\noindent {\bf iii) Color neutrality at saturation}
\smallskip

The vanishing of the charge-charge correlator (see (\ref{mu-sat})),
as $\kk^2$ when $\kk\to 0$ has important consequences for the
infrared behaviour of the CGC effective theory. Consider, the calculation
of the dipole-hadron $S$-matrix element. In the MV model, it was found to
be logarithmically infrared divergent, cf. Eqs.~(\ref{SMV})
or (\ref{SMV2}). If in Eq.~(\ref{mux})
one replaces the MV estimate for the charge correlator $\lambda_A$ by the corresponding quantum expression, 
the ensuing integral over  $\kk$ becomes infrared finite,
due to Eq.~(\ref{lamRPA}). When computed in the quantum effective theory,
the dipole scattering amplitude is infrared safe, and therefore insensitive to the 
non-perturbative physics of confinement. A similar property holds for other gauge-invariant
quantities like the gluon distribution.

Furthermore, the smooth infrared behaviour in Eq.~(\ref{lamRPA}) is 
responsible for $k_\perp$--saturation in the gluon distribution. In what follows, 
we shall argue that this behaviour has a simple physical interpretation--gluonic
color sources are correlated over long distances to ensure that {\it color neutrality} 
is achieved over a transverse area of order $1/Q_s^2(\tau)$ \cite{CARGESE,AM02,GAUSS}.

Consider the total color charge ${\cal Q}^a$ enclosed within a surface
$\Sigma$, as given by  Eq.~(\ref{Qtot}) with $\Delta S_\perp \to \Sigma$.
This is a random quantity with zero average (since $\langle\rho^a(\vec x)\rangle =0$
at any point $\vec x$), so we shall compute the average of the color
charge squared ${\cal Q}^2={\cal Q}^a{\cal Q}^a$.
We have, 
\be\label{QSigma}
\langle {\cal Q}^2 \rangle_\tau\,=\,(N_c^2-1)\int_\Sigma \!{d^2x_\perp}
\int_\Sigma \!{d^2y_\perp}\,\mu_\tau(x_\perp,y_\perp)\,.\ee

In the MV model, where the sources are uncorrelated (cf. Eq.~(\ref{MV-corr})), we have
\be\label{QSigmaMV}
\langle {\cal Q}^2 \rangle_A\,=\,
(N_c^2-1) \Sigma \mu_A^2\,\sim\,\frac{1}{\alpha_s}\,\Sigma Q_A^2\,,\ee
which increases rapidly with $A$, like  $A^{1/3}$.

After including quantum evolution, the charge correlator
acquires a non-trivial momentum dependence ($\mu_A^2 \rightarrow \mu_\tau^2(k_\perp)$), 
and Eq.~(\ref{QSigma}) can be estimated as (up to a color factor): 
\be\label{QSigma0}
\langle {\cal Q}^2\rangle_\tau\,\sim\,\Sigma\,\mu_\tau^2(k_\perp^2
\sim 1/\Sigma)\,.\ee 

For a relatively small area, or moderately high energies, $\Sigma^{-1} \gg Q_s^2(\tau)$, 
and $\mu_\tau^2(k_\perp)$ is given by the BFKL equation 
(\ref{BFKL}). Then Eqs.~(\ref{Fock-high}) and (\ref{muhigh}) imply :
\be\label{QSigmaL}
\langle {\cal Q}^2\rangle_\tau\,\sim\,
\frac{1}{\alpha_s}\,\big(\Sigma Q_s^2(\tau)\big)^\gamma
\qquad {\rm for}\qquad 1/\Sigma \gg Q_s^2(\tau),
\ee
which shows {\it incomplete color shielding}. 
With increasing $\Sigma$, the total charge 
squared enclosed within this surface increases,
but not as fast as the area itself. Thus, the {\it density}
${\langle {\cal Q}^2 \rangle}/{\Sigma}\,\sim\,1/\Sigma^{1-\gamma}$
vanishes in the limit $\Sigma\to\infty$, 
which is consistent with (global) charge conservation.
Furthermore, for fixed $\Sigma$, the total charge $\langle {\cal Q}^2 \rangle_\tau$
increases exponentially with $\tau$.

For larger surfaces, of the order of the  saturation disk $1/Q_s^2(\tau)$ or larger, 
one should rather us eq.~(\ref{mu-sat}), which gives:
\be\label{QSigmaNL}
\langle {\cal Q}^2\rangle_\tau\,\sim\,
\frac{1}{\alpha_s}\,\ln \big(\Sigma Q_s^2(\tau)\big)
\qquad {\rm for}\qquad 1/\Sigma \ll Q_s^2(\tau).\ee
The total charge squared (\ref{QSigmaNL}) is still non-zero, but unlike
Eq.~(\ref{QSigmaL}) it is only 
{\it logarithmically} increasing with both $1/\xx$ and $\Sigma$.
This is {\it complete shielding} : When increasing $\Sigma$, the 
newly included color sources are completely screened by the other sources. 
If the total  charge (squared)  is still increasing with  $\Sigma$
(albeit only slowly), it is because the
longitudinal width $\tau-\tau_s(\Sigma)\sim \ln(\Sigma Q_s^2(\tau))$
of the ``saturated'' piece of the hadron rises logarithmically with $\Sigma$.

This complete shielding, together with the fact that the total charge 
(\ref{QSigmaNL}) is much smaller than the total charge for a system
of uncorrelated color sources with surface density $Q_s^2/\alpha_s$,
cf. eq.~(\ref{QSigmaMV}),
 enables us to speak about {\it color neutrality} already
at the relatively short scale $1/Q_s(\tau)\ll 1/\Lambda_{\rm QCD}$. 
This interpretation is further confirmed by the
fact that the color field created by gluon sources at large distances
$\gg  1/Q_s(\tau)$ cannot be distinguished from a {\it dipolar} field \cite{GAUSS}.

\subsection{A Gaussian effective theory}

Consider the calculation of the dipole-hadron $S$-matrix element (\ref{Ntau})
within the CGC effective theory. After expanding the Wilson lines in
Eq.~(\ref{Ntau}), one is led to evaluate $n$-point functions of the
type $g^n\langle\alpha(\vec z_1)\alpha(\vec z_2)\cdots\alpha(\vec z_n)\rangle_\tau$,
where the transverse arguments $z_{\perp,i}$ are either $x_\perp$ or $y_\perp$.
Each such a $n$-point function receives contributions from either hard 
($k_\perp > Q_s(\tau)$) or semi-hard\footnote{As explained above, gauge invariant 
quantities computed in the effective theory are not sensitive to soft 
($k_\perp \simle \Lambda_{\rm QCD}$) momenta.} ($\Lambda_{\rm QCD} < k_\perp < Q_s(\tau)$) 
momenta. The contributions of the modes with $k_\perp \ll Q_s(\tau)$ are easy to
evaluate. In the saturation regime, the weight function is the Gaussian
(\ref{W-low}), and the only non-trivial correlation is the 2-point function (\ref{lamRPA}).
For hard momenta $k_\perp \gg Q_s(\tau)$, the weight function is strictly speaking
not a Gaussian. In this regime, fields are rather weak ($g\alpha\ll 1$), so the (yet unknown)
contributions of the higher-point correlations are less important than that of the
2-point function. The latter is explicitly known as the solution to the BFKL equation.
Finally, there are the contributions of the modes $k_\perp \sim Q_s(\tau)$,
which we don't know how to evaluate so far. But these contributions should be
relatively unimportant as long as $x_\perp-y_\perp$ is either very small, or very large,
compared to $1/Q_s(\tau)$. To summarize, in order to compute $S_\tau(r_\perp)$ 
for a dipole which is either very large, or very small, it should be enough to use
the previous approximations for the 2-point function, which are valid far away from the
saturation scale.

These considerations can be extended to any quantity which is not very sensitive
to the momenta in the transition regime around $Q_s(\tau)$. To systematically compute such quantities, 
construct a Gaussian approximation to the weight function which encodes 
the limiting behavior of the the 2-point function at high and low momenta and 
which interpolates smoothly between these regimes.
Such a  Gaussian will be as simple to use as the original MV model,
but will extend the latter by including the BFKL evolution at high momenta, and
the physics of saturation and color neutrality at low momenta.

Such a Gaussian weight function has been constructed in Ref. \cite{GAUSS}, and reads:
\be\label{Wgauss}
{W}_{\tau}[\rho]
\,=\,{\cal N}_\tau\,
{\rm exp}\!\left\{-\,{1 \over 2}
\int_{-\infty}^\tau d{\rm y}
\int_{x_\perp,y_\perp}\!\frac{\rho_{\rm y}^a(x_{\perp})
\rho_{\rm y}^a(y_{\perp})}{\lambda_{\rm y}(x_{\perp},y_\perp)}
\right\},\ee
where the kernel $\lambda_{\rm y}(x_{\perp},y_\perp)$ --- the 2-point function
of the color sources (cf. Eq.~(\ref{rhoHM})) --- is such that its Fourier transform
$\lambda_{\rm y}(k_{\perp})$ satisfies the BFKL equation at momenta $k_\perp \gg Q_s(\tau)$
and reduces to Eq.~(\ref{lamRPA}) for momenta $k_\perp \ll Q_s(\tau)$. We can match
these limiting behaviours continuously at $k_{\perp}= Q_s(\tau)$ by imposing
the following  boundary condition on the solution to the  BFKL equation:
\be\label{QSLAM}
\frac{1}{\pi}\,Q_s^2(\tau)\,\simeq\,\lambda_\tau^{\rm BFKL}(k_{\perp}=Q_s(\tau))\,.\ee
Then, the following function provides a smooth interpolation, that we shall use in
Eq.~(\ref{Wgauss}) :
\be\label{barlam}
\lambda_\tau(k_{\perp})\,\equiv\,\frac{
k_{\perp}^2 \lambda_\tau^{\rm BFKL}(k_{\perp})} {k_{\perp}^2+\pi 
\lambda_\tau^{\rm BFKL}(k_{\perp})}\,.\ee
For momenta within the scaling window (\ref{swindow}), the BFKL solution
takes the scaling form (with $\gamma= 0.64$) :
\be\label{lamscaling}
\lambda_\tau^{\rm BFKL}(\kk)\, \simeq\, \frac{1}{\pi}\,  \kk^2
\left(\frac{Q_s^2(\tau)}{\kk^2}\right)^\gamma,
\ee
which allows us to write a simple explicit expression for the kernel
$\lambda_{\rm y}(k_{\perp})$, valid for 
all momenta $\kk\simle Q_s^2(\tau)/\Lambda_{QCD}$:
\be\label{bar2}
    \lambda_{\rm y}(k_{\perp})\,=\,\theta(\tau-{\rm y})\,
\frac{k_{\perp}^2}{\pi}\,\frac{
\left(\frac{Q_s^2({\rm y})}{\kk^2}\right)^\gamma}{1 +
\left(\frac{Q_s^2({\rm y})}{\kk^2}\right)^\gamma}
\,.\ee

\begin{figure}
\begin{center} 
\includegraphics[width=0.7\textwidth]{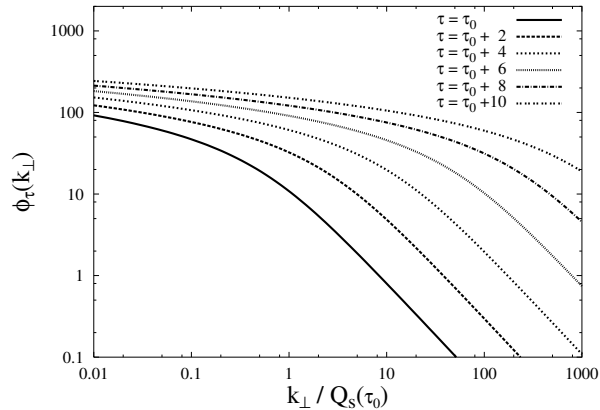}
\caption[]{Energy and momentum dependence of $\varphi_\tau(\kk)$. We have plotted 
  $\varphi_{\tau}(\kk)$  as a function of $\kk/Q_s(\tau_0)$ (with $\tau_0$
  some value of reference) for six values of $\tau$.
  The lines, from the bottom to the top, correspond successively to 
  $\tau=\tau_0,\, \tau_0 + 2,\, \cdots,\, \tau_0+10.$ The increase 
  with $\tau$ is exponential at high momenta (giving equidistant curves 
  in this log--log plot), but only logarithmic at low  momenta. From Ref. \cite{GAUSS}.}
\label{F5} 
\end{center}
\end{figure}

In Sect. 4.3 , Eq.~(\ref{bar2}) will be used to compute  
the dipole--hadron scattering amplitude.
Here, we shall present a different application, namely we shall 
deduce a simple analytic expression for the unintegrated gluon distribution.
Specifically, Eqs.~(\ref{mutau}) and (\ref{rhoHM}) imply $\mu_\tau(k_\perp)=\int d{\rm y}
\lambda_{\rm y}(k_\perp)$. When further combined with 
Eqs.~(\ref{Fock-high}) and (\ref{bar2}), it leads to the following final result:
\be\label{phibar}
\phi_\tau(k_\perp)\,=\, \frac{1}{k_{\perp}^2}\int  d{\rm y}\,
\lambda_{\rm y}(k_\perp)\,=\,\frac{1}{\pi\gamma c\bar\alpha_s}\,
\ln\left(1 \,+\,\left(\frac{Q_s^2(\tau)}{\kk^2}\right)^\gamma\right).\ee
(In performing the integral over y, we assumed Eq.~(\ref{Qs0})
for the saturation scale.) Eq.~(\ref{phibar}) interpolates
smoothly between Eq.~(\ref{Fock-low}) deeply at saturation ($\kk\ll Q_s$)
and Eq.~(\ref{muhigh}) for momenta within the scaling window 
$Q_s \simle \kk\simle Q_s^2/\Lambda_{QCD}$. In Fig. \ref{F5}, we illustrate both
the $\kk$--dependence and the $\tau$--dependence of the function (\ref{phibar}),
which is plotted as a function of $\kk$ for several values of $\tau$.

\section{Deep Inelastic Scattering and the CGC}
\setcounter{equation}{0}

In this section, we shall discuss some applications of the Color Glass 
Condensate picture to Deep Inelastic Scattering. We start by explicitly deriving
a factorization formula introduced in Sect. 2.7,
which allows one to compute the $F_2$
structure function at high energy in terms of the dipole scattering.
We shall then discuss a
phenomenological ``saturation model'', proposed by Golec-Biernat and W\"usthoff,
which is based on similar physical premises, and compares remarably well with
the HERA data. This model will be further reconsidered from the CGC perspective, 
in which the dipole-hadron scattering is computed within QCD. This allows
us to relate the ``geometric scaling'' observed in the HERA data to properties
of the quantum evolution towards saturation.
In addition, we discuss the Froissart bound can be realized conceptually in the 
Color Glass Condensate. We discuss next the relation of saturation to shadowing in 
DIS. Finally, we discuss inclusive and semi-inclusive 
signatures of the Color Glass Condensate in Deep Inelastic Scattering and 
specifically how they may be observed at a future Electron Ion Collider.

\subsection{Structure Functions in the Color Glass Condensate}

We shall show here how one computes structure functions for DIS
in the presence of non-linear effects associated with saturation~\cite{MV99}.
Towards this end, one needs the correlator of the
electromagnetic current in the background of the strong classical gluon field 
representing the CGC. In DIS, the interaction between the hadron and the virtual
photon is encoded in the following tensor expressed
in terms of the forward Compton scattering amplitude $T_{\munu}$~\cite{TB-DIS}:
\be
W^{\munu}(q^2,P\cdot q) &=& 2~{\rm Disc}~ T^{\munu}(q^2,P\cdot q) \nonumber \\
&\equiv&
{1\over {2\pi}}\, {\rm Im} \int d^4 x \, {\rm e}^{iq\cdot x}
 <P|{\rm T}(J^\mu (x) J^\nu (0))|P> \, ,
\label{hadtensor}
\ee
where ``T'' denotes a time-ordered product, $J^\mu={\bar{\psi}}\gamma^\mu
\psi$ is the hadron electromagnetic current and ``Disc'' denotes the
discontinuity of $T_{\munu}$ along its branch cuts in the variable
$P\cdot q$.  Also, $q^2 < 0$ is the transferred momentum
squared (i.e., $q^\mu$ is the momentum of the virtual photon, 
and $Q^2 = -q^2$) and $P^\mu$ is the
momentum of the target.  In the IMF, $P^+\rightarrow \infty$ is the
only large component of the momentum.  
Since 
\be
  < T(J^\mu (x) J^\nu(y))> = <T\left(\overline \psi (x) \gamma^\mu \psi(x)
\overline \psi(y) \gamma ^\nu \psi(y)\right) >\, ,
\ee
the time ordered produced of currents can be expressed, in complete generality, as
\be       <T(J^\mu(x) J^\nu(y))> \,\,= \qquad \qquad \qquad\qquad \qquad \qquad\nn 
=\,{\rm Tr}(\gamma^\mu G_A(x)) {\rm Tr} (\gamma^\nu G_A(y)) +
{\rm Tr}(\gamma^\mu G_A(x,y) \gamma^\nu G_A(y,x) ) \, ,
\label{tad}
\ee
where $G_A(x,y) = -i< \psi(x) \overline \psi (y) >_A$ is the quark
Green's function in the external background gauge field $A^\mu_a$ of 
the hadron.

The first term on the right hand side of Eq.~(\ref{tad}) is a tadpole
contribution without an imaginary part.  It
therefore does not contribute to $W^{\mu\nu}$.  We are thus left with:
\be\label{NOPE}
        W^{\mu \nu}(q^2,P\cdot q)\,=\,{1\over 2\pi}\, {P^+ \over M} \,
{\rm Im} \int d^3X  \,\int d^4x \,e^{iq\cdot x}\,\qquad\qquad\qquad\qquad\qquad\\
\big\langle{\rm Tr}\big( \gamma^\mu G_A (X + x/2,X - x/2)
 \gamma^\nu 
G_A(X- x/2,X + x/2)\big)\big\rangle \, .\nonumber
\ee
The approximation one makes here is to replace the full background gauge field $A_\mu^a$ by the 
classical background field $A_{\mu,{\rm classical}}^a$. In other words, the Green's 
function which, in general, is computed in the full background field of the nucleus, is now computed in 
the saddle--point approximation where $A_{\mu}^a \rightarrow A_{\mu,{\rm classical}}^a$. 
Note that this expression makes no reference to the operator product expansion of DIS~\cite{TB-DIS}. Thus, it
is also valid at small values of x and moderate $Q^2$, where the operator product expansion
is not reliable~\cite{AMueller1} but where the classical approximation is sensible. 

Since the current-current correlator is gauge invariant, one can compute it by using background field propagator
in an arbitrary gauge. For the reasons explained in Sect. 2.4, it is most
convenient to use the covariant gauge, in which the  propagator
reads as follows~\cite{HW,Balitsky1996,MV99}:
\be
       & & G_A(x,y)  =  G_0(x-y) -i \int d^4z \,G_0(x-z)\gamma^- \delta(z^-)G_0(x-z)\nonumber \\
& & \quad\big\{\theta(x^-)\theta(-y^-)
(V^\dagger(z_\perp)-1)  
 - \theta(-x^-)\theta(y^-)(V(z_\perp)-1)\big\}\,,
\label{singprop}
\ee
where $G_0$ is the free propagator and $V$ and $V^\dagger$ are the Wilson lines
of Eq.~(\ref{v}). 
This is obtained by assuming the color source 
to be a $\delta$--function in $x^-$, which is appropriate since this source
is due to relatively fast partons with longitudinal momentum fractions much larger
than the Bjorken $\xx = Q^2/2P\cdot q$ of the collision.

Inserting the fermion propagator (\ref{singprop}) in  Eq.~(\ref{NOPE}) and performing the 
integrations there, one obtains the final result for $W^{\mu\nu}$, which 
is conventionally expressed (for an unpolarized target and for $Q^2<< M_W^2$) in terms of two structure
functions $F_1$ and $F_2$, defined by~\cite{TB-DIS}:
\be
M W^{\munu} &=& -\big(g^{\munu} - {q^\mu q^\nu \over {q^2}}\big) F_1\nonumber \\
&{}&+ \big(P^\mu - {q^\mu (P\cdot q)\over {q^2}}\big)\,\big(P^\nu -
{q^\nu (P\cdot q)\over {q^2}}\big) {F_2\over {(P\cdot q)}} \, . 
\label{strctfn}
\ee

Consider first $F_2$: the corresponding result takes the factorized structure
anticipated in Eqs.~(\ref{F2gamma})--(\ref{sigmadipole}), where $\vert \Psi(z,r_\perp)\vert^2$,  
which signifies the probability that the virtual photon splits into a 
$q\bar q$--pair can be expressed as $\vert \Psi(z,r_\perp)\vert^2 = \vert \Psi_{T}(z,r_\perp)\vert^2 +
\vert \Psi_{L}(z,r_\perp)\vert^2 $, and where each of the terms has the explicit form
\be
\vert \Psi_{T}(z,r_\perp)\vert^2 &=&\frac{3\alpha^{}_{em}}{2\pi^2}
                       \sum_f e_f^2 \left\{
 			(z^2 +(1-z)^2) \bar Q_f^2 K_1^2(\bar Q_f r) + m_f^2 K_0^2(\bar Q_f r)\right\},\nn
\vert \Psi_{L}(z,r_\perp)\vert^2 
&=&\frac{3\alpha^{}_{em}}{2\pi^2}
                       \sum_f e_f^2  \left\{ 
                      4 Q^2 z^2 (1-z)^2 K_0(\bar Q_f r)\right\}\, .
\ee
Above, the sum runs over the quark flavours,
$\bar Q_f^2 = z (1-z)Q^2 + m_f^2$, $m_f$ is the quark mass, 
and $K_0$ and $K_1$ are modified Bessel functions. $\vert \Psi_{T}(z,r_\perp)\vert^2$ 
($\vert \Psi_{L}(z,r_\perp)\vert^2$) denotes the probability that 
a transversely (longitudinally) polarized photon splits into a $q\bar q$-pair. 
This decomposition implies a similar decomposition for $F_2$, namely, $F_2=F_T+F_L$. 
$ F_1$, this is proportional to $F_T$--explicitly, $F_T = 2x F_1$ and the longitudinal 
structure function is $F_L = F_2 -2x F_1$. In the parton model, $F_L=0$--the Callan-Gross 
relation--but is non-zero in QCD and is directly proportional to the gluon distribution. 
An independent measurement of $F_L$ is therefore of great phenomenological interest as 
will be discussed in Sect. 4.6. One can similarly derive expressions for the diffractive 
structure functions as discussed in Ref.~\cite{KovMcLerr}.
Some of the phenomenological implications of the CGC picture for measurements of structure functions were 
discussed in Sects. 2.6 and 2.7.

  \subsection{The Golec-Biernat--W\"usthoff model}

In Ref.~\cite{GBW99}, Golec-Biernat and W\"usthoff introduced a simple phenomenological
model for the dipole-hadron cross-section, Eq.~(\ref{sigmadipole}), which is
generally referred to as the ``Saturation Model'' :
\be\label{Golec}
\sigma_{\rm dipole}(\xx,r_\perp)\,=\,\sigma_0\big(1 - {\rm e}^{-r_\perp^2 Q_s^2(\xx)/4}\big)
\ee
with the parametrization $Q_s^2 (\xx)= Q_0^2 (\xx_0/\xx)^\lambda$.
This shows color transparency at low $r_\perp \ll 1/Q_s(\xx)$ and respects the unitarity
bound, as it approaches a constant value $\sigma_0$ when $r_\perp \gg 1/Q_s(\xx)$.
If one interpretes $\sigma_{\rm dipole}(\xx,r_\perp)/\sigma_0$ as a scattering amplitude
(compare to Eq.~(\ref{sigmadipole})), then this has no dependence upon the impact parameter 
$b_\perp$. Rather, this should be seen as an average of the true scattering amplitude 
over all impact parameters. This amplitude is consistent with the physics of 
saturation (compare to Fig.~\ref{NFIG}), so it is natural to interprete the scale $Q_s(\xx)$ 
in Eq.~(\ref{Golec}) as a ``saturation momentum''. Note however that the
approach towards the unitarity bound ${\cal N}_\tau(r_{\perp})= 1$ for large $r_\perp$
is much faster then predicted by the CGC approach, or the solution to the BK equation
(see Sect. 4.3 below). Besides, at small $r_\perp$, Eq.~(\ref{Golec}) fails to
reproduce, as it should, the leading-twist approximation--it misses the logarithmic factor $\ln{(1/ r_\perp^2\Lambda^2)}$ 
in the exponent of Eqs.~(\ref{SMV2}). Neither does it recover the BFKL prediction (\ref{BFKLsol1}).

By using the dipole cross-section (\ref{Golec}) and the factorization formula
(\ref{sigmagamma}) for DIS, Golec-Biernat and W\"usthoff
were able to fit the HERA data for $\sigma_{\gamma^* p}$ for 
$\xx< 10^{-2}$ and a wide range in $Q^2$ with only three parameters, namely
$\sigma_0=23$ millibarns, $\lambda=0.288$ and $\xx_0=3.04\cdot 10^{-4}$. 
(The reference scale $Q_0^2$ has been fixed as $Q_0^2=1$ GeV$^2$.)
These fits were performed for three light quark flavors; with the addition of 
charm quarks, the best fit was obtained with slightly changed
parameters:  $\sigma_0=29.1$mb, $\lambda=0.277$, and $x_0=0.4\cdot 10^{-4}$. 

The fits are reasonable for $Q^2$ up to $Q^2\sim 20$ GeV$^2$, but are less successful beyond. 
This is related to the above observation that Eq.~(\ref{Golec}) does not have the right
perturbative behaviour at  small $r_\perp$. By replacing Eq.~(\ref{Golec}) with a Glauber--type
formula as shown in Eq.~(\ref{SMV3}), the situation at high $Q^2$
improves considerably and a wider range in $Q^2$ can be fit within the 
framework of this model~\cite{BGK1}. For other phenomenological models of the HERA data 
based on ideas of saturation see Refs.~\cite{Levinetal,FrankStriketal}. 

The diffractive structure function $F_2^D(x_{\cal P},Q^2,\beta)$
 (where $x_{\cal P}= (M^2+Q^2)/(W^2+Q^2)$ and $\beta=Q^2/(M^2+Q^2)$
 where $M$ is the diffractive mass and $W$ is the total energy of the
 virtual photon-proton process) was measured at HERA and several
 striking properties of the diffractive structure function were
 observed.  For instance, the ratio of $\sigma^{diff}/\sigma$ is large
 and is nearly independent of $W$, a feature that was not anticipated in pQCD
 based models. In Ref.~\cite{GBW99}, the simple model that describes
 the inclusive scattering data also describes the diffractive
 structure function data. Interestingly, the form of the diffractive
 cross-section in this model is similar to the inclusive one except
 the dipole-hadron cross-section appearing in the latter is replaced
 by the square of this cross-section~\footnote{This feature of the model can be understood very 
simply in the CGC picture~\cite{KovMcLerr}.}. The parameters appearing in the
 fit of the inclusive cross-section to the data are therefore the same
 as those used in the diffractive fit. The agreement with data is
 quite impressive given these constraints. A version of the model has
 also been applied to study vector meson production with reasonable
 results~\cite{CS,MSM}.

Whereas the behaviour at high $Q^2$ can be improved rather easily, by replacing 
Eq.~(\ref{Golec}) by Eq.~(\ref{SMV3}) as mentioned above,
the impact parameter dependence is problematic. This is expected to be especially important
in a proton. Data on the $t$-dependence of differential cross-sections, 
particularly for vector meson photoproduction, may be useful in understanding parton distributions 
in impact parameter space~\cite{MSM,BartLev}. 

  \subsection{Geometric Scaling in DIS}

The Golec-Biernat--W\"usthoff dipole cross-section, Eq.~(\ref{Golec}), has the remarkable
feature to depends upon the two kinematical variables $\xx$ and $r_\perp$ only via the 
dimensionless combination (the ``scaling variable'') ${\cal T}\equiv r_\perp^2 Q_s^2(\xx)$.
Via the factorization formula (\ref{sigmagamma}), this scaling property transmits
to the virtual photon total cross-section  $\sigma_{\gamma^* p}$ which, in the limit where
the quark masses are negligible, becomes a function of the ratio
$Q^2/Q_s^2(\xx)$ alone (a property usually referred to as ``geometric scaling'').
At a first sight, this may appear an artifact of the simple parametrization (\ref{Golec}) 
specific to the saturation model. Indeed, while the scaling looks natural at saturation
one would na\"{\i}vely expect this scaling to be broken
after generalizing Eq.~(\ref{Golec}) to reproduce the perturbative
behaviour at high $Q^2$ (e.g., the inclusion of the logarithm 
$\ln{(1/ r_\perp^2\Lambda^2)}$ as in Eq.~(\ref{SMV2}) would clearly violate scaling).

\begin{figure} 
\begin{center} 
\includegraphics[width=0.7\textwidth]{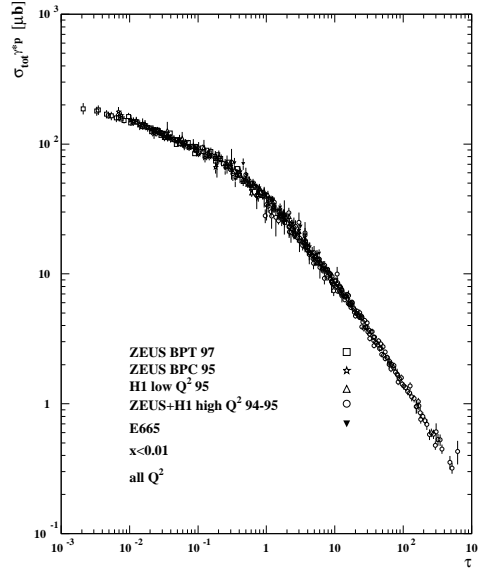} 
\caption[]{HERA data on the cross section for $\gamma^*p$ DIS
from the region $\xx < 0.01$ and $Q^2 < 400 {\rm GeV}^2$ plotted versus the 
scaling variable ${\cal T} = Q^2/Q^2_0(\xx)$ (from Ref. \cite{geometric}). }
\label{gb} 
\end{center}
\end{figure}

It thus appeared as a surprise when Sta\'sto, Golec-Biernat and Kwieci\'nski showed
\cite{geometric} that, to a rather good accuracy,
the HERA data on $\sigma_{\gamma^* p}$ {\em do} show scaling for small enough x ($\xx < 0.01$)
and all $Q^2$ up to $450 {\rm GeV}^2$ (see Fig.
\ref{gb}). Such $Q^2$ are significantly higher than the estimated value of
the saturation scale at HERA, as extracted from the ``saturation model''
fits to $F_2$  \cite{GBW99} : 
$Q_s^2 \simeq 1 \cdots 2\, {\rm GeV}^2$. On the other hand, the data show no scaling
for larger values of x.

Subsequently, some indications of geometric scaling have been found also
in the data for DIS off nuclei \cite{WeigertFreund}, and even in the
particle production at RHIC \cite{JDLR}, although, in these cases, 
the experimental evidence is more uncertain. 

In the light of the previous discussion in Sect. 3, such a scaling
behaviour should not look surprising any more. As explained in Sects. 3.4 (for the
scattering amplitude) and 3.5 (for the gluon distribution), the scaling property at
saturation is preserved by the BFKL evolution up to relatively large $Q^2$, of order
${Q_s^4(\tau)}/\Lambda_{\rm QCD}^2$ (cf. Eq.~(\ref{swindow})). 	If one uses 
the phenomenological values of $Q_s$ alluded to before, and $\Lambda_{\rm QCD} = 0.2 \cdots 0.3
{\rm GeV}$, one finds that the maximum $Q^2$ up to which scaling is expected
is indeed of the order of a few hundred ${\rm GeV}^2$, as observed at HERA \cite{geometric}.

Within the CGC formalism, the scaling properties of the dipole-hadron
scattering amplitude can be explicitly studied by using the Gaussian approximation
to the weight function introduced in Sect. 3.6. The kernel of the Gaussian shows explicit scaling for
all momenta $\kk\simle Q_s^2(\tau)/\Lambda_{QCD}$, cf. Eq.~(\ref{bar2}), and therefore
so does also the $S$--matrix element computed in this approximation. Specifically, a
straightforward calculation yields \cite{GAUSS} (in the case of a fixed coupling, cf.
Eq.~(\ref{Qs0}), for definiteness, and with $\kappa \equiv 4\pi C_F/\gamma c N_c$) :
\be\label{barS}
S_\tau(\rr)\,=\,
\exp\left\{ -\kappa 
\int \!{d^2k_\perp\over (2\pi)^2}\,\frac{1-
{\rm e}^{ik_\perp\cdot r_\perp}}{k_\perp^2}\,\,
\ln\left[1 \,+\,\left(\frac{Q_s^2(\tau)}{\kk^2}\right)^{\!\gamma}\,\right]
\right\}
\ee
Unlike the corresponding prediction of the MV model, Eq.~(\ref{SMV2}), which was infrared sensitive,
and thus dependent upon the non-perturbative scale $\Lambda_{\rm QCD}$, the integral in the
equation above is well behaved both in the infrared and in the ultraviolet, so, clearly, the result
is a scaling function: $S_\tau(\rr)=S(\rr Q_s(\tau))$. 

The limiting behaviour of Eq.~(\ref{barS}) at either small,
or large, transverse size $\rr$ as compared to the saturation length $1/Q_s(\tau)$ are~\cite{GAUSS}: 
i) For a small dipole, $\rr\ll 1/Q_s(\tau)$, 
the dominant contribution to the integral over
$\kk$ comes from {\it hard} momenta $\kk \gg Q_s(\tau)$, from the scattering
off color sources located at $|z_\perp - b_\perp|
\ll 1/Q_s(\tau)$ from the impact parameter $b_\perp$ of the incoming dipole.

If $\gamma < 1$ (in particular, for the BFKL value $\gamma \approx 0.64$), this dominant behaviour
has the power law form in Eq.~(\ref{Nscaling}), with ``anomalous dimension'' equal to $\gamma$.

 If $\gamma = 1$ (the DGLAP value~\cite{DGLAP}, which supplants BFKL evolution at very small $\rr$):
\be
{\cal N}_\tau(\rr)\,\approx\,
\frac{C_f}{4 c N_c}
\,Q_s^2(\tau)\rr^2
\left[ \ln\frac{1}{Q_s^2(\tau)\rr^2}+\big(1+2\psi(2)+2\ln 2\big) 
\right] .\label{short_dist_app}\ee
This is similar to the small--$\rr$ behaviour in the MV model, Eq.~(\ref{transp}), except that 
the infrared cutoff in the logarithm is now the saturation momentum, and not
$\Lambda_{\rm QCD}$.

i) For a large dipole, $\rr\gg 1/Q_s(\tau)$ (but $\rr\ll 1/\Lambda_{\rm QCD}$), 
the dominant contribution comes rather from momenta $\kk$ in
the range $1/\rr\ll \kk \ll Q_s(\tau)$, that is, from scattering off saturated gluon sources,
and reads :
\be\label{Stau_long}
S_\tau(\rr)\propto 
\exp\left\{-\frac{C_F}{2cN_c}\big(\ln {\rr^2Q_s^2(\tau)} \big)^2\right\}\, ,
\ee
in agreement with the results in Refs. \cite{LT99,SAT,AMCARGESE}.
As anticipated, the approach towards the ``black disk'' limit $S_\tau =0$ 
predicted by the CGC is slower than that assumed in the saturation model,
Eq.~(\ref{Golec}).

  \subsection{The Froissart Bound for dipole scattering}

Let us now address the fundamental question of the asymptotic behaviour of the total
cross-section at very high energies. Thus far, in the analysis of the dipole-hadron scattering,
we have neglected the impact parameter dependence. We have assumed the
hadron to be a homogeneous disk with radius $R$.
Then all the theoretical descriptions which include saturation,
from the phenomenological  ``saturation model'' in Eq.~(\ref{Golec}) to the QCD--based
formalisms like the CGC or the Balitsky-Kovchegov equation, lead a dipolar cross-section
which approaches a constant value $\sigma_0 = 2\pi R^2$ as $s\to \infty$.
Although consistent with the Froissart bound \cite{Froissart,Martin}, 
this result cannot be right. From experiments, we know indeed that hadronic cross-sections
(e.g., for $pp$ collisions) keep growing with $s$ up to the highest energies that
have been reached so far. At very high energy, the growth is rather slow, and can be fitted
by either a small power of $s$ (the ``soft pomeron'' $\sim s^{0.08}$), or some power of
$\ln s$, or a combination of them. A recent analysis  \cite{BNetal} of the data for several
high-energy processes appears to favour a dominant behaviour of the 
log--squared type: $\sigma_{\rm tot}(s) \simeq \sigma_0\ln^2s$,
with an universal prefactor $\sigma_0$. If true, this would imply that the Froissart bound
is actually {\it saturated} in nature.

Such a steady growth of total cross-sections is also expected on physical grounds
\cite{Heisenberg,LR90,FB}. As a quantum mechanical bound state, the hadron does not have a sharp edge
but rather a tail. In a theory with a mass gap like QCD, this tail is an exponential 
whose width is fixed by the lowest mass in QCD, namely the pion mass.
Impact parameters which, at some initial energy, are far away in the tail of the distribution
will not contribute much to scattering, since the parton density is low there. But with 
increasing energy, the local gluon density 
will increase rather fast as a power of $s$, since gluon recombination is not 
effective when the density is low. Eventually, for sufficiently large $s$, the local
gluon density will become high enough for these impact parameters
to contribute significantly to scattering. That is,
with increasing energy, the effective interaction radius of the hadron is expected to
grow as well, which then results in an increase of the total cross-section.
In fact, since the local scattering amplitude ${\cal N}_\tau(r_{\perp},b_{\perp})$ 
cannot exceed the unitarity, or ``black'', limit ${\cal N}_\tau=1$, it is clear that,
for sufficiently large energies, the increase of the cross-section with $s$ will proceed
via the expansion of the ``black disk'' (= the central area of the hadron where
the unitarity limit has been reached already).

This general discussion shows that a theoretical description of the
dynamics of the black disk must combine two essential ingredients: i) a mechanism
which ensures the unitarization of the scattering amplitude at {\it fixed} impact parameter,
and ii) a description of the tail of the hadron  wavefunction. In QCD, 
the second issue is certainly related to confinement, and is thus genuinely non-perturbative.
But it has been unclear until recently whether the first issue, that of the unitarization,
can be addressed in perturbation theory or not. Indeed, since gauge interactions are a priori 
long-ranged, it could well be that soft, non-perturbative, interactions are responsible
for the approach towards ``blackness'' at a fixed impact parameter. 
The ``infrared diffusion'' of the BFKL equation may be seen as in argument in that sense.

\begin{figure} 
\begin{center} 
\includegraphics[width=0.85\textwidth]{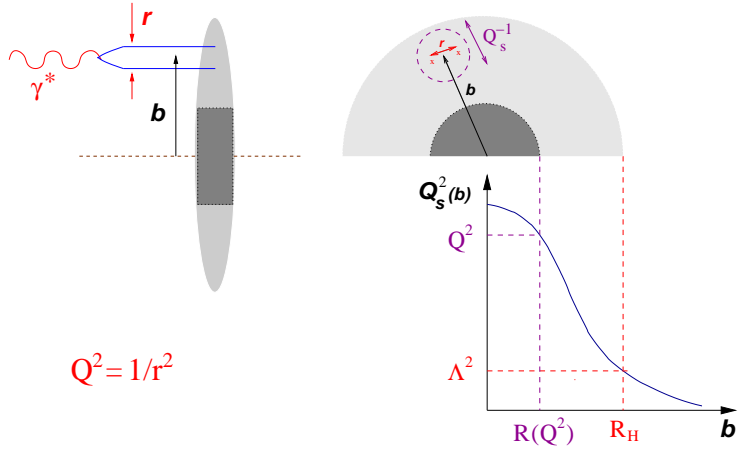}
\caption[]{Dipole-hadron scattering in longitudinal (left) and transverse
right) projections.}
\label{GREYfig}
\end{center}
\end{figure}
However, our present understanding of the physics of high parton densities
shows that the phenomenon of gluon saturation provides a perturbative
framework for the study of unitarization. We have seen indeed that the
perturbative, but {\it non-linear}, evolution equations yield a scattering amplitude 
which respects the unitarity bound (see Fig.~\ref{NFIG} and Eq.~(\ref{barS})).
To study the expansion of the black disk, these equations must be supplemented
with some information about confinement and applied at impact parameters in the tail of
hadron, namely, in the ``grey area'' at $b_\perp > R(\tau,Q^2)$. 

Here, $R(\tau,Q^2)$ is the radius of the black disk for an incoming dipole with
transverse resolution $Q^2\equiv 1/\rr^2$ and relative rapidity $\tau=\ln s/Q^2$.
That is, ${\cal N}_\tau(r_\perp,b_{\perp})$ is of order one for $b_\perp \le R(\tau,Q^2)$, but 
it drops rapidly for $b_\perp > R(\tau,Q^2)$.
Equivalently, $R(\tau,Q^2)$ is such that the {\it local} saturation scale
$Q_s(\tau,\bb)$ --- which is largest towards the center, and decreases 
with $\bb$, so like the gluon density --- becomes equal to $Q^2$ at 
$b_\perp \simeq R(\tau,Q^2)$ (see Fig.~\ref{GREYfig}). Perturbation theory
is valid as long as  $Q_s(\tau,\bb)\gg \Lambda_{\rm QCD}$, that is,
for $\bb < R_H(\tau)$ in the plot in  Fig.~\ref{GREYfig}. This condition leaves
us with a {\it perturbative grey corona} at $R(\tau,Q^2) < \bb < R_H(\tau)$,
which is sufficient, as we shall see, to perform a controllable calculation
of the {\it expansion rate} of the black disk.

Specifically, for impact parameters  within this corona, the dipole undergoes mostly 
{\it hard} scattering, with transferred momenta $\kk^2 \gg Q_s^2(\tau,\bb)$ (cf. the discussion
after Eq.~(\ref{barS})). This implies that it interacts predominantly 
with color sources which are relatively close in impact parameter space,
within a saturation disk of radius $1/Q_s(\tau,\bb)$ centered at $\bb$.
Physically, it is so because of the screening phenomena associated with saturation
(cf. Sect. 3.5) : Color sources which lie further away create only {\it dipolar} fields
at the impact parameter of the incoming dipole; such fields decrease rapidly with distance
(more rapidly then the monopole fields due to the nearby color sources), and thus contribute
less to scattering~\cite{FB,GAUSS}.

Mathematically, this means that scattering in the grey corona is controlled by the linear
BFKL equation (\ref{linearBK}), but with an infrared cutoff of order $Q_s(\tau,\bb)$. This cutoff simulates
the non-linear terms, which, in the full equation (\ref{BK}), would limit the range of the
dominant interactions to $1/Q_s(\tau,\bb)$. Clearly, with such a cutoff, the BFKL equation
is not afflicted by ``infrared diffusion'' any more. As in the previous 
calculation of the saturation scale in Sect. 3.4, the black disk radius can be computed by
first solving the BFKL equation, and 
then imposing a saturation condition similar to Eq.~(\ref{Nsat}) :
\be\label{INT_sat}
{\cal N}_\tau(\rr,\, \bb)\,\simeq\, 1\qquad {\rm for}\qquad b\,\simeq\,R(\tau,Q^2).
\ee

The $\bb$--dependence of the initial condition is determined by the non-perturbative
physics of the confinement, and thus requires a model. However, what we need to know 
about confinement is quite limited, and can be
inferred from general principles. Firstly, what is the typical scale for
transverse inhomogeneity in the hadron: this is clearly $1/\Lambda_{\rm QCD}$. 
Secondly, what is the $\bb$--dependence
of the scattering amplitude in the hadron tail. This is an exponential fall-off
$\propto {\rm e}^{-2m_{\pi}b}$ where twice the pion mass enters since the long
range interactions between the dipole and the (isosinglet) gluons require the exchange
of at least one pair of pions (recall that pions have isospin one).

The first requirement tells us that the inhomogeneity occurs over transverse scales
much larger than the typical range of the interactions, $\simle 1/Q_s(\tau,\bb)$. 
The scattering of the small dipole proceeds {\it quasilocally} in the impact
parameter space\footnote{This would not be true in the absence of the infrared cutoff
generated by the non-linear effects.}. As a consequence, the $\bb$--dependence of 
the solution to the (effective) BFKL equation {\it factorizes}, and is fixed by
the initial condition \cite{FB} :
\be\label{Ngrey}
{\cal N}_\tau(r_\perp,b_{\perp})\,\approx\,{\cal N}_\tau(r_\perp)\,S(\bb),\ee
where ${\cal N}_\tau(r_\perp)$ is the solution to the {\it homogeneous} BFKL equation,
as shown in Eq.~(\ref{BFKLsol1}). The second condition provides the transverse profile
function in the hadron tail: $S(\bb)\propto {\rm e}^{-2m_{\pi}b}$ (the proportionality
coefficient is not important to the present accuracy).

The previous arguments, together with Eq.~(\ref{BFKLsol1}) lead to the following estimate
for the scattering amplitude valid at impact parameters in the grey 
corona\footnote{At very high energies,
we can also neglect the diffusion term in 
Eq.~(\ref{BFKLsol1}).}  (with $Q^2\equiv 1/\rr^2$) :
\be\label{BFKLgrey}
{\cal N}_\tau(Q^2,\, \bb)\big|_{\rm grey}\,\simeq\,
\exp\left\{
-2m_\pi b + \omega\bar\alpha_s\tau - \frac12\ln \frac{Q^2}{\Lambda^2}
\right\}.
\ee
This equation, together with the  blackness condition (\ref{INT_sat}), imply:
\be\label{BD_asy}
R(\tau,Q^2)\, \simeq \, \frac{1}{2m_\pi}\left( \omega \bar\alpha_s \tau 
-\frac12 \ln \frac{Q^2}{\Lambda^2}\right).
\ee
We see that the black disk radius increases linearly with $\tau$. This behaviour
ensures the compensation between the exponential increase
with $\tau$ (due to the perturbative BFKL evolution) and the
exponential decrease with $\bb$ (due to confinement).

Since ${\cal N}_\tau$ is rapidly decreasing at  $\bb\gg R(\tau, Q^2)$,
 the total cross-section is dominated by the black disk:
\be\label{sigmaBD}
\sigma_{\rm dipole}(\tau,Q^2)\,\simeq \,
2\pi R^2(\tau, Q^2).\ee
Together with Eq.~(\ref{BD_asy}), this yields the following
dominant\footnote{The subdominant
behaviour is not completely described by Eq.~(\ref{BD_asy}), since it also receives 
contributions from the grey area~\cite{FB}.} 
behaviour of the cross-section at high energies\cite{FB} (see also Ref. \cite{LR90} for
an early study):
\be\label{sigmaFC}
\sigma_{\rm dipole}(s,Q^2)\,\approx\,
\frac{\pi}{2}\left(\frac{\omega \bar\alpha_s}{m_{\pi}}\right)^2\ln^2s\qquad 
{\rm as\,\,\,\,} s
\to\infty.\ee
This {\it saturates} the Froissart bound, with a {\it universal} coefficient for all hadrons and 
reflects the combined role of perturbative and
non-perturbative physics in controlling the asymptotic behaviour
at high energy. This behaviour is in qualitative agreement
with the phenomenological analysis in Ref.~\cite{BNetal}, but the coefficient
in Eq.~(\ref{sigmaFC}) is too large to fit the data. The difference
with respect to the data can be substantially reduced by using the RG-improved NLO 
estimate for the BFKL intercept, which decreases the leading-order
value $\omega \bar\alpha_s$ by roughly a factor of three \cite{NLBFKL,Salam99}.
In Refs. \cite{KL02}, the result in Eq.~(\ref{sigmaFC}) has been generalized
to $\gamma^*-\gamma^*$ scattering.

From Eq.~(\ref{BFKLgrey}), one can  also estimate the radius
$R_H(\tau)$ where the saturation scale decreases
to $\Lambda_{\rm QCD}$ (beyond which perturbation theory fails to apply).
This is the same 
as the black disk radius for a large dipole with $Q^2\sim\Lambda_{\rm QCD}^2$.
Therefore 
\be\label{RH1}
R_H(\tau)\,\approx\,\frac{\omega \bar\alpha_s }{2m_\pi}\, \tau \, .\ee
and the radial extent of the perturbative corona can be estimated as:
\be\label{RR}
R_H(\tau)\,-\,R(\tau,Q^2)\,\approx\,\frac{1}{4m_\pi}\, 
\ln \frac{Q^2}{\Lambda^2}\,,
\ee
This is independent of $\tau$, and much larger than $1/m_\pi$ (because
of the large logarithm $\ln(Q^2/\Lambda^2)$), which
 demonstrates the self-consistency of the previous calculation:
When $\tau \to \tau + d \tau$ with $\bar\alpha_s d \tau \sim 1$
(the typical rapidity increment at high energy),
the black disk expands from $R(\tau,Q^2)$ to $R(\tau,Q^2)+\omega/2m_\pi\ll R_H(\tau)$,
cf. eq.~(\ref{BD_asy}), and thus remains in the region controlled by perturbation theory.
That is, the expansion of the black disk proceeds within
the perturbative ``grey'' corona for rapidity intervals which are large enough
to allow for the calculation of the  rate of this expansion
\cite{FB}.

The perturbative evolution equation Eq.~(\ref{BK}) should not be used for very large impact 
parameters $\bb\gg R_H(\tau)$, where $Q_s^2(\tau,b)\simle \Lambda^2_{QCD}$.
Such equations lack confinement, so the long-range dipolar fields
generated by the gluons within the black disk can propagate to arbitrarily
large distances, thereby creating power-law contributions to the hadron tail.
At impact parameters within the grey corona, these long-range
contributions are relatively small, and the evolution 
is driven by the short-range interactions, with the conclusions outlined above. 
But for sufficiently large impact parameters, well beyond $R_H(\tau)$,
the power-law contributions will eventually supersede the 
exponentially decreasing one, and the hadron will develop an unphysical power-law tail.
(This is unphysical since, in the real world, it is removed by confinement.)
If one pushes the perturbative expansion until the black disk
enters this power-law tail, then its expansion rate speeds up,
and violates the Froissart bound \cite{KW02}. This is clearly an artifact of using 
perturbation theory outside its range of validity. The only way to circumvent this
difficulty without introducing ad-hoc modifications in the evolution equations
(to account for confinement) is to follow the perturbative evolution for only a 
limited interval of ``time'' $\Delta\tau$, such that perturbation theory
remains valid. As explained above, a limited evolution in time is indeed sufficient 
for the calculation of the expansion {\it rate} of the black disk.
The result of this calculation can be then extrapolated to arbitrarily high 
energies, even though the  perturbative evolution eventually becomes meaningless.

To conclude, while perturbation theory alone appears to be sufficient to
describe unitarization at fixed impact parameter, one still needs some
information about the finite range of the strong interactions
in order to compute total cross-sections. This is reminiscent of 
an old argument by Heisenberg \cite{Heisenberg} which combines unitarity and 
short-range interactions (as modelled by a Yukawa potential) to derive
cross-sections which saturate the Froissart bound. Fifty years later,
our progress in understanding strong interactions allows us
to confirm Heisenberg's intuition, and identify short-range interactions with
confinement, and unitarization with saturation.

  \subsection{Saturation and Shadowing in Deep Inelastic Scattering}

Shadowing is the phenomenon where $F_2^A(x,Q^2)/A F_2^N(x,Q^2) < 1$ at small 
x ($x< 0.1$). It is a large effect in the region where the coherence length of 
the probe (the $q{\bar q}$-pair in DIS) $l_{coh}\sim \frac{1}{2m_N x}$ exceeds the intra-nuclear 
longitudinal distance between any two nucleons in the nucleus. The nuclear parton distribution is 
not merely the sum of nucleon parton distributions but also contains the intereference between 
the parton distributions of the nucleons. When the coherence length is larger than the nuclear 
diameter ($l_{coh}\gg 2 A^{1/3}$, or $x<<1/(4 m_N A^{1/3})$), the $q{\bar q}$--pair interacts coherently 
with the entire nucleus, and the collective effects are expected to be imporatant.

There are several questions about shadowing that have not been unambiguously resolved in the framework of QCD. 

\noindent
$\bullet$ Is shadowing a ``leadig twist'' effect, or it is suppressed by powers of $Q^2$? An empirical 
answer to this question would help settle whether shadowing is an intrinsically leading 
twist phenomenon~\cite{FS0} or whether it is due to weak coupling, higher twist/high parton 
density effects~\cite{GLR,MQ,MV94}. 

\noindent
$\bullet$ What is the relation of shadowing to parton saturation? Does parton saturation provide a 
microscopic understanding of shadowing?

\noindent
$\bullet$ Does the shadowing ratio ``saturate'' at a minimum value for fixed $Q^2$ and $A$ with decreasing 
x? Does it saturate faster for quarks or gluons?

\noindent
$\bullet$ What is the relation of shadowing in nuclei to diffractive scattering of nucleons? 
The relation is well established at low parton densities~\cite{Gribov}
In an interesting recent exercise, it has been shown that diffractive {\it nucleon} data at
HERA could be used to predict the shadowing of quark distributions
observed by NMC~\cite{Kaidalov,FS1}.  
Significant deviations from the simple relation
between shadowing and diffraction, may again suggest the presence of
strong non--linearities. 

\noindent
$\bullet$ Is shadowing universal? For instance, are gluon structure functions extracted from p-A 
collisions at RHIC identical to those extracted from e-A in the same kinematic region? 
The naive assumption that this is true may be incorrect if higher twist effects are 
important.

The answers to these questions will only be conclusively settled by the next generation 
of p-A collider experiments at RHIC and LHC~\cite{CERNyellow} and by e-A collider experiments 
at RHIC (EIC/eRHIC) and/or DESY (HERA III)-see the following sub-section for a 
discussion of e-A collider plans.

From the theoretical perspective, there are several shadowing models which consider only leading twist 
shadowing~\cite{FS1,EKS98,Kumano}. Namely, shadowing effects are put in the non-perturbative initial 
conditions, which are then evolved in $Q^2$ and x using the leading twist DGLAP equations~\cite{DGLAP}. 
Modifications of these leading twist models to include Mueller-Qiu type non-linear contributions 
have also been considered~\cite{EKS02}. 

In the McLerran-Venugopalan saturation model, the gluon distribution (defined as the integral over 
the unintegrated gluon distribution) is additive in A. However, the structure functions computed 
as discussed in section 4.1 will exhibit shadowing for $Q^2<Q_s^2$. There have been a few 
attempts to compute shadowing from the non-linear renormalization group equation. 
One of these calculations, which is based on the non-linear equation derived in Ref.~\cite{JKW99},
predicts that perturbative gluon shadowing
will become large as one goes to smaller $\xx_{Bj}$'s~\cite{JamalWang}. Other more recent computations 
of shadowing have been performed within the context of numerical solutions of the Balitsky-Kovchegov 
equation for both inclusive~\cite{LL,WeigertFreund} and diffractive scattering~\cite{LevinLublinsky}. 
In both cases, 
geometric scaling of the nuclear distributions is claimed but it is not clear that both groups obtain 
the same $A$-dependence for the saturation scale. A very interesting recent theoretical suggestion is that 
the scattering amplitude for high energy scattering for nuclei, at fixed impact parameter, is the same as 
for protons at asymptotic energies!~\cite{MuellerTrianta}. Integrated over impact parameter, this would
suggest that shadowing saturates at very small x. The understanding of shadowing in the saturation picture 
is still preliminary-more detailed global fits of the non-linear equations (as for instance performed for 
the linear DGLAP fits) are needed. In addition, different computations of the $A$-dependence of the saturation 
scale will likely converge as our theoretical understanding improves. Finally, as will be discussed in the 
following, the issue will likely not be resolved conclusively until DIS experiments off nuclei at small x
are performed.

  \subsection{Probing the CGC with an Electron Ion Collider}

A high energy electron--nucleus collider, with a center of mass energy
$\sqrt{s}=60$--$100$ GeV, presents a remarkable opportunity to explore
fundamental and universal aspects of QCD. The nucleus, at these
energies, acts as an amplifier of the novel physics of high parton
densities--aspects of the theory that would otherwise only be explored
in an electron--proton collider with energies at least an order of
magnitude greater than that of HERA. An electon--nucleus collider will
also make the study of QCD in a nuclear environment, to an extent far
beyond that achieved previously, a quantitative science. 
In particular, it will help complement, clarify, and reinforce physics
learnt at high energy nucleus--nucleus and proton--nucleus collisions
at RHIC and LHC over the next decade. For both of these reasons, an eA
collider facility represents an important future direction in high
energy nuclear physics.

We will briefly discuss here experimental observables in
deep inelastic scattering (DIS) which are signatures of the novel
physics of the Color Glass Condensate~\footnote{We will not cover here the interesting physics 
at intermediate and large $x_{Bj}$ that can be studied with an eA collider. 
A nice discussion of these issues can be found in Ref.~\cite{Arneodoetal,EICproceed}.}. 
A more detailed discussion of the following can be found in
Ref.~\cite{RajDIS}.

\begin{figure}[htb]
\begin{center}
\includegraphics[width=0.6\textwidth]{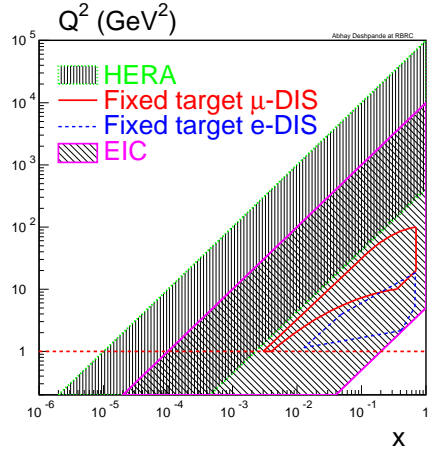}
\caption{The x--$Q^2$ range of the electron ion collider (EIC) compared 
to that of the HERA ep collider and fixed target experiments. The EIC's 
reach would encompass the fixed target regime as well as part of the 
HERA regime.}
\label{xq2}
\end{center}
\end{figure}

The regime of small $x_{Bj}$'s ($x_{Bj}\leq 0.01$) is easily accessed
by an electron--heavy ion collider in the energy range
$\sqrt{s}\approx 60$--$100$ GeV. These energies would be most natural if the Electron Ion Collider (EIC) were 
constructed at BNL-this particular realization is called eRHIC.  The kinematic coverage of EIC/eRHIC is 
shown in Fig.~\ref{xq2}. What is novel about these energies
is that for the first time one can study the physics of $x_{Bj}<<
0.01$ in a nucleus for $Q^2 >> \Lambda_{QCD}^2$, where
$\Lambda_{QCD}\sim 200$ MeV.  Previous (fixed target) experiments such
as NMC and E665 and current ones such as HERMES and COMPASS 
could only access small $x_{Bj}$ at small $Q^2$'s. The center of mass (c.m) energy of the Electron-Ion Collider (EIC) is 
a factor of 10 smaller than that of the current ep-collider at 
HERA (the proposed HERA III plan-which includes e-A scattering would have a center-of-mass energy that's 
roughly 3 times greater than EIC). However, an eA collider has a tremendous advantage--the parton
density in a nucleus, as experienced by a probe at a fixed energy, is
much higher than what it would experience in a proton at the same
energy. Since the parton density grows as $A^{1/3}$, this effect is more
pronounced for the largest nuclei-see Eq.~\ref{universal}.  

Eq.~(\ref{universal}) suggests that $x_{\rm
{proton}} = {x_{\rm {nucleus}}/ \left( A^{1\over
3}\right)^{1/\lambda}}$.  Since the nucleus is dilute and conservatively taking
the effective $A^{1/3}=4$, then, for $\lambda\sim 0.3$, one finds 
$x_{\rm{proton}} \sim x_{\rm{nucleus}}/100$. Thus the same parton density in a nucleus at 
$x_{Bj}\sim 10^{-4}$ and $Q^2\sim$ a few GeV$^2$ is attained in a {\it nucleon} at $x_{Bj}\sim
10^{-6}$ and similar $Q^2$! Impact parameter tagging is feasible by counting knock-out
neutrons~\cite{FS}--if so, the gain in parton density in eA relative to ep may be even more spectacular. 

In the following, we will discuss both inclusive and semi--inclusive 
signatures of the CGC. The latter in particular are very difficult to measure in fixed target DIS

\vskip 0.05in
\noindent\underline{Inclusive signatures of the CGC}
\vskip 0.05in

An obvious inclusive observable is the structure function
$F_2(x_{Bj},Q^2)$ and its logarithmic derivatives with respect to
$x_{Bj}$ and $Q^2$.  The EIC should have sufficient statistical
precision for one to extract the logarithmic derivative of
$F_2$ (and its logarithmic derivative!). Whether the systematic errors at small $x_{Bj}$ will affect the
results is not clear at the moment. The logarithmic derivative $dF_2/d\ln(Q^2)$, at fixed $x_{Bj}$, and
large $Q^2$, as a function of $Q^2$, is the gluon distribution. QCD
fits implementing the DGLAP evolution equations should describe its
behavior at large $Q^2$.  At smaller $Q^2$, one should see a
significant deviation from linear QCD fits--in principle, if the $Q^2$
range is wide enough, one should see a turnover in the
distribution. The $Q^2$ at which the turnover takes place should be
systematically larger for smaller x's and for larger
nuclei~\cite{Levinetal}.
 
At eRHIC one can extract the longitudinal 
structure function $F_L(x_{Bj},Q^2)= F_2 - 2\, x_{Bj} F_1$ at small $x_{Bj}$  
independently since the energy of the colliding beams can be varied significantly. 
In the parton model, $F_L=0$--thus $F_L$ is 
very sensitive to scaling violations. It provides an 
independent measure of the gluon distribution~\cite{Wilczeketal} and in particular of 
higher twist saturation effects which may be
prominent in both $F_L$ and $F_T$ but may cancel in the sum~\cite{Bartels}.  

The extended kinematic range of EIC may help determine whether shadowing is entirely a leading twist
phenomenon, or if there are large higher twist perturbative corrections.  
As also discussed previously, there is a close relation between
shadowing and diffraction. At EIC the validity of this relation can be
explored directly--different nuclear targets are available, and the diffractive structure function may also be measured
independently.

\vskip 0.05in
\noindent\underline{Semi--inclusive signatures of the CGC}
\vskip 0.05in

A striking semi--inclusive measurement is hard
diffraction wherein the virtual photon emitted
by the electron fragments into a final state $X$, with an invariant
mass $M_X^2\gg \Lambda_{QCD}^2$, while the proton emerges unscathed in
the interaction. A large rapidity gap--a region in rapidity essentially
devoid of particles--is produced between the fragmentation region of
the electron and that of the proton.  In pQCD, the probability of a
gap is exponentially suppressed as a function of the gap size. At HERA
though, gaps of several units in rapidity are unsuppressed; one finds
that roughly 10\% of the cross--section corresponds to hard
diffractive events with invariant masses $M_X > 3$ GeV. 

Hard diffraction probes the color singlet object (the ``Pomeron'') within the proton that
interacts with the virtual photon. It addresses, in a novel
fashion, the nature of confining
interactions within hadrons. The mass of
the final state is large and one can reasonably ask questions about the
quark and gluon content of the Pomeron. A diffractive structure function 
$F_{2,A}^{D(4)}$ can be defined~\cite{BereraSoper,Collins} (analogous to $F_2$) as
\be
{d^4 \sigma_{eA\rightarrow eXA}\over {dx_{Bj} dQ^2 d\xp dt}} & &=
A\cdot {4\pi \alpha_{em}^2\over x Q^4}\,
\left\{ 1-y + 
{y^2\over 2[1+ R_A^{D(4)}(\beta,Q^2,\xp,t)]}\right\}\nonumber\\
& & F_{2,A}^{D(4)}
(\beta,Q^2,\xp,t) \, ,
\ee
where $y=Q^2/s x_{Bj}$ and   
$R_A^{D(4)}=F_L^{D(4)}/F_T^{D(4)}$. Also, $Q^2=-q^2$ and $x_{Bj}=Q^2/ (2P\cdot q)$ 
have the usual DIS definitions and 
\be
\xp = {q\cdot (P-P^\prime)\over q\cdot P}\,\,;\,\, t = (P-P^\prime)^2 \,\,;\,\,\beta = {x_{Bj}\over \xp}\, .
\ee
Here $P$ is the initial nuclear momentum, $P^\prime$ the net momentum of the fragments $Y$ in the proton 
fragmentation region and $M_X$ the net momentum of the 
fragments $X$ in the electron fragmentation region. An illustration of 
the hard diffractive event is shown in Fig.~22.
Unlike $F_2$ however, $F_2^{D(4)}$ is not truly universal--it cannot be applied, for instance, 
to predict diffractive cross--sections in $p$--$A$ scattering; it can 
be applied only in other lepton--nucleus scattering studies~\cite{Collins}.

\begin{figure}
\begin{center}
\includegraphics[width=0.6\textwidth]
{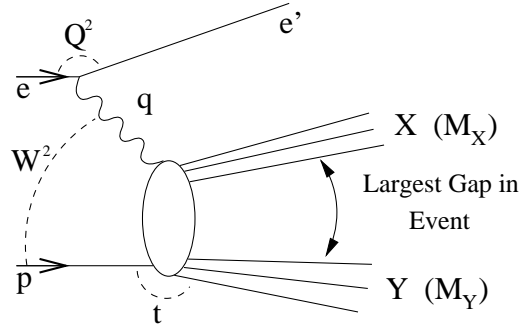}
\caption[]{The diagram of a process with a rapidity gap between the systems 
X and Y. The projectile nucleus is denoted here as p. Figure from Ref.~\cite{Arneodoetal}.}
\end{center}
\label{madif}
\end{figure}

In practice the structure function $F_{2,A}^{D(3)}=\int F_{2,A}^{D(4)} dt$ is measured, where 
$|t_{min}|<|t|<|t_{max}|$, where $|t_{min}|$ and $|t_{max}|$ and the limits of the empirically measurable 
momentum transfer to the nucleus. The ratio for two nuclei $A1$ and $A2$
\be
R_{A1,A2}(\beta,Q^2,\xp) = {F_{2,A1}^{D(3)}(\beta,Q^2,\xp)\over 
F_{2,A2}^{D(3)}(\beta,Q^2,\xp)} \, ,
\ee
can be measured with high accuracy~\cite{Arneodoetal}. If $R_{A1,A2} =1$,  
the structure of the Pomeron is universal, and one has an 
$A$--independent Pomeron flux. If $R_{A1,A2} = f(A1,A2)$, then albeit 
a universal Pomeron structure, the flux is A--dependent. Finally, if 
Pomeron structure is $A$--dependent, some models argue that 
$R_{A1,A2} = F_{2,A1}/F_{2,A2}$.

The ratio of 
$R_D=\sigma_{\rm{diffractive}}/\sigma_{\rm{total}}$ at HERA is 
$\sim 10$\% for $M_X > 3$ GeV. The systematics of 
hard diffraction at HERA can be understood in saturation models~\cite{GBW99}. 
For eA collisions at EIC energies, saturation
models predict that the ratio $R_D^A$ can be much higher--on the order 
of $30$\% for the largest nuclei~\cite{LevMaor,FS1}. 

An important semi--inclusive observable in eA DIS at high 
energies is coherent (or diffractive) and inclusive vector meson production.  
For instance, the forward vector meson diffractive leptoproduction cross--section off 
nuclei is~\cite{Brodsky}
\be
{\frac{d\sigma}{dt}}|_{t=0}(\gamma^{*}A\rightarrow VA) \propto \alpha_S^2(Q^2)\,\left[G_A(x,Q^2)\right]^2 
\, ,
\ee
for large $Q^2$ and is therefore a sensitive probe of gluon saturation/shadowing. 
It is very important to measure inclusive and diffractive open charm  and jets since they provide 
useful and complementary data to those of vector mesons.

Due to large color fluctuations at small $x_{Bj}$, on can expect: a) a broader rapidity 
distribution in larger nuclei relative to lighter nuclei and protons, 
b) Enhanced anomalous multiplicity distributions where anomalous  
multiplicity in one rapidity interval in an event would be accompanied by 
an anomalous multiplicity in rapidity intervals several units 
away~\cite{KLM}, and c) a correlation between the central multiplicity with the multiplicity 
of neutrons in a forward neutron detector~\cite{FS}.

In pA scattering at RHIC one also has the
opportunity to study the high parton densitiy phenomena-see section 5.5 for a discussion. 
Some of the differences between pA and eA are as follows. In the pA Drell--Yan process, it is very hard to 
reliably extract distributions in the region below the $\Psi^\prime$ tail--
namely, one requires $Q^2 > 16$ GeV$^2$. In the $x$ region of interest, 
one expects saturation effects to be important at lower $Q^2$ of 
$1$--$10$ GeV$^2$. For $Q^2=16$ GeV$^2$, one might have to go to significantly 
smaller x's to see large saturation effects. Secondly, the survival 
probability of large rapidity gaps is smaller in pA relative to eA. This is 
because in pA (unlike eA) the gap is destroyed due to secondary interactions between 
``spectator'' partons in the proton and the ``Pomeron'' from the nucleus. 
Thus one expects that diffractive vector meson and 
jet production in pA should be qualitatively different than what one 
will see in eA.

\section{Melting the CGC in Nucleus-Nucleus and Proton-Nucleus Collisions}  
\setcounter{equation}{0}

In general, the problem of high energy hadron--hadron collisions is very difficult. 
As we discussed in the introductory 
section, a full understanding of  multi-particle production in high energy collisions is one of the outstanding problems in 
high energy QCD. The approach developed in previous sections however suggests that classical methods may be useful in 
studying multi-particle production in high energy collisions. In this section, we will discuss applications of
this approach in describing the initial stages of very high energy heavy ion collisions, proton (or deuteron)--nucleus 
collisions and briefly, peripheral nucleus-nucleus collisions. 

Understanding the initial conditions for heavy ion collisions is of the utmost importance in studying the high energy 
heavy ion experiments performed at the Relativistic Heavy Ion Collider (RHIC) at BNL  
and (to be performed) at the Large Hadron Collider (LHC) at CERN. The initial conditions are crucial in determining 
whether the quark-gluon matter produced immediately after the collision will equilibrate to briefly form a 
thermalized Quark Gluon Plasma (QGP). Classical methods are applicable because the occupation numbers of gluons 
in the nuclear wavefunctions (and immediately after the collision) are of order $1/\alpha_S$. One can therefore study the 
space-time evolution of partons which are ``frozen'' in the Color Glass Condensate through their release in the 
collision and their subsequent evolution. Unfortunately, it is not possible in this approach to follow their evolution 
all the way to equilibration. This is because the gluon occupation numbers become small well before equilibration thereby 
signalling the breakdown of the classical field approach.

Since the classical field equations are fully non-linear, they cannot be solved analytically~\footnote{For recent 
analytical work in this direction, see Ref.~\cite{Yuri2001}}. The solutions can however 
be determined numerically and the initial energy and number distributions can be obtained. A particular feature of our 
formulation of the scattering problem is that the classical Yang-Mills field equations are boost invariant-the 
equations are independent of the space-time rapidity. 
The boost invariance leads to a significant simplification-the classical Yang-Mills equations are only functions 
now of the two transverse spatial directions and the proper-time. 

In order to maintain gauge invariance, the numerical problem is formulated on the lattice and solved as a function 
of proper time. In the following, we will discuss briefly the numerical procedure and the results of the simulations. 
The non-perturbative results can be compared to the results of analytical computations at large 
transverse momenta and we will show that they agree with these computations. 

The results of these numerical computations can now be interpreted in light of the recent data from RHIC. We will 
discuss some of the phenomenological consequences of our results-in particular the implications for the importance 
of final state interactions of partons (beyond the initial stage of high occupation numbers) and their 
subsequent thermalization. Quantum effects, such as the non-trivial geometric scaling discussed earlier, may also 
play an important role in interpreting the very interesting and in some instances very puzzling RHIC data. 

As we will discuss, a contentious issue in deciphering the RHIC data is whether the phenomena observed are primarily 
due to initial state or final state effects. Experiments underway at RHIC on Deuteron-Gold collisions should help settle 
the issue since one expects final state effects in these collisions to be relatively unimportant~\footnote{Scattering 
effects that are often thought of as ``final state'' effects in proton-nucleus effects can, in a different gauge, 
be interpreted as initial state effects~\cite{KM98}. The final state effects emphasized here are those involving 
re-scattering of on-shell partons off each other.}
Recently, there 
have been several papers studying various final states at RHIC energies in the CGC model. We will discuss these 
works and their predictions for the RHIC data. Finally, we will also discuss briefly the predictions of the 
CGC model for particle production in peripheral nucleus-nucleus collisions.

\subsection{Classical Picture of Nuclear Collisions}

The classical picture of nuclear collisions was first formulated by Kovner, McLerran and Weigert~\cite{KMW} 
in the framework of the McLerran-Venugopalan model for a single nucleus. 
At very high energies, $P^+\rightarrow \infty$ of one of the nuclei (and $P^-\rightarrow \infty$ for the other), 
the hard valence quark (and gluon) modes are highly
Lorentz contracted, static sources of color charge for the wee parton,
Weizs\"acker--Williams, modes in the nuclei. For simplicity, we will first consider only central collisions 
of cylindrical nuclei with uniform matter distributions. The case of realistic nuclei and non-central collisions 
will be discussed later. The valence sources are then described
by the current
\begin{equation}
J^{\nu,a}(r_\perp) = \delta^{\nu +}\rho_{1}^a (r_\perp)\delta(x^-) + \delta^{\nu -}
\rho_{2}^a (r_\perp) \delta(x^+) \, ,
\label{sources}
\end{equation}
where $\rho_{1(2)}$ correspond to the color charge densities of the
hard modes in nucleus 1 (nucleus 2) respectively. The $\delta$-function in $x^-$ for the valence parton 
current of one nucleus (or $\delta$-function in $x^+$ for the other) implies that we are literally assuming 
the nuclei to move at the speed of light. In reality of course this condition should and can be relaxed. We 
will discuss this issue later on in this section. In the collision, these valence partons are assumed to 
be Eikonal sources-they continue on their straight line trajectories along the light cones. They 
may acquire a phase due to the rotation of the color charge in the scattering~\cite{KovRischke}. However, in the 
gauge we will use, no such phase will appear in our treatment of the classical scattering problem.

For each color configuration in each of the nuclei, the classical field describing the small x modes in the 
Effective Field Theory (EFT) is obtained by solving the Yang--Mills equations in the presence of the two sources. 
We then have 
\begin{equation}
D_\mu F^{\mu\nu} = J^\nu \, .
\label{yangmill}
\end{equation}
To compute a physical quantity $\langle O\rangle$, the gauge field configurations have to be 
averaged over the respective Gaussian path integrals of the two nuclei, 
\begin{equation}
\langle O\rangle_\rho = \int d\rho_{1}d\rho_{2}\, O(\rho_1,\rho_2) 
 \exp\left( -\int d^2 r_\perp {\left[\rho_1^a(r_\perp) \rho_1^a(r_\perp)+\rho_2^a(r_\perp) \rho_2^a(r_\perp)
\right]\over {2g^4\mu_A^2}}\right) \, .
\end{equation}
For instance, the small x gluon distribution is simply related to the Fourier
transform $A_i^a (k_\perp)$ of the solution to Eq. (~\ref{yangmill}) by $<A_i^a(k_\perp) A_i^a(k_\perp)>_\rho$, where the 
subscript denotes the average above.
Here have assumed identical nuclei with equal Gaussian weights $g^4\mu_A^2$, where $\mu_A^2$ is the average 
color charge squared of a nucleus, defined in Eq.~\ref{MV-corr}. We will henceforth use the variable 
$\Lambda_s^2=g^4\mu_A^2$. It is simply related to the saturation scale $Q_s$ by the relation 
\begin{equation}
   Q^2_s =  {N_c \Lambda_s^2\over 4\pi}
            \log\left({\Lambda_s^2\over \Lambda^2_{QCD}}\right)\, .
\label{eq:qsls}
\end{equation}
In practice, $\Lambda_s\sim Q_s$.

In general, one can make the following ansatz for the gauge 
fields as a function of proper time $\tau=\sqrt{2x^+x^-}$ in the different light cone regions: 
\begin{eqnarray}
\label{eq:ansatz}
 A^i &=& \alpha^i_3(\tau,x_{T})\theta(x^-)\theta(x^+) 
      + \alpha^i_1(\tau,x_{T})\theta(x^-)\theta(-x^+) \nonumber\\
       &+& \alpha^i_2(\tau,x_{T})\theta(-x^-)\theta(x^+),\\
 A^{\pm} &=& \pm x^{\pm} \alpha(\tau,x_\perp)\theta(x^-)\theta(x^+).
\end{eqnarray}
where $\alpha_3^i$ and $\alpha$ are respectively the transverse and longitudinal components of the gauge 
field the forward light cone while $\alpha_{1,2}^i(r_\perp)$ ($i=1,2$ are the two transverse Lorentz indices)
are {\it pure gauge} fields defined through the gauge transformation
parameters $\Lambda_{q}(\eta,r_\perp)$~\cite{MikLar}
\begin{equation}
\alpha_{1,2}^i(r_\perp) = {1\over i}\left(Pe^{-i\int_{\pm \eta_{\rm proj}}^0
d\eta^\prime
\Lambda_{1,2}(\eta^\prime,r_\perp)}\right)
\nabla^i\left(Pe^{i\int_{\pm \eta_{\rm proj}}^0 d\eta^\prime 
\Lambda_{1,2}(\eta^\prime,r_\perp)}\right) \, .
\label{puresoln}
\end{equation}
Here $\eta=\pm\eta_{\rm proj}\mp\log(x^{\mp}/x_{\rm proj}^{\mp})$ is
the rapidity of the nucleus moving along the positive (negative) light
cone with the gluon field $\alpha_{1(2)}^i$. The
$\Lambda_{1,2}(\eta,r_\perp)$ are determined by the color charge
distributions $\Delta_\perp\Lambda_q=\rho_q$ (q=1,2) with
$\Delta_\perp$ being the Laplacian in the perpendicular plane.

We work in the Fock-Schwinger (or ``radiation'') gauge $$A^{\tau}\equiv x^+ A^- + x^- A^+ =0\,,$$ 
which is the interpolation between two light cone gauges. Fixing this gauge however 
does not fix the gauge completely. 
The residual gauge freedom can be fixed by imposing the Coulomb gauge condition 
$\nabla_\perp\cdot A_\perp =0$ in the two transverse dimensions. 

Substituting Eq.~(\ref{eq:ansatz}) in Eq.~(\ref{yangmill}), and re-writing the equations 
in terms of the transverse coordinates $x_\perp$, the proper time $\tau$ and the 
space-time rapidity $\eta$, one observes that the Yang-Mills 
are independent of $\eta$, namely, they are boost invariant. The fields $\alpha^i_3$ and $\alpha$ are 
functions of $x_\perp$ and $\tau$ only. The boost invariance of the solutions of the Yang-Mills equations 
is entirely due to our approximation-using $\delta$-function sources. Smearing the sources in rapidity would 
destroy the boost-invariance of the solutions. Nevertheless, one expects that the boost-invariance 
approximation is a reliable one especially at central rapidities in a nuclear collision. 
This point will be discussed further.

The initial conditions for the solution of the Yang--Mills equations in the forward light cone 
are determined by matching the solutions
in the space-like and time-like regions at $\tau=0$. Requiring that the gauge fields be regular at $\tau=0$,
the coefficients of the singular pieces of the equations $D_{\mu i} F^{\mu i} = 0 $ and $D_{\mu \pm} F^{\mu \pm} = J^\pm$ 
(for $x^-,x^+ \to 0$) have to be set to zero. These give the boundary conditions at $\tau=0$:
\begin{eqnarray}
 \alpha^i_3(0,x_\perp) &=& \alpha^i_1(0,x_\perp) + \alpha^i_2(0,x_\perp),\\
 \alpha(0,x_\perp) &=& {i\over 2} [\alpha^i_1(0,x_\perp),\alpha^i_2(0,x_\perp)].
\label{eq:matching}
\end{eqnarray}
These conditions, first formulated for infinitely large nuclei, 
are the same for finite nuclei as well. The boundary conditions remain the same even when the fields
$\alpha_{1,2}^i$ before the collision are smeared out in rapidity 
properly account for singular contact terms in the equations of motion~\cite{MikLar}.
Further, since the equations are very singular at $\tau=0$, the
only condition on the derivatives of the fields that would lead to regular
solutions are $\partial_\tau \alpha|_{\tau=0},\partial_\tau \alpha_\perp^i
|_{\tau=0} =0$.

Perturbative solutions of the Yang--Mills equations to order $\rho^2$
in the color charge density (or equivalently to second order in
$\Lambda_s/k_\perp$) were found, and at late times, after averaging over
the Gaussian sources, the number distribution of classical gluons was
found to be~\cite{KMW,MikLar,KovRischke}
\begin{equation} 
{dN\over {dyd^2 k_\perp}} = \pi R^2 {2g^6 \mu^4\over {(2\pi)^4}} {N_c (N_c^2-1)
\over k_\perp^4} L(k_\perp,\lambda) \, ,
\label{GunBer}
\end{equation}
where $L(k_\perp,\lambda)$ is an infrared divergent function at the scale
$\lambda$. This result agrees with the quantum bremsstrahlung formula
of Gunion and Bertsch~\cite{GunionBertsch}. 

The distributions are very
sensitive to $L(k_\perp,\lambda)$. What is novel about the classical approach is that, at sufficiently high
energies, the non--linearities in the Yang--Mills fields
self--consistently regulate this infrared divergence. To confirm this
claim, one needs to solve the Yang--Mills equations to all orders in
$\Lambda_s/ k_\perp$. This is very difficult to do analytically. 
Fortunately, as we will discuss in section 5.2, the classical problem can be solved numerically. 

The discussion here has been extended to treat the collision of finite, ultrarelativistic nuclei with 
realistic nuclear matter distributions~\cite{AYR02b}. For the case of finite nuclei, the issue of 
color neutrality, as discussed in sections 2.4 and 3.10, becomes very important. A global neutrality 
constraint at the level of the nucleus is insufficient to ensure that there are no large field strengths outside 
the nuclear radius-color neutrality must be imposed at the nucleon level. The practical implementation of color 
neutrality  will be discussed further in the following.

As mentioned earlier, a major simplification occurs in the classical approach when boost invariance is 
assumed. It is likely a good approximation, especially in the central region, but the effects of relaxing 
this condition are unclear. A first step in answering this question (before performing a fully 3+1-dimensional 
simulation) is to study the stability of the 2+1-dimensional results with respect to a perturbation in the 
$\eta$-direction~\cite{AYR03a}.

\subsection{Numerical Gluodynamics of Nuclear Collisions}

Classical real time numerical solutions of gauge theories were first discussed in the context of 
sphaleron transitions during the electroweak phase transition in the early universe~\cite{AmbjornKrasnitz}. 
Similar techniques can be applied to discuss the problem at hand. It is most convenient to follow the 
Hamiltonian approach, namely, to construct the appropriate lattice Hamiltonian and solve Hamilton's equations 
on the lattice with the lattice analog of the initial conditions discussed in section 5.1. These will be discussed
below. We will subsequently discuss the results of our classical numerical simulations.

\bigskip
\noindent {\bf i) Numerical Solution of the Yang-Mills Equations}
\smallskip

The QCD action for gauge field in the $\tau, \eta, \vec{r}$
coordinates reads
\begin{equation}
 S_{QCD} = \int \tau d\eta d\tau d^2 r \left[
            -{1\over 8} \Tr(g^{\mu\alpha}g^{\nu\beta}F_{\alpha\beta}F_{\mu\nu})
        \right],
\end{equation}
where
$F_{\mu\nu}=\partial_{\mu}A_{\nu}-\partial_{\nu}A_{\mu}-ig[A_{\mu},A_{\nu}]$
and the metric is diagonal with
 $g^{\tau\tau}=-g^{xx}=-g^{yy}=1$ and $g^{\eta\eta}=-1/\tau^2$.
$A_{\mu} \equiv A_{\mu}^a t^a$, and $t^a$
represent a gauge group matrices with the normalization
of $\Tr(t^at^b) = 2\delta_{ij}$.
The Lagrangian density in $A^{\tau}=0$ gauge is 
\begin{equation}
 {\cal L}  =  \Tr\left( 
            {\tau\over4}(\partial_{\tau}A_i)^2
           +{1\over4\tau}(\partial_{\tau}A_{\eta})^2
           -{\tau\over8} F_{ij}^2 
           -{1\over 4\tau}F_{\eta i}^2 \right) \,,
\end{equation}
where $i,j$ runs over transverse coordinate $x$ and $y$.

Now let us assume $\eta$ independence of the fields. As discussed previously, the 
Yang-Mills equations have this property if the sources are strictly $\delta$-function 
sources on the light cone. We have 
\begin{equation}
\label{adjscalar}
  A_i(\tau,\eta,{\vec r}) = A_i(\tau,{\vec r}), \qquad
  A_{\eta}(\tau,\eta,{\vec r}) = \Phi(\tau,{\vec r}) \, ,
\end{equation}
resulting in $F_{\eta i} = -D_i\Phi$,
where $D_i = \partial_i - ig[A_i,\cdots]$ is the covariant derivative
in the adjoint representation.
Defining the conjugate momenta $E_i =  \tau\partial_{\tau}A_i$ and 
$p_{\eta}  =  {1\over\tau}\partial_{\tau}A_{\eta}$, one finds that the 
boost invariant Yang-Mills Hamiltonian is the QCD Hamiltonian 
in 2+1 dimensions coupled to an adjoint scalar~\cite{AR99}:
\begin{equation}
 H = \int d^2r \Tr \left\{ 
            {1\over4\tau} E_i^2
           +{\tau\over4} p_{\eta}^2
           +{\tau\over8}F_{ij}^2
           +{1\over 4\tau}(D_i\Phi)^2
        \right\} \,.
\end{equation}

In order to realize numerically the solutions to the equations of motion 
in the previous section, while maintaining the gauge symmetry, we introduce the link variables
at the site $i$
\begin{equation}
    U_{j,i} = \exp\left[ i a A_j(i) \right],\quad (j=x,y),
\end{equation}
where, $a$ is a lattice spacing.
Defining the plaquette $U_{\Box} \equiv U_{l,j}U_{m,j+l}U^{\dagger}_{l,j+m}U^{\dagger}_{m,j}$, 
the Hamiltonian on the lattice is 
\begin{eqnarray}
  & &H_L = 
    {1\over 4\tau}\sum_{\ell\equiv (j,\hat{n})}{\rm Tr}E_{\ell}^2
+ {\tau\over 2}\sum_{\Box}(N_c-{\rm ReTr}U_{\Box}) \nonumber\\
&+& {1\over 4\tau}\sum_{j,\hat{n}}{\rm Tr}
           (\Phi_j - U_{\hat{n},j}
           \Phi_{j+\hat{n}} U^{\dagger}_{\hat{n},j})^2 
+ {\tau \over 4}\sum_{j}{\rm Tr}p_j^2 \, ,
\label{eq:HDL}
\end{eqnarray}
where the convention for the generators of the SU(3) color group is ${\rm Tr}(\tau^a \tau^b) = 2\delta^{ab}$. 
For $g=2$, one obtains the correct normalization of the Hamiltonian in the 
continuum limit. 
Lattice equations of motion follow directly from $H_L$ of Eq.~\ref{eq:HDL}.  For
any dynamical variable $v$, with no explicit time dependence, ${\dot
v}=\{H_L,v\}$, where ${\dot v}$ is the derivative with respect to $\tau$, and
$\{\}$ denote Poisson brackets. We take $E_l$, $U_l$, $p_j$, and $\Phi_j$ as
independent dynamical variables, whose only nonvanishing Poisson brackets are
$$\{p_i^a,\Phi_j^b\}=\delta_{ij}\delta_{ab}; \ \ 
\{E_l^a,U_m\}=-i\delta_{lm}U_l\sigma^a; \ \
\{E_l^a,E_m^b\}=2\delta_{lm}\epsilon_{abc}E_l^c$$
(no summing of repeated indices). 
The initial conditions for the transverse gauge field and the adjoint scalar field 
on the lattice can be obtained in complete analogy to the procedure followed in 
the continuum. The details of this procedure and the expression for the 
initial conditions can be found it Ref.~\cite{AR99}.

We impose periodic boundary conditions on an $N\times N$ transverse lattice, where 
$N$ denotes the number of sites. 
The physical linear size of the system is $L=a\,N$, where $a$ is the
lattice spacing. It
was shown in Ref.~\cite{AR99} that numerical computations on a
transverse lattice agreed with lattice perturbation theory at large
transverse momentum.

\bigskip
\noindent {\bf ii) Numerical Method for Finite Nuclei}
\smallskip

In early studies~\cite{AR99,AR00,AR01,AYR01}, 
nuclear collisions, for simplicity, were idealized as central collisions of 
infinite, cylindrical nuclei. 
The color charge squared $\Lambda_s^2$ 
was taken to be a constant for the uniform cylindrical nuclei. Furthermore, 
color neutrality was imposed only in 
a global sense~\cite{RajGavai}, namely, the color charge distribution over the entire nucleus was constrained to be 
zero. While very useful in obtaining first estimates of the space-time 
evolution of the produced gluonic matter, 
these studies did not make predictions for realistic nuclear collisions. 
In addition, studies of the distributions 
in peripheral collisions, in particular of the azimuthal anisotropy associated with elliptic flow, require 
finite nuclei and realistic nuclear matter distributions within each nucleus. These requirements were 
discussed in Refs.~\cite{AYR02a,AYR02b}.

\begin{figure}[htb]
\begin{center}
\includegraphics[width=0.5\textwidth]{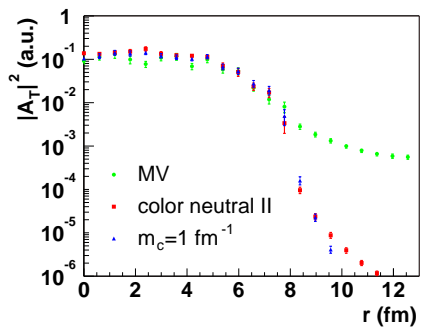}
\caption{Gluon field as a function of radial distance.
Original MV model is shown by circles,
while squares correspond to the Color Neutral II prescription (see text).
The triangles represent results from solution of Poisson's equation with a 
screening mass. The results are for  $\Lambda_{s0}R = 37$.
}
\end{center}
\label{magfield}
\end{figure}

The problem with color neutrality for a finite nucleus can be stated simply as follows. 
If we impose the simple and obvious constraint that the color charge distribution must 
be zero outside the nucleus, the solution of Poisson's equation 
can still give a non-zero gluon distributions outside the nucleus.
In two dimensions, the fall-off of the gluon field is rather slow as shown 
in Fig.~\ref{magfield}.
This slow fall-off is a problem for a finite nucleus since the gluon field is 
associated with a non-zero field strength. Clearly the simple prescription 
for color neutrality 
is not sufficiently stringent.

A more realistic prescription would be to apply the color neutrality
constraint already at the nucleon level.  Our numerical procedure to implement
the constraint for finite nuclei is as follows.  We first sample $A$ nucleons
on a discrete lattice requiring that they satisfy a Woods-Saxon nuclear density profile in the transverse 
plane.  Note
that this procedure generates the same distribution in the continuum as 
$\Lambda_s^2 (\bm{x}_\perp) = \Lambda_{s0}^2 T_A(\bm{x}_\perp)$ 
where $T_A(\bm{x}_\perp) = \int^{\infty}_{-\infty} dz\,
\kappa(\bm{r})$ is a thickness function, $\bm{x}_\perp$ is the transverse coordinate
vector (the reference frame here being the center of the nucleus), $\kappa(\bm{r})$ is the Woods-Saxon nuclear
density profile, and $\Lambda_{s0}^2$ is the color charge squared per unit
area in the center of each nucleus. The only external dimensional variables in
the model are $\Lambda_{s0}$ and the nuclear radius $R$.

Next, Gaussian color charge distributions are generated on the lattice. The probability 
distribution of color charge in a nucleon is expressed as 
\begin{equation}
P[\rho] = \exp\left(-\sum_j^N { \rho_j^{2}
   \over  2\Lambda_{n,j}^2} \right) \, ,
\label{eqn2}
\end{equation} 
where $\Lambda_{n,j}^2$ is the color charge distribution squared, per unit area, of a nucleon at a 
lattice site $j$ and $N$ is the number of lattice sites that comprise a nucleon. 
$\Lambda_{n,j}^2$ is obtained from $\Lambda_{s0}^2$ by assuming that 
the color charges of the nucleons add incoherently.
There are two versions of the subsequent step. In the first (which we term 
Color Neutral I), we subtract from every $\rho_j$ the spatial average
$\sum_j \rho_j/N$ in order to guarantee color neutrality 
$\langle \rho \rangle = 0$ for each 
nucleon.
In the second (termed 
Color Neutral II), the dipole moment $\bf d$
of each nucleon is eliminated in a similar manner.

\begin{figure}[htb]
\begin{center}
\includegraphics[width=0.5\textwidth]{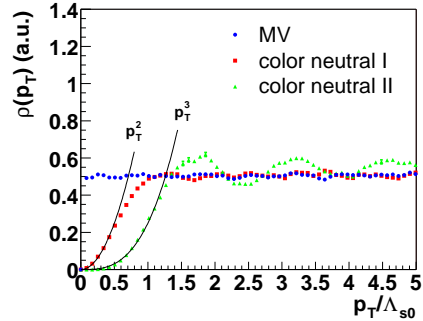}
\caption{Color Charge Correlator in momentum space.  Original MV model is
shown by circles, while squares correspond to the result from color neutrality
condition I and triangles correspond to color neutrality condition II (see
text).  The color charge correlator is plotted versus $p_\perp$ in units of
$\Lambda_{s0}$.  }
\end{center}
\label{colcorr} 
\end{figure}

In Fig.~\ref{colcorr}, we plot the Fourier transform of 
the charge correlator, which in the continuum is defined as 
\begin{equation}
\tilde \rho(p_\perp) = \,\int d^2 x_\perp \exp(i\bm{p}_\perp\cdot\bm{x}_\perp)
 \langle \rho^a(x_\perp) \rho^a(0)\rangle \, ,
\end{equation}
for the MV model and for the two variants
 which impose color neutrality on the nucleon level. 
In the MV model, $\tilde \rho (p_\perp)$ is a constant everywhere
except at $p_\perp=0$ where it is constrained to be zero
from the global charge constraint. In the Color Neutral I (II) variant,
 we see that $\tilde \rho (p_\perp)\sim p_\perp^2$ ($\sim p_\perp^3$) 
for small momenta $p_\perp < \Lambda_{s0}$ and is constant at larger momenta.
The oscillatory behavior seen for Color Neutral I and II is due to the fact that
the correlator in coordinate space is not strictly a delta-function.
In the coordinate space the charge correlator for the two models,
Color Neutral I and II, falls off rapidly, as $\sim 1/x_\perp^4$ and $\sim 1/x_\perp^5$ respectively, at larger distances.

It is an interesting coincidence that the behavior of $\tilde \rho(p_\perp)$ in our model is similar to 
the behavior expected from the renormalization group (RG) evolution of 
color charges in the McLerran-Venugopalan model as discussed here in Section 3.10.
In Refs.~\cite{SAT,Mueller2002}, it is shown that the screening of color charges
due to the RG evolution gives a behavior
$\tilde \rho(p_\perp)\sim p_\perp^2$ for $p_\perp\leq \Lambda_{s0}$
(and $\tilde\rho(p_\perp)$=constant for $p_\perp >\Lambda_{s0}$).

\bigskip
\noindent {\bf iii) Numerical Results for Distributions in Energy and Number in Central Collisions}
\smallskip

We will now discuss results for central collisions of very large cylindrical nuclei. We will see later 
that they are not very different from those for central collisions of realistic nuclei. 
There are only two free parameters for the problem of nuclear collisions as formulated 
int the EFT. One is the saturation scale $\Lambda_s$ while the other is the nuclear radius 
$R$~\footnote{Strictly speaking, the saturation scale is not a free parameter since it can be determined from 
the gluon density as in Ref.~\cite{AGL,KM98}. However, for the momentum scales of interest, there is much uncertainity 
in the gluon density. Moreover, for nuclei, gluon shadowing contributions are not under control. 
Our results will therefore be non-perturbative formulae valid for a wide range of $\Lambda_s$.} 
(For an infinite cylindrical nucleus, one has $\pi R^2 = L^2$, where $L$ is the lattice size.)
Any dimensional quantity $P$, well defined within the EFT, can be
written in terms of the physically relevant parameters $\Lambda_s$ and $R$ 
as $\Lambda_s^d\,f_P (\Lambda_s R)$, where $d$ is the dimension of
$P$. The non--trivial physical information is therefore contained
in the dimensionless function $f_P (\Lambda_s R)$.  On the lattice, $P$
will generally depend on the lattice spacing $a$; this dependence can
be removed by taking the continuum limit $a\rightarrow 0$.  
The broad range of physically relevant values of $\Lambda_s$ for RHIC and LHC
energies are $\sim$ 1-2 GeV and 2-4 GeV respectively-corresponding to $\Lambda_s R\approx 30$-$120$ 
approximately\footnote{If we extrapolate from the Golec-Biernat 
HERA parametrization, we get $\Lambda_s=1.4$ GeV for RHIC and $\Lambda_s=2.2$ GeV for LHC.}. Also,
for central Au--Au collisions, we obtain $L=11.6$ fm as the
physical linear dimension of our square lattice.  

\begin{figure}[htb]
\begin{center}
\includegraphics[width=0.5\textwidth]
{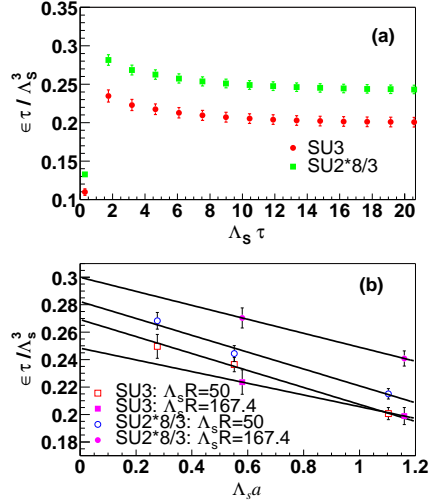}
\caption{
(a) $\varepsilon\tau/\Lambda_s^3$ as a function of $\tau\Lambda_s$ for
$\Lambda_s R = 167.4$.
(b) $\varepsilon\tau/\Lambda_s^3$ as a function of $\Lambda_s a$ for
$\Lambda_s R = 167.4$ (squares) and 50(circles), where $a$ is
the lattice spacing.
Lines are fits of the form $a-bx$.
}
\end{center}
\label{etau}
\end{figure}
For the transverse energy of gluons we get, on purely dimensional grounds,  
\begin{equation}
 {1\over \pi R^2}{dE_T\over d\eta}|_{\eta=0}
     = {1\over g^2}f_E(\Lambda_s R)\Lambda_s^3,
 \label{eq:dedy}
\end{equation}
The function $f_E$ is  
determined non-perturbatively as follows. In Figure~\ref{etau} (a),
 we plot the Hamiltonian density,
for a particular fixed value of $\Lambda_s R=167.4$ (on
a $512\times 512$ lattice) in dimensionless units as a function of the
proper time in dimensionless units. We note that $\varepsilon\tau$ converges very rapidly to a constant
value. The form of $\varepsilon \tau$ is well parametrized by the
functional form $\varepsilon\tau= \alpha + \beta\exp(-\gamma \tau)$. Here
$dE_T/d\eta/\pi R^2 = \alpha$ has the proper interpretation of being
the energy of produced gluons per unit area per unit rapidity, while
$\tau_D=1/\gamma/\Lambda_s$ is the ``formation time'' of the produced
glue. 

In Figure~\ref{etau} (b),
 the convergence of $\alpha$ to the continuum limit
is shown as a function of the lattice spacing in dimensionless
units for two values of $\Lambda_s R$. 
In Ref.~\cite{AR00}, this convergence to the continuum limit was 
studied extensively in an SU(2) gauge theory for very large lattices (up to $1024\times 1024$ sites) 
and shown to be linear. The trend is the same for the SU(3) results-thus, 
despite being further from the continuum limit for SU(3) (due to the 
significant increase in computer time) a linear extrapolation is justified. 
We can therefore extract the continuum value for $\alpha$. 

Numerical results for the physically relevant RHIC and LHC initial conditions were discussed in several 
papers~\cite{AR00,AR01,AYR01,AYR02a,AYR02b}. Very recently, Lappi pointed out that the overall normalization 
in Eq.~\ref{eq:dedy} was incorrect and the $f_E$ computed in Refs.~\cite{AR00,AYR01} is too large by a 
factor of two~\cite{Lappi,AYR03b}. As we will discuss below, the normalization for the number is correct. 
Wherever feasible, the corrected results will be discussed below. 
We find $f_E(50)=0.269$ and $f_E(167.4)=0.248$.
The RHIC and LHC values likely lie in this 
wide range of $\Lambda_s R$.
The SU(2) value is approximately half the SU(3) value. 
Note that the variation of $f_E$ as a function of $\Lambda_s R$ is extremely weak.The formation time 
$\tau_D=1/\gamma/\Lambda_s$ is essentially the same for for both the SU(3) and SU(2) cases-for 
$\Lambda_s R = 167.4$, $\gamma=0.362\pm 0.023$ and . As discussed 
in Ref.~\cite{AR00}, it is $\sim 0.3$ fm for RHIC and $\sim 0.13$ fm for LHC 
(taking $\Lambda_s=2$ GeV and $4$ GeV respectively-for the values of $\Lambda_s$ extracted using the Golec-Biernat 
fit, the corresponding times are of course larger).

We now combine our expression in Eq.~(\ref{eq:dedy}) with our 
non-perturbative expression for the formation time to obtain a 
non-perturbative formula for the initial energy density~\cite{AYR03b},
\begin{equation}
\varepsilon = {0.08\over g^2}\, \Lambda_s^4
\label{eq:edens}
\end{equation}
This formula gives a rough estimate of the initial energy density, at 
a formation time of $\tau_D = 1/\bar{\gamma}/\Lambda_s R$ where we have 
taken the average value of the slowly varying function $\gamma$ to be 
$\bar{\gamma}\sim 0.3$. For $\Lambda_s=1.4$ GeV, one obtains $\varepsilon\sim 10$ GeV/fm$^3$.

We now report our results for the initial multiplicity of gluons
produced at central rapidities. First consider
a free field theory whose Hamiltonian in momentum space has the form
\be
H_f = {1\over 2}\sum_k\,\left(|\pi(k)|^2 + \omega^2 (k)\,|\phi(k)|^2\right)\, ,
\ee
where $\phi(k)$ is the $k$th momentum component of the field, $\pi(k)$
is its conjugate momentum, and $\omega(k)$ is the corresponding
eigenfrequency.  The average particle number of the $k$-th mode is
then
\begin{equation}
N(k)=\omega(k)\langle|\phi(k)|^2\rangle
=\sqrt{\langle|\phi(k)|^2|\pi(k)|^2\rangle},\label{nfree} \, .
\end{equation}
In our case, the average $\langle\rangle$ is over the initial
conditions.

We use two different generalizations of the particle number to an
interacting theory. We have verified that the two definitions agree in
the weak-coupling regime at late times~\cite{AR00}.  Our first
definition is based on the behavior of a free-field theory under
cooling. We obtain~\cite{AR01}
\begin{equation}
N=\sqrt{8\over\pi}\int_0^\infty{{{\rm d}t}\over\sqrt{t}}\,V(t)\,,\label{ncool}
\end{equation}
where $t$ is the cooling time ({\it not to be confused with real or proper time}), 
and $V(t)$ is the potential energy of the relaxed free field after cooling. 
The relaxed potential $V(t)$ is
gauge-invariant-hence so is this definition of the
particle number.  This is an attractive feature of the cooling 
method. Unfortunately, it presently only permits
determination of the total particle number and cannot be used to find
the number distribution $N(k_\perp)$.

Our second definition of the multiplicity will enable us to 
compute $N(k_\perp)$.  We impose
the Coulomb gauge condition in the transverse plane,
${\vec\nabla}_\perp\cdot{\vec A}_\perp=0$, and substitute the momentum
components of the resulting field configuration into
Eq.~(\ref{nfree}). We
can determine $N(k_\perp)$ from the rightmost expression of
Eq.~(\ref{nfree}); the middle expression of Eq.~(\ref{nfree}) can then
be used to obtain $\omega(k_\perp)$. 

In Fig.~26(a), we plot the normalized gluon transverse momentum
distributions versus
$k_\perp/\Lambda_s$ with the value $\Lambda_s R = 167.4$, for both the SU(3) and 
SU(2) gauge theories~\footnote{Our results for the number distribution agree with Ref.~\cite{Lappi} if 
$\Lambda_s \rightarrow \Lambda_s/2$ in the ensuing discussion of the $p_\perp$ distributions. See Ref.~\cite{AYR03b} 
for details.}. Clearly, we see that the normalized result for SU(3) is
suppressed relative to the SU(2) result in the low momentum region.
In Fig.~26 (b), we plot the same quantity over a wider
range in $k_\perp/\Lambda_s$ for two values of $\Lambda_s R$. At large
transverse momentum, we see that the distributions scale exactly as
$N_c^2-1$, the number of color degrees of freedom. This is as expected 
since, at large transverse momentum, the modes are nearly
those of non--interacting harmonic oscillators. At smaller momenta,
the suppression is due to non-linearities, whose effects, we have 
confirmed, are greater
for larger values of the effective coupling $\Lambda_s R$. 
\begin{figure}[htb]
\begin{center}
\includegraphics[width=0.5\textwidth]{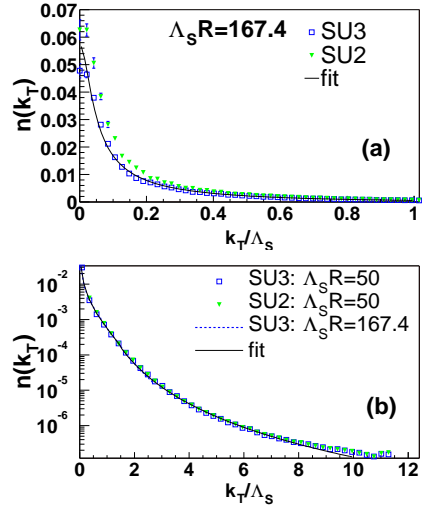}
\caption{
Transverse momentum distribution of gluons, normalized to the color degrees
of freedom, $n\,(k_\perp) = {\tilde f}_n/(N_c^2-1)$ (see Eq.~(\ref{momdis}))
as a function of $\Lambda_S R$ for SU(3) (squares)
and SU(2) (diamonds).
Solid lines correspond to the fit in Eq.(\ref{eq:fit}). 
}
\end{center}
\label{dNdpt}
\end{figure}
The SU(3) gluon momentum distribution can be fitted by
the following function,
\begin{equation}
 {1\over \pi R^2}{dN\over d\eta d^2k_{T}}
     = {1\over g^2}{\tilde f}_n(k_{T}/\Lambda_s)\, ,
\label{momdis}
\end{equation}
where ${\tilde f}_n(k_{T}/\Lambda_s)$ is
\begin{equation}
{\tilde f}_n = \left\{
  \begin{array}{ll}
     a_1\left[\exp\left(\sqrt{k_\perp^2+ m^2}/ T_{\rm eff}\right) -1\right]^{-1}
                       & (k_{T}/\Lambda_s \leq 1.5) \\ \\                

    a_2\,\Lambda_s^4\log(4\pi k_{T}/\Lambda_s)k_\perp^{-4}
                       & (k_{T}/\Lambda_s > 1.5) \\
\end{array} \right.
\label{eq:fit}
\end{equation}
with $a_1=0.118$, $m=0.034\Lambda_s$, $T_{\rm eff}=0.47\Lambda_s$,
and $a_2=0.0087$.
At low momenta, the functional form is approximately that of a Bose-Einstein
distribution in two dimensions even though the underlying dynamics is that 
of classical fields. The functional form at high momentum is motivated
by the lowest order perturbative calculations~\cite{MikLar,KMW,KovRischke}.

Integrating our results over all momenta, we obtain for the gluon 
number per unit rapidity, the non-perturbative result,  
\begin{equation}
 {1\over \pi R^2}{dN\over d\eta}|_{\eta=0}
     = {1\over g^2} f_N(\Lambda_s R)\Lambda_s^2 \, .
 \label{eq:dNdy}
\end{equation}
We find that $f_N(167.4)=0.3$. Our results for $f_N$ are in reasonable agreement with those of 
Lappi~\cite{Lappi,AYR03b}.
We have checked for an SU(2) gauge theory that the results for a wide range of $\Lambda_s R$ vary 
on the order of $10$\%. The results from the Cooling and Coulomb methods also show very good 
agreement (less than $10$\%) especially for larger values of $\Lambda_s R$. 

If we take the ratio of Eq.~(\ref{eq:dedy}) to Eq.~(\ref{eq:dNdy}), we find that the {\it initial} 
transverse energy per gluon is $dE_T/d\eta/dN/\eta|_{\eta=0}=\frac{f_E}{f_N}\, \Lambda_s\equiv 0.88\, \Lambda_s$. 
If we take $\Lambda_s = 1.4$ GeV, we find that the energy per gluon is $\sim 1.23$ GeV-about a factor of 2 
larger than the value for charged hadrons measured at RHIC. 

The topological charge generated in the initial stages of a heavy ion collision can also be computed in 
the classical CGC framework. An interesting result is that if strict boost invariance is a good 
assumption sphaleron transitions are suppressed~\cite{ADR}. This is because the Chern--Simons number in this 
case is invariant under all rapidity independent gauge transformations. 
The primary mechanism for the generation of topological charge at the
early stage is then by fluctuations of the color electric and magnetic fields. 
It was shown in Ref.~\cite{ADR} that the topological charge generated in this manner, at the early stage, 
are small. These results may be relevant for 
the formation of $P$ and $CP$-odd metastable states in the late stages of heavy ion 
collisions~\cite{KharzeevPisarski}. 

\bigskip
\noindent {\bf iv) Numerical Results for Centrality Dependence of Energy, Multiplicity and Elliptic Flow}
\smallskip

For realistic nuclei, the non-perturbative relations discussed in section 5.2.3 
are less simple. One can write Eq.~\ref{eq:dNdy} more generally as 
\begin{equation}
{dN_g\over d\eta}
    = {f_N(b) \over g^2} {\Lambda_{s0}^2\over \rho_0}N_{part}(b),
\label{eq:npart}
\end{equation}
where $\rho_0 = {\tilde{\rho}}(0,0) = 4.321 { fm}^{-2}$
 and $N_{part}=\int d^2 x_\perp {\tilde{\rho}}(b,x_\perp)$.
For explicit expressions for $f_N(b)$ for different values of $\Lambda_{s0}$, see Ref.~\cite{AYR02b}.
One can similarly compute $dE_g/d\eta$ for finite nuclei. 

\begin{figure}[htb]
\begin{center}
\includegraphics[width=0.5\textwidth]{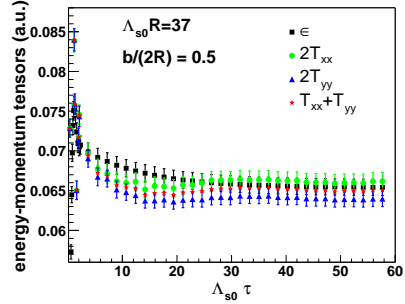}
\caption{The two components of the transverse pressure
 ($T_{xx}$ and $T_{yy}$) and the energy density 
$\epsilon$ plotted as a function of $\tau$ in dimensionless units.
 The results are for a impact parameter $b=R$
 and dimensionless coupling $\Lambda_{s0}R=37$.
Also shown is the sum of the two transverse pressures.}
\end{center}
\label{stress}
\end{figure}

The azimuthal anisotropy in the transverse momentum distribution
is a sensitive probe of the hot and dense matter
produced in ultra-relativistic heavy ion collisions~\cite{Ollitrault,Sorge}. 
A measure of the azimuthal anisotropy is the second Fourier coefficient of 
the azimuthal distribution, the elliptic flow parameter $v_2$. 
Its definition is~\cite{Voloshin} 
\begin{eqnarray}
v_2 = \langle \cos(2\phi) \rangle 
= \frac{ \int^{\pi}_{-\pi} d\phi \cos(2\phi) \int d^2p_{T}
                \frac{d^3N}{dy d^2p_{T}d\phi}}
       {\int^{\pi}_{-\pi} d\phi  \int d^2p_{T}
                \frac{d^3N}{dy d^2p_{T}d\phi}  }.
\label{eqn1}
\end{eqnarray}

The classical Yang--Mills approach may be applied to compute elliptic
flow in a nuclear collision. For peripheral nuclear collisions, the 
interaction region is a two-dimensional almond shaped region, with the $x$ axis lying 
along the impact parameter axis and the $y$ direction perpendicular to it and to the 
beam direction. Even though 
large electric and magnetic fields (and the corresponding transverse components of the 
pressure in the x and y directions) are generated over very short time scales $\tau\sim 1/\Lambda_s$, 
the significant differences in the pressures, responsible for elliptic flow, 
are only built up over much longer time scales $\tau \sim R$. This can be seen in  
Fig.~27 where we plot the two transverse components of the pressure ($T_{xx}$ and $T_{yy}$) 
and the energy density as a function 
of proper time (in units of $\Lambda_{s0}$) for a peripheral nuclear collision. Moreover, the elliptic flow 
is generated by soft modes $p_{T}\sim \Lambda_s/8$. Our result has important consequences for 
the theoretical interpretation of the RHIC data-these will be discussed later in the text. 

The elliptic flow, defined by Eq.~(\ref{eqn1}), can be computed, as in the case of the gluon multiplicity, 
in two different ways; directly in Coulomb Gauge (CG)
and by solving a system of relaxation (cooling) equations for the fields.
It is easy to show that $v_2\,N$, $N$ being the total gluon number, can
be reconstructed from the cooling time history of $T_{xx}-T_{yy}$, just as $N$
can be reconstructed from that of the energy functional~\cite{AR01}:
\begin{equation}
   v_2\, N = \sqrt{\frac{2}{\pi}}\,
\int^{\infty}_0 \frac{dt}{\sqrt{t}} (T^{xx}(t) - T^{yy}(t)) \, . 
\label{v2cool}
\end{equation}
This expression for $v_2\, N$ is manifestly gauge invariant.

\begin{figure}[htb]
\includegraphics[width=0.5\textwidth]{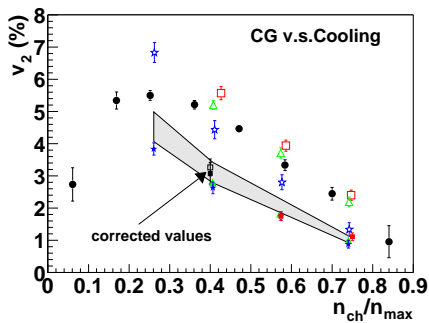}
\caption[]{The centrality dependence of 
$v_2$ at the earliest times  
is computed using cooling (open symbols) and CG 
(filled symbols). Results are for 
$\Lambda_{s0}R$ spanning the RHIC-LHC range, specifically, 
$\Lambda_{s0} R=37$ (squares), 74 (triangles), and 148 (stars). 
Full circles denote {\it preliminary} STAR data~\cite{Snellings}. 
The band denotes the estimated value of $v_2$ when extrapolated to 
very late times. ``Corrected values'' denotes the late time cooling and CG 
result for $\Lambda_{s0}R=37$ at one centrality value. 
}
\label{v2CentDep}
\end{figure}

In Fig.~\ref{v2CentDep} we plot $v_2$ reconstructed from the cooling
time history of only the potential terms in 
$T_{xx}-T_{yy}$, along with the CG values (also including potential terms only)
as a function of $n_{ch}/n_{tot}$ for different 
values of $\Lambda_{s0} R$ as discussed in the figure. The systematic 
errors represented by the band (for $\Lambda_{s0}=37$ 
are primarily due to limited resources 
available to study the slow convergence of the cooling and CG  
computations. We have studied the late time behavior of $v_2$
for one impact parameter-the results are shown in the figure.

The asymptotic values of $v_2$, as predicted by the model, undershoot the
data. For a fixed impact parameter, the model predicts that, as
$\Lambda_{s0}R\rightarrow \infty$, the classical
contribution to the elliptic flow goes to zero. This is because 
increasing $\Lambda_{s0} R$ is equivalent to increasing
$R$ for fixed $\Lambda_{s0}$ and therefore reducing the initial anisotropy.

\begin{figure}[htb]
\begin{center}
\includegraphics[width=0.5\textwidth]{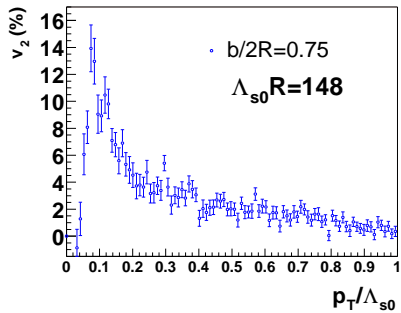}
\caption{$v_2(p_\perp)$ as a function of transverse momentum
 in dimensionless units for $\Lambda_{s0}R=148$.
}
\end{center}
\label{v2dndpt512}
\end{figure}

In Fig~\ref{v2dndpt512}, $v_2 (p_\perp)$ is plotted for
$b/2R=0.75$ for $\Lambda_{s0}R=148$.  Our calculations show that the
elliptic flow rises rapidly and is peaked for $p_\perp\sim
\Lambda_{s0}/8$ before falling rapidly.  The theoretical
prediction~\cite{DerekRaju} is that for $p_\perp \gg \Lambda_{s0}$,
$v_2(p_\perp)\sim \Lambda_{s0}^2/p_\perp^2$.  The lattice numerical data
appear to confirm this result-better statistics are required to
determine the large momentum behavior accurately. The dominant contribution of very
soft modes to $v_2$ helps explain why the cooling and CG 
computations differ until very late times. The soft gluon modes have
large magnitudes and therefore continue to interact strongly until
very late proper times. Concomitantly, the occupation number of these modes 
is not small and the classical approach may be adequate to describe these 
modes even at the late times considered. 

\subsection{Melting the Color Glass Condensate at RHIC}

\begin{figure}[htb]
\begin{center}
\includegraphics[width=0.5\textwidth]{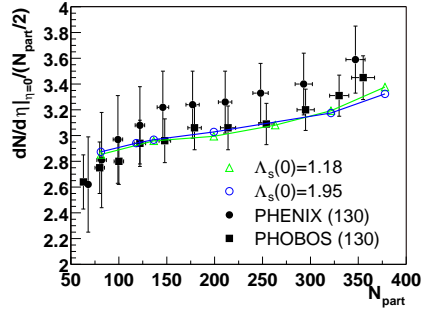}
\caption[]{Comparison of the centrality dependence of the gluon distribution
from SU(3) lattice results to data from
experiments~\cite{phenix_mult,phobos130}.  The strong coupling constant is
fixed to the value $g^2=4$.  The lattice results for $\Lambda_s(0)=1.18$ GeV
and $\Lambda_s(0)=1.95$ GeV
are multiplied by a factor 2.4 and 1.1, respectively.
}
\end{center}
\label{multcentdep}
\end{figure}
%
%
In Fig~\ref{multcentdep},
we plot the computed centrality dependence of gluons together with the experimental data
from PHOBOS~\cite{phobos130} and PHENIX~\cite{phenix_mult}.
We assume here that the charged particle multiplicity is
two thirds of the gluon number. 
The classical computation is performed for fixed $\alpha_s$;
the centrality dependence, as seen from Eq.~(\ref{eq:npart}),
comes from the dependence of $f_N$ on the impact parameter.
In Ref.~\cite{AR01},
it was shown that $f_N\equiv f_N(\Lambda_s R)$ increases slowly
with $\Lambda_s R$-hence one expects it to vary with impact parameter.
We see that the results agree reasonably well with the data.

\begin{figure}[htb]
\begin{center}
\includegraphics[width=0.5\textwidth]{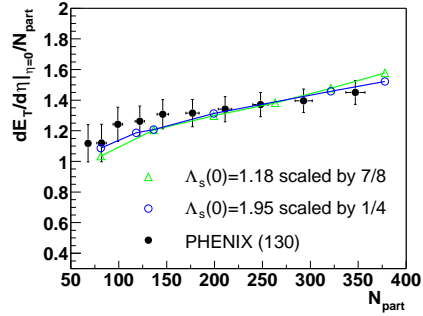}
\caption[]{Comparison of the centrality dependence of the
gluon transverse energy distribution from SU(3) lattice results
to data from experiments~\cite{phenix_et}.
The strong coupling constant is fixed  to the value $g^2=4$.
The lattice results for $\Lambda_{s}(0)=1.95$ GeV
are scaled by ${1\over 4}$ while those for $\Lambda_s(0) =1.18$ are scaled 
by ${7\over 8}$.
}
\end{center}
\label{etcentdep}
\end{figure}

The centrality dependence of the transverse energy is studied in Fig.~\ref{etcentdep}.
As in the case of the multiplicity, even though the absolute normalization is 
strongly dependent on one's choice of $\Lambda_s$, the centrality dependence is 
very similar for the two $\Lambda_s$'s and shows reasonable agreement with the data.

Let us now compare our results with those derived previously by Kharzeev and Nardi in Ref.~\cite{KN} and discussed
further in Ref.~\cite{BMSS2}.  In these works, one obtains in terms of
${\bar{Q_s}}$, the average saturation scale, the result 
\begin{eqnarray}
{dN_g\over d\eta} = c_N \frac{N_c^2-1}{4\pi^2 N_c} \int d^2x_\perp {Q_s^2(b,x_\perp)
\over \alpha_s}\approx c_N xG(x,{\bar {Q}_s}^2(b))\frac{N_{part}}{2} \, .  
\end{eqnarray} 
In the leading logarithmic
approximation, if $c_N$ is constant, one obtains a logarithmic dependence on
the centrality entirely from $xG(x,{\bar{Q}_s^2(b)})$.  One could thus
attribute the logarithmic behavior at the classical level to fixed $\alpha_s$
and leading logarithmic behavior of the gluon distribution function or
equivalently, at higher order, to the one loop running of $\alpha_s$. In the
physically interesting regime, it is difficult to distinguish between the two.
A reasonable agreement with the data is also seen in this formulation of the
problem.

The relation between the two formulations is as follows. 
What we simulate numerically is the color charge squared per unit area,
$\Lambda_s^2$~\cite{AR01}. The saturation scale, on the other hand, is
a scale determining the behaviour of the gluon number
distribution~\cite{KM98}. The relation between the two is given in Eq.~\ref{eq:qsls}. 
The relation between $c_N$ and $f_N$ is then 
$c_N=4\pi^2 f_N / (N_c^2-1)/ \ln(Q_s^2/\Lambda_{QCD}^2)$. Therefore,
if $c_N$ is to be a constant, $f_N$ increases logarithmically with
$Q_s$. A weak rise in $f_N$ is seen in our simulations. If the
infrared scale, as argued in Ref.~\cite{KIIM}, is a number of order
$O(1/Q_s)$, we would have $\Lambda_s\approx Q_s$. This is indeed the case in practice~\cite{AYR02b}.
 
The initial transverse 
energy per particle is $E_T/N\sim 0.88 \Lambda_s$, which for the Golec-Biernat value of $\Lambda_s=1.4$ GeV 
(also the value favored by Refs.~\cite{KN,KLN}), gives $E_T/N=1.23$ GeV. This is a factor of 2 larger
than the value for measured charged hadrons~\cite{phenix_et}. For this value of $\Lambda_s$, the initial number 
of gluons is approximately half the multiplicity of hadrons at central rapidity.  
The excess ratio of $E_T/N$ in the initial state 
can be reduced by a) inelastic fragmentation of partons, which increases $N$ and b) thermalization and hydrodynamic 
expansion expansion which increases $N$ and decreases $E_T$ respectively. Thermalization increases the 
particle number because it is driven primarily by inelastic partonic processes~\cite{BMSS}.
It is very conceivable that the multiplicity increases by a factor of two due to either a) or b) or both. 
Thermalization increases the 
particle number because it is driven primarily by inelastic partonic processes~\cite{BMSS}.

The $p_\perp$ distributions from the melting of the CGC were discussed previously. The distributions obtained numerically 
from purely classical considerations can be used as initial conditions in parton transport equations which describe 
the evolution of the system when the description in terms of classical fields is inadequate. The parton distributions 
exhibit geometric scaling-they are functions of $p_\perp/\Lambda_s$ alone. (At large $p_\perp$ ($p_\perp>> \Lambda_s$), one 
expects this classical geometric scaling break down and distributions to be described by perturbative QCD.)
It has been argued that the RHIC data demonstrate scaling behavior in a broad kinematic region in 
$p_\perp$~\cite{JDLR,HwaYang}. A very similar scaling is seen in the string percolotion framework~\cite{Percolation}.  
In Ref.~\cite{DLL}, a novel mechanism for nuclear collisions was suggested, where, due to quantum evolution of saturation 
effects in the wavefunctions of the incoming nuclei, the $p_\perp$ window, where geometric scaling holds, may extend to 
$p_\perp > Q_s^2/\Lambda_{QCD}$. This mechanism has precisely the same origin as the geometric scaling discussed at 
length in Sect. 3.4 and which was suggested in Sect. 4.3   
to be an explanation the large scaling window observed in Deeply Inelastic Scattering at HERA. 
As a consequence of geometric scaling, the 
$p_\perp$ distributions in AA-collisions at RHIC (and more strikingly at LHC) scale with centrality $\propto N_{part}$. The 
boundary of this region is estimated to be $\sim 4$ GeV at RHIC and $\sim 9$ GeV at LHC. The CGC+geometric scaling 
picture therefore explains the qualitative behavior of the quenching seen in central collisions at RHIC~\cite{PHENIXPT}. 
For peripheral collisions, the saturation scale is small: the window for geometric scaling shrinks and the binary 
scaling of pQCD is restored just as observed in the data. 

This CGC+geometric scaling picture of Ref.~\cite{DLL} is in striking contrast to the 
QGP+energy loss description of the quenching 
of spectra seen in central AA-collisions~\cite{GyulassyVitev}. The two scenarios, at face value, make very different 
predictions for quenching in d-Au collisions. The former predicts quenching in roughly the same $p_\perp$ range with 
the centrality dependence $(N_{part})^{1/2}$; the latter predicts a ``Cronin-enhancement'' just as seen in lower 
energy p-A collisions. However, multiple-scattering of the Cronin variety is also, in principle, present in the 
CGC picture. The magnitude of this effect must be quantified before we rush to any conclusions about the RHIC data. 
We will discuss p-A collisions further in Section 5.5.

We now turn to the theoretical interpretation of the RHIC $v_2$ data in the 
CGC approach. It is clear from Fig.~\ref{v2CentDep} that our result for $v_2$ contributes 
only about 50$\%$ of the measured $v_2$ for various centralities. 
Our $p_\perp$ distributions also clearly disagree with experiment~\cite{STARv2,Voloshin:2002ii}.
Naively, one could argue that the classical Yang-Mills approach is only applicable at early times 
so additional contributions to $v_2$ will arise from later stages of the collision. While there is 
merit in this statement, it is also problematic as we will discuss below.
The reason the situation is complex is as follows. We observed that it takes a long time $\tau\sim R$ 
to obtain a significant elliptic flow in the CGC. At these late times, one would expect that the classical approach 
would be inapplicable due to the rapid expansion of the system. On the other hand, we have seen that 
$v_2$ in the classical approach is dominated by soft modes which are strongly interacting and don't 
linearize even at time scales $\tau\sim R$. Clearly, the soft modes cannot be treated as on-shell partons 
even at times $\tau\sim R$! This conceptual problem will be discussed further in Section 5.4. The observation 
in Ref.~\cite{AYR02a} that the CGC alone is unable to explain the low $p_\perp$ behavior of the RHIC $v_2$ data is, 
thus far, the strongest argument for the importance of final state effects at RHIC.

Finally, we note that $v_2$ is extracted   
from a variety of techniques-in addition to a reaction plane method, two and four particle cumulant
analyses are used~\cite{Nicola}. As we noted previously, the CGC $p_\perp$ distribution at high $p_\perp$ decays as $1/p_\perp^2$ 
while the data is flat up to $p_\perp\sim 10$ GeV. Recently, non-flow 
two particle correlations in the CGC model were suggested as an explanation of the $v_2$ data~\cite{YuriKyrril}. 
It is unclear at present whether non-flow correlations can explain
other features of the measured azimuthal anisotropy. In the approach reported previously,
a procedure very similar to the experimental approach can be followed
and two and four particle correlations can be determined~\cite{AYR03a}. It is not yet clear whether the results 
of this study would affect our above conclusion about the relevance of final state effects in heavy ion collisions. 

  \subsection{Equilibration and the Quark Gluon Plasma}

An outstanding issue in the physics of heavy ion collisions is whether the hot and dense matter formed 
equilibrates to form a Quark Gluon Plasma (QGP). Hydrodynamic models, which assume local equilibrium, 
have been used to study the evolution of the QGP and are successful in describing some of the data at 
RHIC~\cite{hydro}. It is however not easy to show from first principles that thermalization does occur in the extreme 
conditions of a high energy heavy ion collision. Even though the initial density is very high, the system is 
expanding and becoming dilute very rapidly; moreover, if the momenta of the gluons is large, asymptotic 
freedom dictates that the cross-section of the gluons is small. 

In the CGC approach, we have argued that the energy densities right
after the collision are very high-$\varepsilon\sim 10$ GeV/fm$^3$ at
RHIC and $\varepsilon\sim 60$ GeV/fm$^3$ at LHC energies. These
energy densities are formed very rapidly after the collision-about
$\tau_{formation}=0.4$ fm for RHIC and $\tau_{formation}=0.25$ fm at
LHC using the values of $Q_s$ from the Golec-Biernat parametrization.
At the high energies of interest, a heavy ion collision in the
classical picture is boost invariant~\footnote{This assumption is
modified by quantum effects-however, at central rapidities, boost
invariance is a good approximation.}.  The dynamics of particle
production at early times is therefore purely transverse. Consider a
narrow slice in space-time rapidity, around the central rapidity
$\eta=0$. Partons with any significant longitudinal momentum $p_z$
will not be found in this slice (over time scales of interest)-they
have only transverse momentum $p_\perp\sim Q_s$. Our classical dynamics
only describe the initial transverse dynamics of partons. In
Fig.~27, we see that at a time of $\sim 20
\Lambda_{s0}\tau$ ($\sim 2.8$ fm), $\varepsilon = T_{xx}+T_{yy}$. This
suggests that the partons that were strongly interacting in the
transverse plane are free-streaming by this time.

It is unlikely however that the classical approach is valid at these late times. Shortly after partons are formed 
$\sim 1/Q_s$, they begin to scatter off each other and off the transverse plane. The physics of this process is 
not contained in the classical picture. A first guess is to see if small angle elastic $2\rightarrow 2$ scattering 
will lead to thermalization. Analytical estimates~\cite{Mueller2000} suggest and numerical solutions of the 
Boltzmann equation confirm~\cite{JeffRaj} that the time scale for equilibration due to small angle scattering 
is $\sim \exp(1/\sqrt{\alpha_s})/Q_s$ parametrically. In practice, this time scale is not much smaller than the 
lifetime of the system. Also, as pointed out in Ref.~\cite{AdrianMiklos}, the pressure computed 
from these simulations is smaller than that of an equilibrated QGP.

It was suggested by Baier et al.~\cite{BMSS} that inelastic $2\rightarrow 3$ scattering may provide the dominant 
mechanism driving the system towards equilibrium. Naive power counting suggests that 
this process is suppressed relative to elastic scattering by a power of $\alpha_s$-however, the $2\rightarrow 3$ process 
is more efficient in re-distributing momenta and changing particle number. This scenario for thermalization, 
termed ``bottom-up'' on account of our particular initial conditions, can be outlined briefly as follows. 
As we discussed previously, the classical fields can be linearized
into partons for $\tau > (Q_s)^{-1}$.  The parton distribution can
be expressed as in Eq.~\ref{momdis} with the functional form of the
distribution given in Eq.~\ref{eq:fit}. This description, in weak
coupling, is valid (parametrically) up to a proper time $\tau\sim
(\alpha_s)^{-3/2}/Q_s$. This is the time at which, on account of the
longitudinal expansion, the occupation number of partons $f < 1$. In
a small slice in rapidity, $p_z\sim 1/\tau$ which is small for $\tau >
1/Q_s$. However, $p_z$ is built up gradually through random multiple
scattering: $p_z^2=m_D^2 N_{coll}$, where $m_D$ is the Debye mass and
$N_{coll}$ is the collision rate. Simple estimates of these give
self-consistently $p_z \sim Q_s/(Q_s\tau)^{1/3}$. Thus even though
$p_z$ decreases with $\tau$, it does so at a slower rate than given by
free streaming. The occupation number of the hard gluons (with
$p_\perp\sim Q_s$) is then
\begin{equation}
f={dN_h\over dz S_A Q_s^2 p_z} \approx {Q_s^3\over \alpha_s (Q_s\tau)}
\, {1\over Q_s^2}\, {(Q_s\tau)^{1/3}\over Q_s} \,,
\end{equation}
which gives $f<1$ for $\tau > \alpha_s^{-3/2}/Q_s$. This estimate was
made assuming one had only elastic $2\rightarrow 2$ scattering. There
is also inelastic $2\rightarrow 3$ scattering going on but the number
of these soft gluons $N_s << N_h$ at these times. When $f<1$, this
inequality continues to hold but the soft gluons begin to dominate the
contribution to the Debye screening mass. One obtains $N_s\sim N_h$ at
$Q_s\tau =(\alpha_s)^{-5/2}$. 

For $Q_s\tau > (\alpha_s)^{-5/2}$, most of the gluons are soft. These collide frequently with 
each other and achieve an equilibrated distribution with a temperature $T$. However, there are still 
some hard gluons left which continue to transfer energy to the soft gluons thereby acting as a heat bath. 
Indeed, as a consequence, the temperature of the soft gluons increases linearly initially even though the 
system as a whole is expanding. The temperature finally stops increasing when the hard gluons have lost 
all their energy. This happens at a time $Q_s\tau= (\alpha_s)^{-13/5}$ at a temperature of $T=\alpha_s^{2/5}Q_s$. 
The temperature subsequently decreases as $\tau^{-1/3}$ as one would expect for a fluid undergoing one dimensional 
expansion.

The bottom-up scenario is an attractive one and the authors in Ref.~\cite{BMSS2} have show that the 
centrality dependence of the RHIC experimental data can be fit successfully. Nevertheless, a fully successful 
application of the idea to phenomenology is still remote.
For the various stages of this scenario to be realized, $\alpha_s$ must be very small-likely, much smaller 
than may be realized at RHIC and perhaps even LHC energies. In the bottom-up scenario itself, the final temperature 
and the equilibration time are not determined up to a constant-however this number can in principle be determined 
within the theoretical framework itself. A number that needs to be determined externally is the liberation 
coefficient $c_N$. As we described previously in section 5.2.3 and 5.2.4, we obtain $c_N=0.3-0.5$ depending on 
what $Q_s$ is. From empirical considerations, a larger value $c_N\sim 1$ is favored in the bottom-up picture~\cite{BMSS2}. 
Even for a favorable choice of parameters which fit the data, one finds $\tau_{equil} \approx R/2$ where 
$R=6.8$ fm is the radius of a Gold nucleus. While still smaller than the size of the system, a necessary condition 
for thermalization, it is not smaller by a large enough margin to make it appear inevitable.

The long equilibration times discussed here are problematic for understanding the RHIC data on 
$v_2$. Hydrodynamical models that fit the data require early thermalization times 
of $\tau_{equil}\sim 0.6$ fm. The $v_2$ generated by the CGC alone is not sufficient to explain the data. 
The correct way to treat the theoretical problem may be as follows. Hard modes with $k_\perp\geq \Lambda_{s}$ 
linearize on very short time scales $\tau\sim 1/Q_s$. Their subsequent evolution is treated incorrectly in 
the classical approach, which has them free streaming in the transverse plane. In actuality, as discussed here, they are 
scattering off each other via elastic $gg\rightarrow gg$
 and inelastic $gg\leftrightarrow ggg$ collisions 
which drive them towards an isotropic distribution~\cite{BMSS}. This dynamics would indeed provide an 
additional {\it pre-equilibrium} contribution to $v_2$ and is calculable. An effect to consider here would 
be the possible screening of infrared divergences in the hard scattering by the time dependent classical field. 
More complicated is the effect of these hard modes on the classical dynamics of the soft modes and on 
their possible modification of the contribution of the latter to $v_2$. One has here a little explored 
dynamical analog to the interplay of hard particle and soft classical modes in the 
kinetic theory of Hard Thermal Loops~\cite{BI01,JSRJ}. The overlap between the classical field and Boltzmann pictures 
in this context has been discussed recently in Ref.~\cite{MuellerSon}. A practical issue of interest is to use the 
results of the classical field simulations as the initial input to parton cascade models which simulate the later 
stages of heavy ion collisions~\cite{Bass}.

  \subsection{p--A \& Peripheral A--A collisions at RHIC and LHC}

Proton-Nucleus collisions at RHIC and LHC will provide an excellent
probe of the properties of the CGC.  Deuteron-Gold collisions with the
center of mass energy of $\sqrt{s}=200$ GeV/nucleon at RHIC started in
January 2003 and Proton-Nucleus collisions are an important part of
the program at the upcoming LHC collider at CERN~\cite{CERNyellow}. A
big difference between p/D-A collisions and A-A collisions is that
final state interactions are more important in the latter than in the
former. Thus p/D-A collisions provide an important benchmark to
disentangle novel phenomena such as the Quark Gluon Plasma, which
arise as a consequence of strong final state interactions, from
initial state phenomena which may result from the Color Gluon
Condensate.

A case in point is the remarkable observation of the suppression of
the single particle spectra, as a function of $p_\perp$, in Au-Au
collisions relative to that in pp-collisions (per binary collision) at
$\sqrt{s}=130$ GeV/nucleon and 200 GeV/nucleon~\cite{PHENIXPT}. An
interpretation is that high $p_\perp$ partons traversing a QGP suffer
significant energy loss resulting in fewer high $p_\perp$
partons~\cite{energyloss}. If correct, the observed suppression is
evidence that a QGP has been created in high energy heavy ion
collisions. In contrast, in p/D-Au collsions, it is argued that one
will see the Cronin effect-the ratio of d-Au relative to pp-collisions
will show an enhancement at moderate $p_\perp$, peaking at $p_\perp\sim 3-4$
GeV before going down to unity at larger
$p_\perp$~\cite{GyulassyVitev}. The argument is that energy loss in the
``cold matter" of d-Au collisions is quite small and the $p_\perp$
broadening due to multiple scattering causes the Cronin effect.  On
the other hand, if no Cronin effect is seen and a suppression is seen
instead it may be an indication that initial state effects are
responsible for the phenomenon in both d-Au and Au-Au collisions.

In this section, we will briefly summarize recent work on proton-nucleus collisions in the CGC framework. 
Proton-Nucleus collisions in this framework were 
first considered by Kovchegov and Mueller~\cite{KM98}. The proton in this case was modelled by a gauge invariant 
gluonic current. They showed that one obtains a Glauber-type formula with a saturation scale $Q_s$ (of the nucleus) for the inclusive gluon cross-section. The problem was looked at in more detail in Ref.~\cite{DumitruMcLerran}. The 
problem of proton-nucleus scattering was considered in a manner similar to the classical fields treatment of nuclear 
scattering~\cite{KMW,AR99}, except now they introduced two saturation scales $Q_{s1}$ and $Q_{s2}$ for the nucleon and 
nucleus respectively with $Q_{s1}<<Q_{s2}$. They were able to obtain analytical solutions for classical gluon 
production in the regions $k_\perp > Q_{s2}> Q_{s1}$ as well as for $Q_{s2}> k_\perp > Q_{s1}$. In the former, one 
obtains $dN/d^2 k_\perp d^2 b \propto Q_{s1}^2 Q_{s2}^2/k_\perp^4$ while in the latter case they obtained $dN/d^k_\perp d^2b \propto 
Q_{s1}^2 \ln(k_\perp^2/Q_{s1}^2)/k_\perp^2$. In this latter regime, one is solving the classical equations to all orders in 
$Q_{s2}^2/k_\perp^2$. No analytical solution is available for $k_\perp<Q_{s1}<Q_{s2}$, even when $k_\perp> \Lambda_{QCD}$. The classical problem discussed here can also be formulated numerically~\cite{AYR02b}. The $k_\perp$ dependence of the two 
kinematical regions is seen clearly in the numerical result. It was also argued in Ref.~\cite{DumitruMcLerran} that 
since the rapidity distributions, for fixed $k_\perp$, in one or the other regime are so different, these rapidity 
distributions could be used to isolate the Renormalization Group (RG)-evolution of the partons in both the proton and 
the large nucleus. 

The proton fragmentation region provides an excellent probe of the CGC. Final states measured in this region are 
produced by the scattering of high x partons in the proton off very small x partons in the nucleus. The scattering 
can be described as the convolution of the probability to find a quark in the proton $q(x_1,Q^2)$ times the 
probability for the quark to scatter off the classical field of the nucleus characterized by a saturation scale 
$Q_{s2}(x_2)$~\cite{DumitruJJM,GelisJJM}. As discussed in Ref.~\cite{DumitruMcLerran}, the $k_\perp$ distribution 
for this scattering is modified from the usual tree level pQCD distribution-this modified distribution, when 
convolved with the appropriate fragmentation functions will be reflected in hadronic final 
states~\cite{DumitruJJM,Lenaghan}. Electromagnetic final states such as photons and di-leptons are a particularly sensitive 
probe of saturation dynamics in the proton fragmentation region~\cite{GelisJJM}. Interestingly, the modified 
$p_\perp$ distributions lead to $p_\perp$ broadening of the final state and the Cronin effect~\cite{DumitruJJM,GelisJJM}. This 
effect is also seen in other (dipole) models of saturation~\cite{Raufeisen}.

Though the CGC picture gives rise to the Cronin effect (which can be represented by multiple scattering tree level 
diagrams), quantum effects in the wavefunction can modify this picture significantly. In sections 4.3 and 5.3 
respectively, we discussed 
the CGC+geometric scaling mechanism which was first applied to understand the HERA DIS data and subsequently the 
suppression of  $p_\perp$ spectra of charged hadrons in Au-Au collisions at RHIC. In the geometric scaling regime, 
the anomalous dimensions for the evolution are very close to BFKL anomalous dimensions~\cite{KIIM,MuellerTrianta}-this 
change in the anomalous dimensions (from the DGLAP one) is what causes the suppression of the $p_\perp$ spectra in the 
initial state CGC scenario. A similar suppression must then persist in p/D-A collisions. 
In Ref.~\cite{DLL}, it is 
predicted that the dependence of semi-hard processes on the number of participating nucleons of the nucleus 
in D-A collisions will be $\sim (N_{part}^{Au})^{1/2}$. The number of participants are determined a la the 
Glauber calculation of Kharzeev and Nardi~\cite{KN}. Thus more quantitatively, the suppression in D-Au collisions 
(relative to $pp$ ) at $\sqrt{s}=200$ GeV/nucleon due to the geometric scaling initial state effect 
translates into a 25-30\% of moderately high $p_\perp$ particles in the top 15\% centrality range. This conclusion 
will be modified by the Cronin final state scatterings of Refs.~\cite{DumitruJJM,GelisJJM}--these will change the behavior 
and one may even recover an enhancement. A quantitative calculation including both geometric scaling and 
final state scattering is urgently required.

Very recently, in 
Ref.~\cite{KLN212}, quantitative predictions (along the lines of the qualitative ones in Ref.~\cite{DumitruMcLerran}) 
were made for hadron multiplicities in Deuteron-Gold scattering at RHIC-in particular for the rapidity and centrality 
dependence. The results of the current Deuteron-Gold run should therefore help determine whether a Color Glass Condensate 
is formed already at the moderately high energies at RHIC. 

An important part of the RHIC program is the study of peripheral nuclear collisions. Very intense electromagnetic 
fields are created at these energies and a variety of photon-photon and photon-Pomeron final states can be 
studied~\cite{PERI}. Of particular interest to us is inclusive and diffractive $Q{\bar Q}$-in the color field of 
a nucleus~\cite{GelisPeshier}. The problem is analogous to the photo-production of heavy quark pairs in Deeply 
Inelastic Scattering. The transverse momentum and invariant mass distribution of quark pairs can be shown to 
depend sensitively on the saturation scale $Q_s$. Thus ultra-peripheral nuclear collisions provide an independent 
method to extract properties of Color Glass Condensate.

\section*{Acknowledgements}

We would like to thank our colleagues and friends, J. P. Blaizot,
A. Dumitru, E. Ferreiro, F. Gelis, K. Itakura, J. Jalilian-Marian,
D. Kharzeev, Y. Kovchegov, A. Krasnitz, A. H. Mueller, Y. Nara,
D. Teaney and H. Weigert for their insights and collaborations over
the years. We are especially grateful to L. McLerran whose approach to
physics has had a large influence on our work in general and on this
review in particular.  R.V.'s research was supported by
DE-AC02-98CH10886 and by the RIKEN-BNL Research Center.

\end{document}